\documentclass[12pt,preprint]{aastex}

\slugcomment{}
\shorttitle{FK Comae: COCOA PUFS}
\shortauthors{T.R.\ Ayres et al.}

\begin{document}

\title{FK Comae Berenices, King of Spin\footnote{Trimble et al.\ (2007) have argued that FK~Com (lit.\ star FK belonging to constellation {\em Berenice's Locks}\,) should rather be called ``Queen of Spin,'' given the grammatical gender of the Latin constellation name ({\em Coma}\/ [{\em f.}]= hair), and perhaps also because Berenice herself was a queen.  However, the (grammatical) gender of a star does not follow from its constellation; for example, Canopus ({\em m.}) of Carina ({\em f.}).  Because ``FK'' provides no gender guidance to the contrary, we feel justified retaining ``King.''}{~:}\\ The COCOA-PUFS\footnote{{\normalsize\bf CO}ordinated {\normalsize\bf C}ampaign of {\normalsize\bf O}bservations and {\normalsize\bf A}nalysis, {\normalsize\bf P}hotosphere to {\normalsize\bf U}pper Atmosphere, of a {\normalsize\bf F}ast-rotating {\normalsize\bf S}tar}\,\,\,Project}

\author{Thomas R.\ Ayres}

\affil{Center for Astrophysics and Space Astronomy,\\
389~UCB, University of Colorado,
Boulder, CO 80309;\\ Thomas.Ayres@Colorado.edu}

\author{V.~Kashyap and S.~Saar}

\affil{Harvard-Smithsonian Center for Astrophysics, Cambridge, MA}

\author{D.~Huenemoerder}

\affil{Massachusetts Institute of Technology, Cambridge, MA}

\author{H.~Korhonen}

\affil{Finnish Centre for Astronomy with ESO (FINCA), University of Turku, V{\"a}is{\"a}l{\"a}ntie 20, FI-21500 Piikki{\"o}, Finland\\
and\\
Niels Bohr Institute and Centre for Star and Planet Formation, 
University of Copenhagen, {\O}ster Voldgade 5, 1350 Copenhagen K, Denmark}

\author{J.~J.~Drake, P.~Testa, O.~Cohen, and C.~Garraffo}

\affil{Harvard-Smithsonian Center for Astrophysics, Cambridge, MA}

\author{T.~Granzer and K.~Strassmeier}

\affil{Leibniz-Institut f{\"u}r Astrophysik Potsdam (AIP), An der Sternwarte 16, D-14482 Potsdam, Germany}

\begin{abstract}
COCOA-PUFS is an energy-diverse, time-domain study of the ultra-fast spinning, heavily spotted, yellow giant FK Comae Berenices (FK Com: HD117555; G4 III).  This single star is thought to be a recent binary merger, and is exceptionally active by measure of its intense ultraviolet (UV) and X-ray emissions, and proclivity to flare.  COCOA-PUFS was carried out with Hubble Space Telescope ({\em HST}\,) in the UV (1200--3000 \AA), using mainly its high-performance Cosmic Origins Spectrograph (COS), but also high-precision Space Telescope Imaging Spectrograph (STIS); {\em Chandra}\/ X-ray Observatory in the soft X-rays (0.5--10 keV), utilizing its High-Energy Transmission Grating Spectrometer (HETGS); together with supporting photometry and spectropolarimetry in the visible from the ground.  This is an introductory report on the project.  

FK Com displayed variability on a wide range of time scales, over all wavelengths, during the week-long main campaign, including a large X-ray flare; ``super-rotational broadening'' of the far-ultraviolet (FUV) ``hot-lines'' (e.g., \ion{Si}{4} 1393 \AA; $8\times10^4$ K) together with chromospheric \ion{Mg}{2} 2800~\AA\ and \ion{C}{2} 1335~\AA\ (1--3$\times10^4$~K); large Doppler swings suggestive of bright regions alternately on advancing and retreating limbs of the star; and substantial redshifts of the epoch-average emission profiles.  These behaviors paint a picture of a highly extended, dynamic, hot ($\sim 10$~MK) coronal magnetosphere around the star, threaded by cooler structures perhaps analogous to solar prominences, and replenished continually by surface activity and flares.  Suppression of angular momentum loss by the confining magnetosphere could temporarily postpone the inevitable stellar spindown, thereby lengthening this highly volatile stage of coronal evolution.  
\end{abstract}


\section{INTRODUCTION}

FK Comae is an ultra-fast rotating ($\upsilon\sin{i}\sim 160$ km s$^{-1}$), single yellow giant, thought to be a recent W~Ursa Majoris binary merger (see Eggen \& Iben 1989, and references to previous work therein).   The star is heavily ``spotted'' (see Strassmeier [2009] review of starspots), leading to extraordinary levels of UV and X-ray emissions, and proclivity to flare regularly at high energies.  Not surprisingly, then, FK~Com has a long history of attention at optical, ultraviolet, and X-ray wavelengths.  Its high-energy characteristics rank FK~Com a coronal powerhouse on par with the most extreme of the better known activity heavyweights: short-period RS Canum Venaticorum binaries (e.g., Walter et al.\ 1980).  But, as a single object, FK~Com has clear advantages as a laboratory for exploring the outer limits of magnetospheric activity among the coronal late-type stars. 

The present paper introduces COCOA-PUFS ({\bf CO}ordinated {\bf C}ampaign of {\bf O}bservations and {\bf A}nalysis, {\bf P}hotosphere to {\bf U}pper Atmosphere, of a {\bf F}ast-rotating {\bf S}tar), a project to conduct intensive, contemporaneous, multi-wavelength measurements of FK~Com over several rotational cycles.  The initial section outlines the observational strategy and how it was carried out.  Subsequent sections summarize preliminary results, focusing mainly on variability of UV lines, broad-band coronal X-rays, and the visible photometry.  Future papers will describe the optical supporting observations in more detail, including surface spot models and magnetic maps derived by Zeeman Doppler Imaging (ZDI: e.g., Semel 1989); consider the UV and X-ray line spectra in greater depth; and present theoretical magnetospheric simulations to help better understand the extraordinary properties of this ultra-fast spinning yellow giant.

In the COCOA-PUFS study, the photometric ephemeris of Jetsu et al. (1993) was adopted:
\begin{equation}
HJD=\,\,(2,439,252.895{\pm}0.010)\, +\, (2.4002466{\pm}0.0000056)\,\phi,
\end{equation}
where $\phi$ is the photometric cycle number, and phase 0.0 was the photometric minimum in the original epoch.  (In what follows, $\phi$ is used interchangeably as ``cycle number'' and ``folded phase,'' guided by the context.)   Also, in subsequent diagrams and tables, the photometric cycles are defined relative to cycle~6841 in the Jetsu et al.\ ephemeris, which corresponds roughly to the beginning of the main program of joint {\em HST/Chandra}\/ observations during 2011~April 21--29.

Appendix A is a Glossary of Selected Abbreviations used in the text.  All laboratory wavelengths $< 3200$~\AA\ are quoted in vacuum.

\section{OBSERVATIONS}

\subsection{Previous High-Energy Studies}

The photometric and spectropolarimetric properties of FK~Com in the visible have been described in numerous previous publications (see Vida et al.\ 2015 and references therein), so only the high-energy face of the yellow giant will be summarized here.

In the UV, FK~Com was well observed by the {\em International Ultraviolet Explorer}\/ ({\em IUE}\,) in the 1980's and 1990's, at low resolution ($\sim 6$~\AA) in the FUV, and moderate echelle resolution in the NUV ($\sim 30$~km s$^{-1}$).  Sanad \& Bobrowsky (2014) published integrated flux measurements from the archival {\em IUE}\/ material, and some of the {\em HST}\/ COS pointings described here.  Their conclusions were limited to assessments of the temporal variability, typically factors of a few and generally increasing with ionization temperature (e.g., from \ion{Mg}{2} to \ion{C}{4}: $10^4$--$10^5$~K), qualitatively matching the behavior of the large body of previous ground-based H$\alpha$ measurements.

A relatively short (13~kilosecond [ks]) pointing on FK~Com in 2004 by the {\em Far-Ultraviolet Spectroscopic Explorer}\/ ({\em FUSE}\,) discovered ultra-broad, highly redshifted profiles of \ion{C}{3} 977~\AA\ and \ion{O}{6} 1031~\AA\,+\,1037~\AA\ ($T\sim 0.8$--$3{\times}10^{5}$~K) (Ayres et al.\ 2006).  Unfortunately, the singular observation could not be repeated, owing to subsequent degradation of the satellite's attitude-control system.  Further, the FUV spectra were taken just a few months before {\em Hubble's}\/ Space Telescope Imaging Spectrograph (STIS) suffered an electrical failure in 2004, so at the time there was no opportunity to turn the more powerful gaze of STIS on the remarkable yellow giant.

Prior to COCOA-PUFS, FK~Com had been recorded at least once by all the major X-ray observatories, including {\em ROSAT}\/ (Welty \& Ramsey 1994), {\em ASCA}\/ (Huenemoerder 1996), {\em XMM-Newton}\/ (Gondoin et al.\ 2002), and {\em Chandra}~HETGS (Drake et al.\ 2008).  In all cases, FK~Com not only displayed a high coronal X-ray luminosity ($L_{\rm X}\sim 1{\times}10^{31}$ erg s$^{-1}$; about $10^4\,L_{{\rm X,} \odot}$), but also episodes of flaring on top of the already substantial quiescent level, typically factors of several.  In the HETGS study, there even were hints of Doppler shifted $\sim 1$~keV Fe L-shell lines, although the resolving power was not sufficient to definitively measure excess broadening in the $T\sim 10$~MK coronal features. 

Figure~1 places FK~Comae in the context of other active single F--K giants.  The diagram compares the coronal soft X-ray emission (0.2--2~keV) against the integrated intensity of the subcoronal \ion{Si}{4} 1400~\AA\ doublet ($T\sim 8\times10^4$~K) for a sample of early-F to early-K giants taken from Ayres et al.\ (1998) and Ayres et al.\ (2007).  (\ion{Si}{4} is preferred over [usually brighter] \ion{C}{4} for such flux-flux diagrams of giants, because the carbon abundance can be affected by deep mixing in evolved stars.)  The figure is adapted from Fig.~2 of the first study, adding several stars from the second, and with some updates to the X-ray and FUV fluxes based on more recent archival material (including the {\em Chandra}\/ and {\em HST}\/ measurements of FK~Com described later).  

The yellow wedge approximates the power-law correlations traced by G and K Main sequence stars, with the warmer dwarfs along the lower edge, cooler dwarfs along the upper edge, more active stars to the upper right, and low-activity objects to the lower left.  In addition to ``FK'' for FK~Com, the Sun symbol refers to an average over the past two solar cycles (based on UV and XUV irradiances from the LISIRD database\footnote{See http://lasp.colorado.edu/lisird/ﬁsm}); ``EK'' refers to the hyper-active young solar analog EK~Draconis (G2~V) from Ayres (2015a); ``A'' and ''B'' refer to averages of lower activity $\alpha$~Centauri A (G2~V) and B (K1~V) from Ayres (2015b); and ``24'' is for the yellow giant 24~Ursa Majoris (G4~III-IV according to  Cenarro et al.\ 2009), similar in physical properties to FK~Com (except rotation) and utilized later as a spectral comparison.  The exact positions of the points are not important, but rather the fact that the cooler giants (G1--K0) tend to cluster along the same relationship obeyed by the cooler dwarfs, whereas the warmer yellow giants (F0--G0) tend to fall below the MS G-dwarf trend into an ``X-ray-deficient'' zone (blue oval), although there are a few exceptions that might represent transitions from the X-ray-deficient locus to the cool giant zone (noting that the yellow giants quickly evolve into cooler objects during their rapid transit of the Hertzsprung gap: Ayres et al.\ 1998).  

FK~Com sits in the upper right corner of the diagram, in a region normally reserved for the extremely active, short-period, tidally-synchronized binaries of the RS~CVn class mentioned earlier.  Note, however, that FK~Com falls away from a simple extrapolation of the active cool giant locus, and instead seems to lie on a continuation of the X-ray-deficient trend.  Such behavior among the most active objects often signals a ``saturation'' of their coronal X-rays with respect to subcoronal species like \ion{Si}{4} and \ion{C}{4} (e.g., Vilhu \& Rucinski 1983).  Aside from that detail, the comparison emphasizes the exceptional high-energy status of the ultra-fast spinning yellow giant.

\begin{figure}[ht]
\figurenum{1}
\vskip  0mm
\hskip 12mm
\includegraphics[scale=0.75]{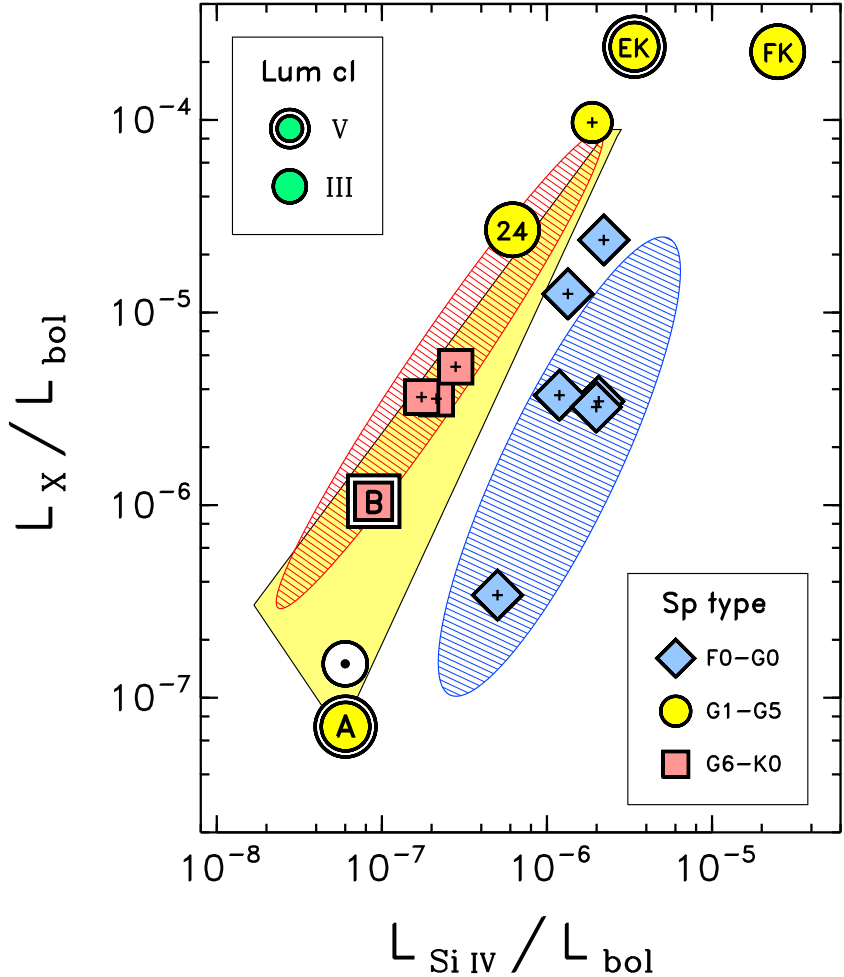} 
\vskip -7mm
\figcaption[]{\small Flux-flux diagram for single giant stars pitting coronal X-rays (0.2--2~keV) against integrated \ion{Si}{4} 1393~\AA\,+\,1402~\AA\ fluxes, both normalized to the stellar bolometric luminosity.  The bolometric normalization allows a fair comparison among stars of different sizes and distances.  Spectral type ranges and luminosity classes are color/symbol-coded according to the legends.  The yellow wedge highlights the trends followed by Main sequence stars, cooler on the upper edge and warmer along the lower.  The red hatched oval approximates the correlation followed by the cooler active giants (G1--K0), and is coincident with that obeyed by the late-G/early-K dwarfs.  The blue hatched oval delimits the ``X-ray-deficient'' zone where fast-rotating late-F/early-G Hertzsprung gap giants often are found.  The Sun sign refers to the average Sun; ``EK'' represents the hyper-active young solar analog EK~Dra; ``A'' and ``B'' are for the low-activity sunlike dwarfs of $\alpha$~Centauri; ``FK'' is for FK~Com, subject of this study; and ``24'' is for comparison yellow giant 24~UMa.
}
\end{figure}

\clearpage
\subsection{COCOA-PUFS}
                                                                                    
The dynamic optical hydrogen emissions of FK~Com recorded from the ground, and the high degree of temporal variability observed in the UV and X-rays from space, were strong motivations to carry out a focused campaign of coordinated multi-wavelength photometry and spectroscopy on this unusual yellow giant, hence COCOA-PUFS.  The top-level objectives of the campaign were to: (1) catalog -- in the visible, UV, and X-rays -- high-energy surface features on FK~Com by means of ``Doppler imaging;'' (2) determine the relationship of these ``active regions'' to the photospheric magnetic spots; (3) measure persistent kinematic properties of the chromospheric and subcoronal line profiles, analogous to the bimodal, dominantly redshifted hot lines of active solar-type stars like EK~Dra; and (4) record transient variability associated with the surface activity, including any large flare events that serendipitously might be captured during the intensive coordinated multi-facility campaign.  
                                                                                   
The COCOA-PUFS project was a combination of two Guest Observer programs, one {\em HST,}\/ the other {\em Chandra;}\/ both joint with the other mission.  Initially, GO-12279 was approved by the {\em HST}\/ Telescope Allocation Committee (TAC), awarded 11 {\em HST}\/ orbits and 120~ks of joint {\em Chandra}\/ time.  In parallel, a Large Project (LP) proposal had been submitted to {\em Chandra}\/ to undertake a similar X-ray/UV time-domain study of FK~Com, but on the larger scale possible for an LP.  A few weeks after the {\em HST}\/ peer review, and with knowledge of the outcome, the {\em Chandra}\/ TAC awarded the LP proposal 240~ks of {\em Chandra}\/ time and 11 additional joint {\em HST}\/ orbits.  The combination of the two joint programs allowed a more comprehensive observational approach than would have been possible with the original GO-12279, ``King of Spin: the Movie,'' alone.  The {\em HST}\/ part of the {\em Chandra}\/ LP was designated GO-12376.                       

COCOA-PUFS exploited the high sensitivity of {\em HST}\/ Cosmic Origins Spectrograph (COS), to record kinematically resolved UV spectra of FK~Com; partnered with the powerful High-Energy Transmission Grating Spectrometer (HETGS) of {\em Chandra,}\/ to capture a diverse range of hot ($\sim 10$~MK) coronal emissions in the 1--30~\AA\ (10-0.5~keV) soft X-ray region.  Contemporaneous ground-based photometry and Zeeman Doppler Imaging supplied a history of the surface arrangement of the dark, magnetized starspots in the weeks leading up to the main campaign, as well as during and after the joint spacecraft operations.  Also, about six months prior to the main campaign, a snapshot FUV spectrum was taken with {\em Hubble's}\/ Space Telescope Imaging Spectrograph (STIS) to assess the \ion{H}{1} Ly$\alpha$ intensity for COS detector safety reasons.  The advance STIS spectrum proved to be crucial later, in a cross-comparison with the COS spectra from the main campaign, for identifying systematic errors in the COS FUV wavelength scales, such as described in the recent STIS/COS study of EK~Dra (Ayres 2015a).

The characteristics and operational capabilities of {\em HST}\/ COS have been detailed in a number of previous publications, especially Green et al.\ (2012); and companion UV spectrograph STIS, by Woodgate et al.\ (1998) and Kimble et al.\ (1998).  The {\em Chandra}\/ HETGS has been described by Canizares et al.\ (2005), among others.

Table~1 is a catalog of observations for {\em HST,}\, and Table~2 for {\em Chandra.}\/  Figure~2 sketches the {\em HST/Chandra}\/ timeline, and Figure~3 is a phase/time map of the main coordinated campaign (2011 April 21--29).  These observing programs, including an overview of the supporting optical work, are described below. 

\begin{deluxetable}{cccclcrlrl}
\rotate
\tabletypesize{\footnotesize}
\tablenum{1}
\tablecaption{{\em HST}\/ Observation Log}
\tablewidth{0pt}
\tablecolumns{10}
\tablehead{
\colhead{Dataset} &  
\colhead{Start Date} & 
\colhead{MJD} &  
\colhead{Cycle} &
\colhead{Mode/Setting} &
\colhead{Aperture} & 
\colhead{$t_{\rm exp}$} & 
\colhead{FP-POS} &
\colhead{Detector} &
\colhead{Notes}\\[3pt]                
\colhead{(1)} &  
\colhead{(2)} &
\colhead{(3)} &  
\colhead{(4)} &
\colhead{(5)} &  
\colhead{(6)} &               
\colhead{(7)} &               
\colhead{(8)} &               
\colhead{(9)} &               
\colhead{(10)}               
}                
\startdata   
obme01010 & 2010-11-21.134 &  55521.134 &  \nodata & G230MB--2836 & 
$52{\times}0.2$ &   60 &  \nodata               & NUV &  \\
obme01020 & 2010-11-21.141 &  55521.141 &  \nodata & E140M--1425  & 
$0.2{\times}0.2$ & 1612 & \nodata               & FUV &  Ly$\alpha$ safety check \\
\hline
lbme02010 & 2011-03-07.042 &  55627.042 &  \nodata & G285M--2676  & 
PSA             &   63 & 3               & NUV & COS ACQ test \\
lbme02020 & 2011-03-07.046 &  55627.046 &  \nodata & G160M--1577  & 
PSA             &  600 & 3,4             & FAB &  \\
lbme02030 & 2011-03-07.063 &  55627.063 &  \nodata & G130M--1291  & 
PSA             &  660 & 3,4             &FA &  \\
\hline
lbme03010 & 2011-04-21.568 &  55672.568 &  0.036 & G285M--2676  & 
PSA             &   64 & 3               & NUV &  \\
lbme03020 & 2011-04-21.571 &  55672.571 &  0.037 & G185M--1900  & 
PSA             &  685 & 3               & NUV &  \\
lbme03030 & 2011-04-21.582 &  55672.582 &  0.042 & G160M--1577  & 
PSA             &  900 & 3,4,1           & FAB &  \\
lbme03040 & 2011-04-21.629 &  55672.629 &  0.061 & G130M--1291  & 
PSA             & 5079 & 3,4,1,2,3,4,1,2 &FA &  \\
\hline
lbkm01010 & 2011-04-21.767 &  55672.767 &  0.119 & G285M--2676  & 
PSA             &   51 & 3               &   NUV   &  \\
lbkm01020 & 2011-04-21.776 &  55672.776 &  0.123 & G130M--1291  & 
PSA             & 1620 & 3,4             &FA &  \\
\hline
lbkm02010 & 2011-04-22.387 &  55673.387 &  0.377 & G285M--2676  & 
PSA             &   51 & 3               &   NUV   &  \\
lbkm02020 & 2011-04-22.427 &  55673.427 &  0.394 & G130M--1291  & 
PSA             & 1620 & 3,4             &FA &  \\
\hline
lbkm03010 & 2011-04-22.965 &  55673.965 &  0.618 & G285M--2676  & 
PSA             &   51 & 3               &   NUV   &  \\
lbkm03020 & 2011-04-22.974 &  55673.974 &  0.622 & G130M--1291  & 
PSA             & 1620 & 3,4             &FA &  \\
\hline
lbkm04010 & 2011-04-23.315 &  55674.315 &  0.764 & G285M--2676  & 
PSA             &   51 & 3               &   NUV   &  \\
lbkm04020 & 2011-04-23.359 &  55674.359 &  0.782 & G130M--1291  & 
PSA             & 1620 & 3,4             &FA &  \\
\hline
lbkm05010 & 2011-04-23.963 &  55674.963 &  1.034 & G285M--2676  & 
PSA             &   51 & 3               &    NUV  &  \\
lbkm05020 & 2011-04-23.972 &  55674.972 &  1.037 & G130M--1291  & 
PSA             & 1620 & 3,4             &FA &  \\
\hline
lbkm06010 & 2011-04-24.562 &  55675.562 &  1.283 & G285M--2676  & 
PSA             &   51 & 3               &    NUV  &  \\
lbkm06020 & 2011-04-24.571 &  55675.571 &  1.287 & G130M--1291  & 
PSA             & 1620 & 3,4             &FA &  \\
\hline
lbme04010 & 2011-04-24.762 &  55675.762 &  1.367 & G285M--2676  & 
PSA             &   64 & 3               & NUV &  {Mg}~{\scriptsize II}, {Si}~{\scriptsize IV}\\
lbme04020 & 2011-04-24.766 &  55675.766 &  1.368 & G185M--1900  & 
PSA             &  685 & 3               & NUV &  \\
lbme04030 & 2011-04-24.777 &  55675.777 &  1.373 & G160M--1577  & 
PSA             &  900 & 3,4,1           & FAB &  \\
lbme04040 & 2011-04-24.823 &  55675.823 &  1.392 & G130M--1291  & 
PSA             & 5079 & 3,4,1,2,3,4,1,2 &FA &  \\
\hline
lbkm07010 & 2011-04-25.627 &  55676.627 &  1.727 & G285M--2676  & 
PSA             &   51 & 3               &    NUV  &  \\
lbkm07020 & 2011-04-25.636 &  55676.636 &  1.731 & G130M--1291  & 
PSA             & 1620 & 3,4             &FA &  \\
\hline
lbkm08010 & 2011-04-26.486 &  55677.486 &  2.085 & G285M--2676  & 
PSA             &   51 & 3               &    NUV  &  \\
lbkm08020 & 2011-04-26.495 &  55677.495 &  2.089 & G130M--1291  & 
PSA             & 1620 & 3,4             &FA &  \\
\hline
lbkm09010 & 2011-04-26.758 &  55677.758 &  2.198 & G285M--2676  & 
PSA             &   51 & 3               &    NUV  & {Mg}~{\scriptsize II}, {Si}~{\scriptsize IV} \\
lbkm09020 & 2011-04-26.767 &  55677.767 &  2.202 & G130M--1291  & 
PSA             & 1620 & 3,4             &FA &  \\
\hline
lbkm10010 & 2011-04-27.557 &  55678.557 &  2.531 & G285M--2676  & 
PSA             &   51 & 3               &     NUV &  \\
lbkm10020 & 2011-04-27.566 &  55678.566 &  2.535 & G130M--1291  & 
PSA             & 1620 & 3,4             &FA &  \\
\hline
lbkm11010 & 2011-04-29.447 &  55680.447 &  3.319 & G285M--2676  & 
PSA             &   51 & 3               &     NUV & {Si}~{\scriptsize IV} \\
lbkm11020 & 2011-04-29.482 &  55680.482 &  3.333 & G130M--1291  & 
PSA             & 1620 & 3,4             &FA &  \\
\hline
lbme05010 & 2011-04-29.755 &  55680.755 &  \nodata & G285M--2676  & 
PSA             &    0 & 3               & NUV & FGS ACQ failure \\
lbme05020 & \nodata &  \nodata &  \nodata & G185M--1900  & 
PSA             &    0 & 3               & NUV &  \\
lbme05030 & \nodata &  \nodata &  \nodata & G160M--1577  & 
PSA             &    0 & 3,4,1           & FAB &  \\
lbme05040 & \nodata &  \nodata &  \nodata & G130M--1291  & 
PSA             &    0 & 3,4,1,2,3,4,1,2 &FA &  \\
\hline
lbme55010 & 2011-06-01.503 &  55713.503 &  \nodata & G285M--2676  & 
PSA             &   64 & 3               & NUV & lbme05 repeat \\
lbme55020 & 2011-06-01.506 &  55713.506 &  \nodata & G185M--1900  & 
PSA             &  685 & 3               & NUV &  \\
lbme55030 & 2011-06-01.517 &  55713.517 &  \nodata & G160M--1577  & 
PSA             &  900 & 3,4,1           & FAB &  \\
lbme55040 & 2011-06-01.564 &  55713.564 &  \nodata & G130M--1291  & 
PSA             & 5080 & 3,4,1,2,3,4,1,2 &FA &  \\
\enddata
\vskip -7mm
\tablecomments{Col.~(1) prefix ``o'' is for STIS, ``l'' for COS.  Program ID ``BKM'' refers to GO--12279, ``BME'' to GO--12376.  Next two digits are the visit number and final three digits are the exposure number.  Col.~(3) is JD$-$2,400,000.5 in days.  Col.~(4) photometric cycles are based on Jetsu et al.\ (1993), relative to cycle~6841 in that ephemeris.  Col.~(6) aperture designation for STIS is $h(\arcsec){\times}w(\arcsec)$; for COS, ``PSA'' is the 2.5\arcsec-diameter Primary Science Aperture.  Col.~(7) exposure times are in seconds.  Col.~(8) COS FP-POS steps are in order of execution (FP-POS=\,3 is the ``home'' position of the grating wheel).  Program BME exposure time per FP-POS step is total $t_{\rm exp}$ (Col.~[7]) divided by number of FP-splits.  STIS BKM E140M observation was a pair of 806~s sub-exposures.  BKM COS G130M FP-POS sub-exposures were 900~s and 720~s, respectively.  Col.~(9) ``FAB'' refers to both COS FUV detector segments (A and B) on; ``FA'' is for side-A, only.  Col.~(10) ``{Mg}~{\scriptsize II}'' indicates elevated peak count rates at 2800~\AA, and similarly for ``{Si}~{\scriptsize IV}'' at 1400~\AA.
}
\end{deluxetable} 

\begin{deluxetable}{ccccrc}
\tablenum{2}
\tablecaption{{\em Chandra}\/ Observation Log}
\tablewidth{0pt}
\tablecolumns{6}
\tablehead{
\colhead{ObsID} &  
\colhead{Start Date} & 
\colhead{$t_{\rm exp}$} & 
\colhead{MJD} &  
\colhead{Cycle} &
\colhead{$\Delta$\,Phase} \\[3pt]                
\colhead{(1)} &  
\colhead{(2)} &
\colhead{(3)} &  
\colhead{(4)} &
\colhead{(5)} &  
\colhead{(6)}              
}                
\startdata   
12297  &  2011-04-21.470  &  39.5   &    55672.470 &  $-0.005$ &  0.190 \\
12356  &  2011-04-22.272  &  62.9   &    55673.272 &  0.329    &  0.303 \\
13251  &  2011-04-23.883  &  94.7   &    55674.883 &  1.000    &  0.457 \\
12298  &  2011-04-25.398  &  39.1   &    55676.398 &  1.631    &  0.189 \\
13259  &  2011-04-26.355  &  44.0   &    55677.355 &  2.030    &  0.212 \\
12357  &  2011-04-27.387  &  34.8   &    55678.387 &  2.460    &  0.168 \\
12299  &  2011-04-29.224  &  39.1   &    55680.224 &  3.225    &  0.189 \\
\enddata
\vskip -5mm
\tablecomments{All pointings with the High-Energy Transmission Grating Spectrometer and ACIS-S readout.  Col.~(3) exposure durations are in kiloseconds (ks).  Col.~(4) is JD$-$2,400,000.5 in days.  Col.~(5) cycle number at exposure start was based on Jetsu et al.\ (1993), relative to cycle~6841 in that ephemeris.  Col.~(6) is the phase duration of the pointing ($P_{\rm rot}= 2.40$~d).
}
\end{deluxetable}

\clearpage
\begin{figure}
\figurenum{2}
\vskip  0mm
\hskip  -5mm
\includegraphics[scale=0.75,angle=90]{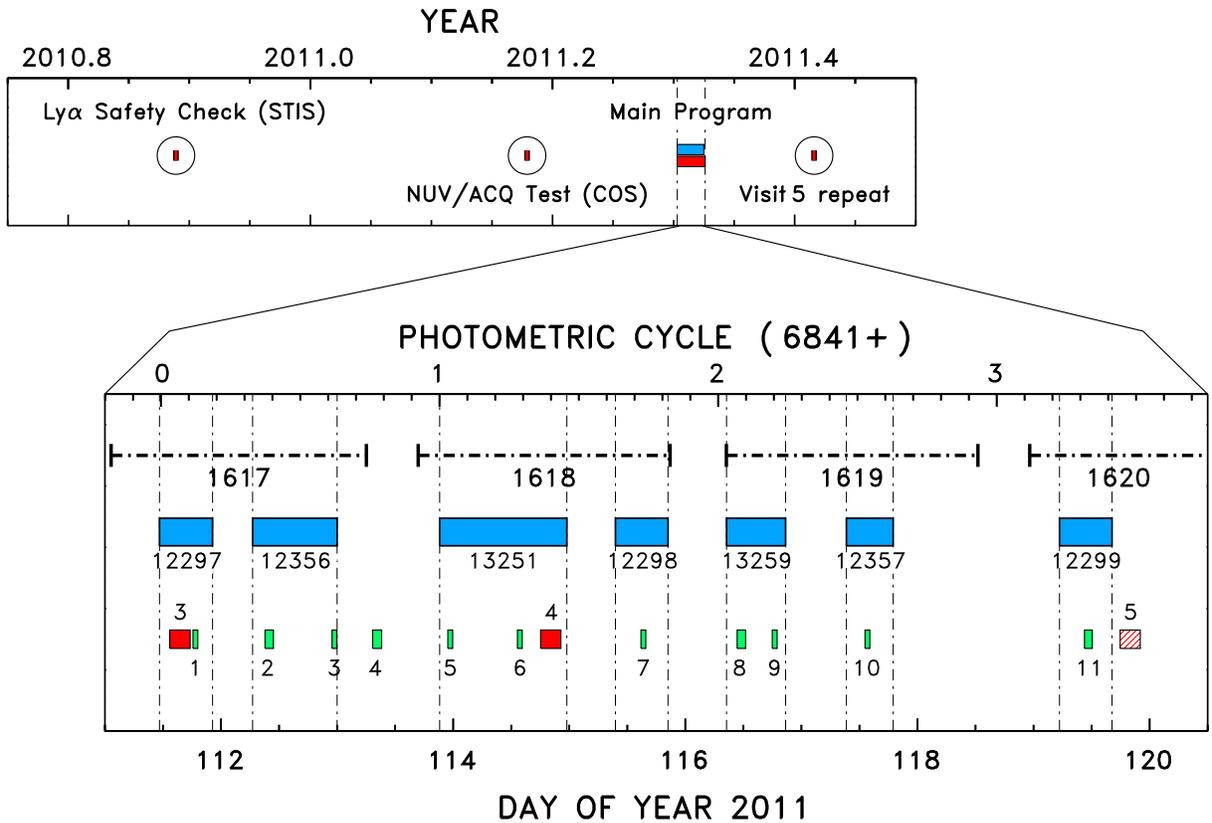} 
\vskip 0mm
\figcaption[]{\small Overview of the FK~Com COCOA-PUFS project.  The upper panel sketches major elements of the campaign.  The lower panel focuses on the intensive main program in 2011 April: the {\em Chandra}\/ pointings (blue-- ObsIDs annotated), the three ``long program'' 3-orbit {\em HST}\/ Visits of GO-12376 (red rectangles-- Visit numbers annotated), and the interspersed eleven 1-orbit ``short program'' {\em HST}\/ Visits of GO-12279 (green-- Visit numbers annotated).  The red hatched GO-12376 Visit~5 failed (owing to a Guide Star acquisition issue), and was repeated two months later, in 2011 June.  Dotted-dashed bars in the upper part of the panel depict orbital visibility periods for {\em Chandra;}\/ gaps represent perigee passes when {\em Chandra}\/ could not observe.
}
\end{figure}

\begin{figure}
\figurenum{3}
\vskip  0mm
\hskip  6mm
\includegraphics[scale=0.875]{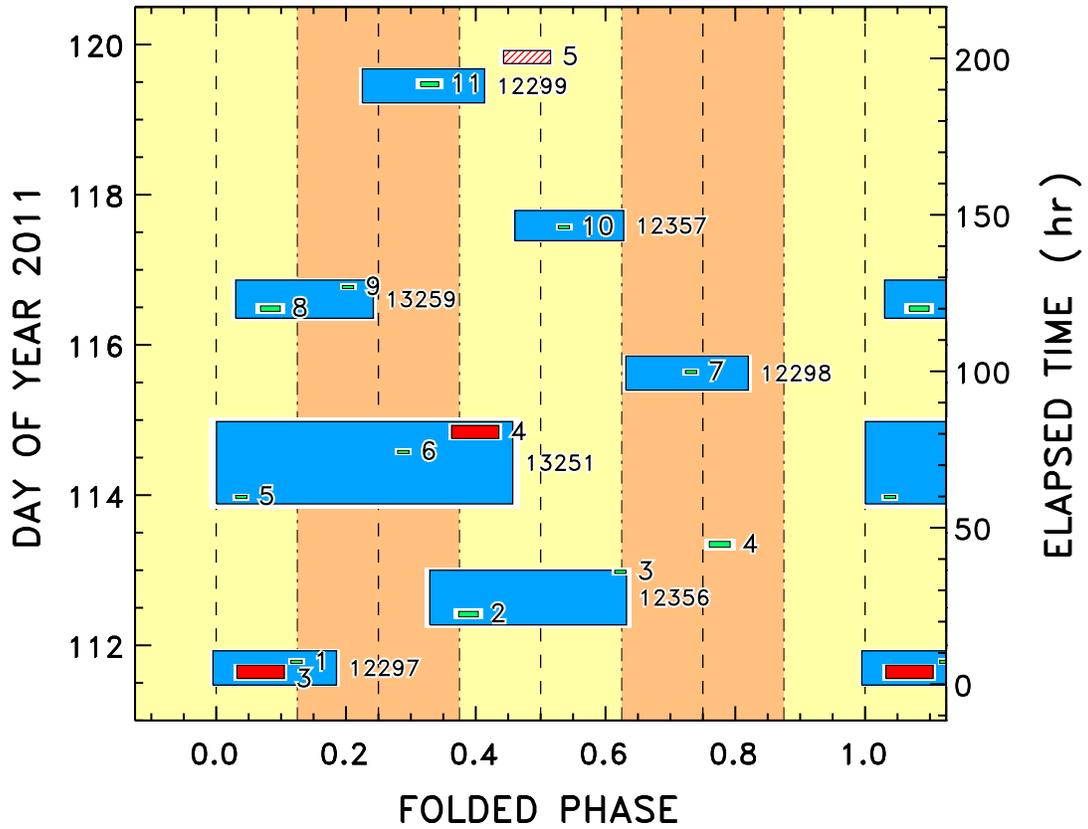} 
\vskip 0mm
\figcaption[]{\small Phase/time map of the COCOA-PUFS main program.  Time runs from the bottom of the diagram to the top; photometric (folded) phase runs from left to right.  Sizes of the observation rectangles represent the respective durations of the {\em Chandra}\/ HETGS pointings (blue) and the {\em HST}\/ COS visits (red and green).  {\em Chandra}\/ ObsIDs and {\em HST}\/ Visit numbers are annotated.  {\em HST/Chandra}\/ phase coverage is best for $\phi= 0.0$--0.8.
}
\end{figure}

\clearpage
\subsubsection{HST Visit 1: STIS Ly$\alpha$ Safety Check}

The sensitive COS detectors can be harmed by excessive UV illumination, and are subject to strict safety rules.  Even though FK~Com is a distant, rather faint, 8th magnitude star, its high levels of chromospheric and coronal activity were of concern, specifically bright \ion{H}{1} 1215~\AA\ Ly$\alpha$.  {\em IUE}\/ low-resolution FUV spectra of FK~Com suggested that the Ly$\alpha$ peak intensity was within a factor of two of the COS local count rate limit, albeit on the low side; but close enough to warrant further consideration, especially since FK~Com was known to flare from time to time.  If Ly$\alpha$ were ruled in violation of the COS bright limits, there would be unfortunate repercussions: the ``blue'' segment of the COS detector (side B) would have to be deactivated when using grating G130M.  Consequently, not only would key Ly$\alpha$ itself be lost, but also other important diagnostics, such as the \ion{N}{5} resonance doublet at 1240~\AA\ ($T\sim 2\times10^5$~K), hottest of the readily accessible features in the FUV. 

However, low-resolution {\em IUE}\/ could not fully resolve the bright limit issue.  Given the obvious benefit of having side B activated for the G130M exposures, it was decided to spend one of the {\em HST}\/ orbits to definitively measure Ly$\alpha$ with less sensitive STIS, well in advance of the main COS campaign.  In that way, not only would the detector concerns be resolved, one way or the other, but also at least one profile of Ly$\alpha$ would be secured, even if FK~Com failed the COS safety check.  A side benefit was that a few sharp interstellar lines (such as \ion{O}{1} 1302~\AA\ and \ion{C}{2} 1334~\AA) would be captured on top of bright FUV chromospheric emissions, and these could be used, given the high precision of the STIS wavelengths, to help remove systematic errors in the COS scales.

This initial single-orbit STIS Visit~1 was carried out 2010 November under program GO-12376.  The target was acquired, and centered, using the STIS CCD direct imaging mode (MIRVIS) through the ND3 filter.  Next, a 60~s exposure was taken with medium-resolution grating G230MB-2836 ($R\equiv {\lambda}/{\Delta\lambda}\sim 6000$: 2758--2914~\AA) and the CCD, through the 52$\times$0.2$^{\prime\prime}$ slit, to record the important chromospheric \ion{Mg}{2} 2800~\AA\ resonance doublet.   Filling out the remaining visibility period, a medium-resolution FUV echelle exposure was taken with E140M-1425 ($R\sim 4\times10^4$: 1150--1700~\AA), through the $0.2^{\prime\prime}\times$0.2$^{\prime\prime}$ ``photometric'' aperture, split into a pair of 806~s sub-exposures.  Disappointingly, the processed E140M spectrum revealed that the Ly$\alpha$ peak emission was too close to the COS screening limit to permit use of G130M side B.  

Figure~4 is a schematic view of the STIS E140M-1425 observation.  The FUV emission lines of FK~Com are unusually broad, mirroring those of the previous {\em FUSE}\/ spectrum.  For example, the \ion{C}{2} 1335~\AA\ multiplet normally appears as a clearly resolved pair of components in more slowly rotating yellow giants (as illustrated later), but in FK~Com it is a single broad feature.

\begin{figure}[hb]
\figurenum{4}
\vskip  -10mm
\hskip  10mm
\includegraphics[scale=0.875]{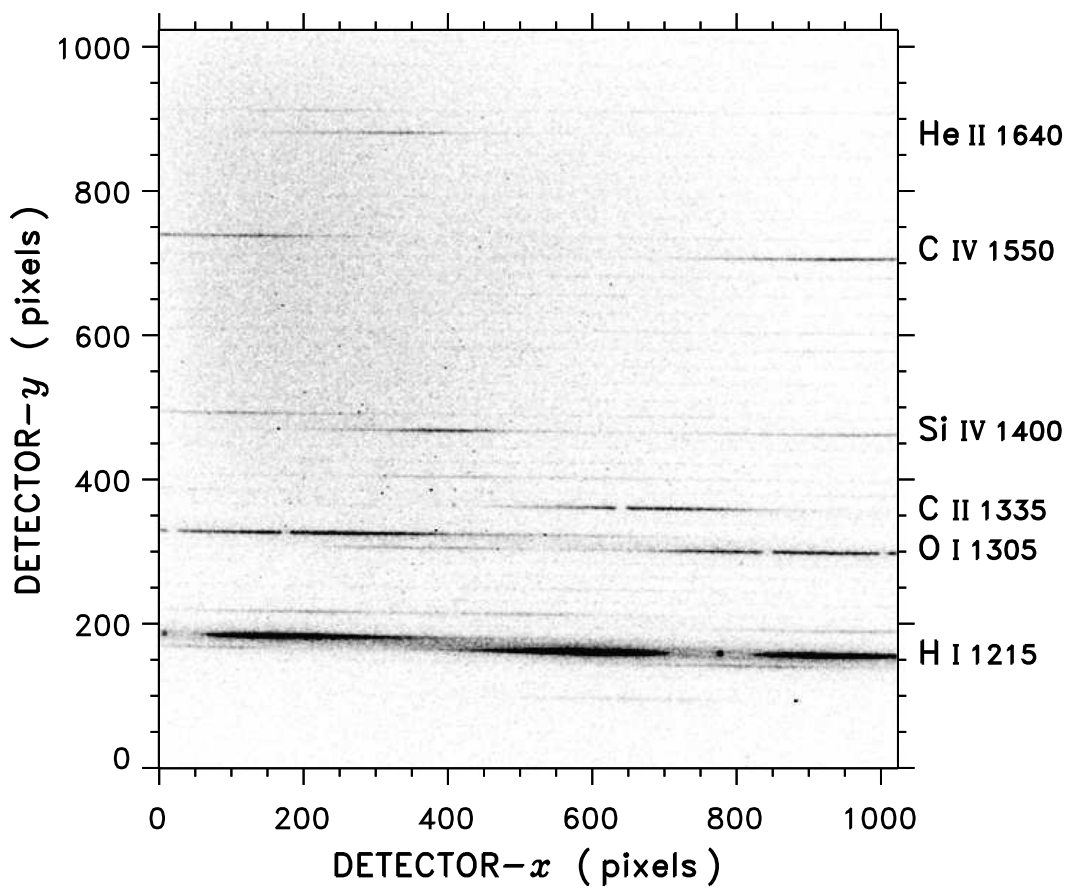}
\figcaption[]{\small STIS E140M-1425 raw echellegram of FK~Com, displayed in negative.  The mainly horizontal, but slightly tilted, echelle orders can be seen faintly in the continuum emission at the longer wavelengths (toward the top of the diagram).  The conspicuous doubled feature at the lower right is the very broad, intense \ion{H}{1} Ly$\alpha$ 1215~\AA\ chromospheric emission, bisected by a strong interstellar atomic hydrogen absorption.  The dot between the blue and red lobes of the stellar Ly$\alpha$ feature is geocoronal hydrogen emission.  Other smaller dots scattered across the image are hot pixels.  The red lobe of the stellar Ly$\alpha$ feature is repeated in the next echelle order up, to the left side of the image.  Prominent stellar features are noted at the right side of the diagram.  Several of these also are repeated in adjacent echelle orders.  The broad chromospheric blends of \ion{O}{1} and \ion{C}{2} carry sharp interstellar absorptions.   
}
\end{figure}

\clearpage
\subsubsection{HST Visit 2: COS Acquisition Test}

Not only was FK~Com over-bright at Ly$\alpha$, but it also exceeded safety limits for the normal NUV imaging acquisition (Primary Science Aperture [PSA] and MIRROR-B) used for moderately faint objects ($V\gtrsim 9$).  An alternative was the Bright Object Aperture (BOA) plus MIRROR-A, a combination that nevertheless was not considered ideal because the BOA image is somewhat distorted, and the impact on acquisitions had not yet been fully assessed at the time (still relatively early in the COS operations).  Other approaches were investigated, including several types of dispersed light ACQ/SEARCH followed by cross-dispersion (XD) and in-dispersion (D) peak-ups\footnote{Raster scans to centroid the source location.}.  But, the NUV imaging ACQ was fastest, a key advantage for a campaign involving a large number of short (1 orbit) COS pointings (for the GO-12279 part).  

It was therefore decided to schedule GO-12376 Visit~2 to vet the COS acquisition strategy well before the main program, to allow for adjustments if a more conservative approach was merited.  The single orbit test utilized the BOA NUV imaging acquisition described earlier, followed by brief exposures with G285M-2676, G160M-1577, and G130M-1291 to fill out the orbit.  These were the principal COS grating settings utilized, in various combinations, during the main campaign.  The acquisition and associated test exposures were successful, so no adjustments to the observing scenario were necessary.                

\subsubsection{HST Subsequent Visits: COS Main Campaign Long and Short Programs}

All the COS exposures of the main program were taken in time-tag (T-TAG) mode, which delivers a time-stamped record of photon coordinates that can be processed in a number of ways to provide temporal histories of fluxes and/or line profiles.  The remaining nine orbits of GO-12376 were divided into three 3-orbit visits, the ``long program.''  These led off with the NUV/BOA imaging acquisition; then a short (64~s) exposure of the 2800~\AA\ region with G285M-2676, at FP-POS\footnote{The FP-POS are small grating steps intended to mitigate fixed pattern noise by shifting the spectrum onto different sets of detector pixels.}~3 through the PSA, to capture the key chromospheric \ion{Mg}{2} h (2803~\AA) \& k (2796~\AA) features (on stripe ``C'').  This was followed by a deeper exposure (685~s) with G185M-1900 to reach the subcoronal \ion{Si}{3} 1892~\AA\ and \ion{C}{3} 1909~\AA\ intersystem lines (stripe B), and a few important chromospheric emissions that happened to fall on stripes A and C.  Then, three G160M-1577 sub-exposures of 300~s each, at FP-POS~3, 4, and 1, were taken to fill out the first $\sim 50$~minute visibility period.  The second orbit featured four G130M-1291 sub-exposures of 635~s each, at the standard FP-POS steps (1--4), but on detector side A only.  The third orbit was identical to the second.  The \ion{Si}{4} 1393~\AA\ feature -- important because it is the most isolated of the bright ``hot lines'' and thus least affected by rotationally-exaggerated blending -- was captured in both the G130M/A and G160M/B segments, providing temporal continuity through the three orbits.  The two orbits with G130M achieved 5~ks of exposure depth at key coronal forbidden line [\ion{Fe}{21}] 1354~\AA\ ($T\sim 10$~MK), nearly 3 times that expected for the single-orbit visits (``short program'') of companion GO-12279.  The long-program exposures were arranged so that the reconfiguration overhead (deactivating detector side B) was ``hidden'' during the occultation period between orbits 1 and 2. 

There were eleven additional single orbits, from program GO-12279, interspersed among the long-program visits.  Each of these short-program orbits began with the NUV/BOA imaging acquisition, followed by a 51~s exposure of G285M-2676, and concluding with two G130M-1291 integrations, again only on side A, at FP-POS~3 (900~s) and 4 (720~s).  This pair of exposures captured the key subcoronal \ion{Si}{4} 1400~\AA\ doublet, as well as the important chromospheric multiplet of \ion{C}{2} at 1335~\AA\ and coronal [\ion{Fe}{21}] 1354~\AA.  FP-POS~3 has the cleanest profiles of these features (avoiding grid-wire shadows), so was allocated slightly more exposure time.

The intent was to balance long multiple-setting exposures at a few phases with short effectively single-setting exposures at many phases.  The combination of the deep and shallow strategies probes the range of time scales expected for various phenomena of interest, especially rotational modulations and transient flares.

Figure~5 is an example of a COS NUV G285M-2676 raw image, containing the important chromospheric \ion{Mg}{2} h and k lines near 2800~\AA.  Figure~6 is an overview of the spectral range covered by the paired G130M-1291 and G160M-1577 settings of program GO-12376.  

\begin{figure}
\figurenum{5}
\vskip  0mm
\hskip  12mm
\includegraphics[scale=0.875]{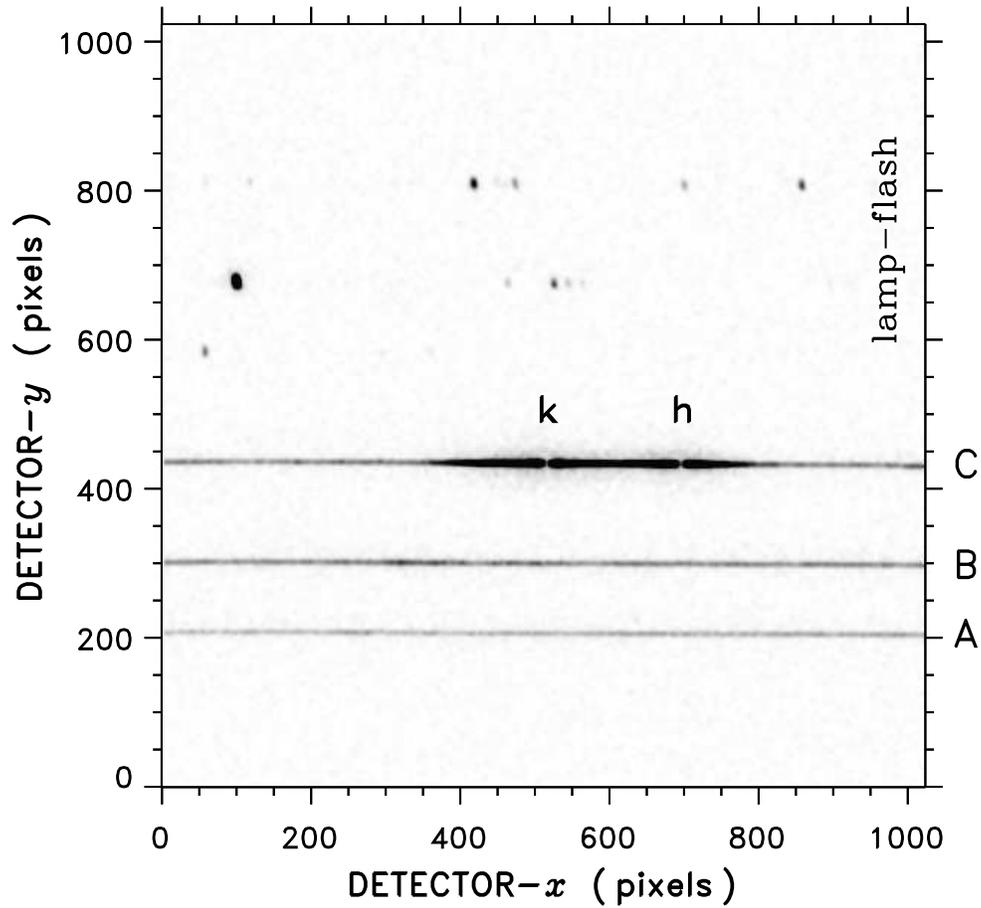} 
\vskip 0mm
\figcaption[]{\small Schematic COS NUV G285M-2676 raw image, in negative: observation lbme03010 from GO-12376 Visit~3, the leading exposure of the first sequence of the main campaign.  The three spectral stripes can be seen in the lower half of the image (labeled A--C), while the corresponding ``lamp-flash'' spots are at the top (for calibration of the wavelength scale zero-point shift).  The key \ion{Mg}{2} hk lines are marked.  Unlike STIS E140M, the COS NUV mode is not a true cross-dispersed echelle design, but rather utilizes a simple optical system to multiplex the three short segments from a Rowland-circle spectrum delivered by a first-order disperser.
}
\end{figure}

\begin{figure}[ht]
\figurenum{6}
\vskip  0mm
\hskip  -3mm
\includegraphics[scale=0.75,angle=90]{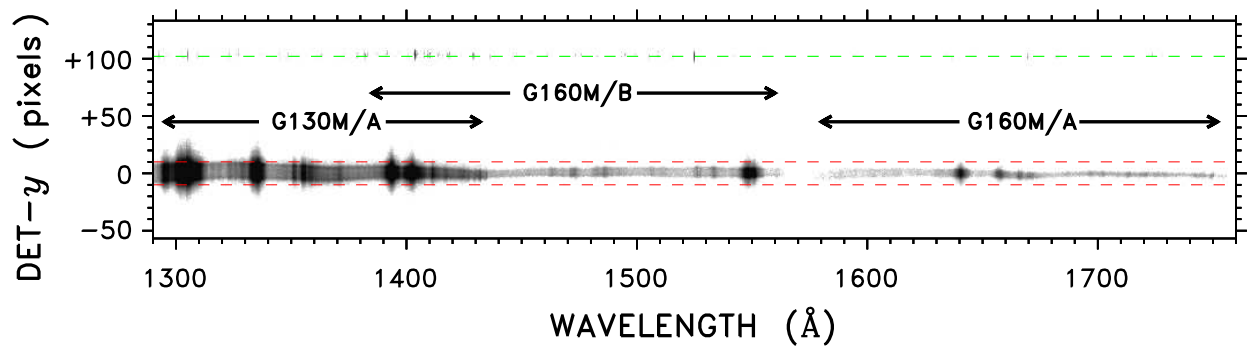} 
\vskip 0mm
\figcaption[]{\small Schematic COS FUV spectral image of FK~Com, in negative, from the 2011 main campaign.  The G130M and G160M settings overlap in the important 1400~\AA\ region.  Note the detector gap between G160M-1577 sides A and B at 1560-1580~\AA, and also that only side A of G130M-1291 could be activated, owing to the bright limit violation by Ly$\alpha$ on side B.  The image was constructed from the photon event lists of GO-12376 Visits 3 and 4, with adjustments in the detector $y$ centroids to align the three separate segments, for display purposes.  The faint narrow features under the green dashed line in the upper portion of the panel are from the lamp-flash spectrum, which is used to determine the zero-point wavelength shift.  A number of bright stellar emissions are apparent in the main spectral stripe, including the blends of chromospheric \ion{O}{1} 1305~\AA\ and \ion{C}{2} 1335~\AA; the (partially separated) \ion{Si}{4} doublet at 1400~\AA; the (blended) \ion{C}{4} doublet at 1550~\AA; and isolated \ion{He}{2} 1640~\AA.  The G160M segments are fainter overall than G130M side A in this representation, because the latter had more than five times the exposure duration for the long-program visits illustrated here. 
}
\end{figure}

\clearpage
\subsubsection{{\em Chandra}\/ HETGS Pointings}

Seven pointings with {\em Chandra}\/ HETGS were carried out over the main campaign period 2011 April 21--29.  ACIS-S was utilized as the grating readout, capturing medium- and high-energy spectral channels: MEG (2.5-31~\AA) and HEG (1.2-15~\AA), respectively.  The resolving power ($R\equiv {\lambda}/{\Delta\lambda}$) varied between 700--1100, increasing with wavelength in each grating arm.  The exposures ranged from 35~ks to 95~ks (see Table~2 and Figs.~2 and 3).  As with FUV COS, the HETGS event lists can be decomposed into time series of broad-band fluxes, or of individual bright lines if desired.  Here, mainly the integrated X-ray photometry will be considered, as a context for the {\em HST}\/ UV spectroscopy.  However, for reference purposes, Figure~7 illustrates an HETGS image of FK~Com.  It was assembled from the event lists of all the {\em Chandra}\/ pointings of the COCOA-PUFS main campaign, screened (exploiting the energy resolution of ACIS-S) to isolate the plus and minus first orders of the HEG and MEG spectral arms.  

\begin{figure}[hb]
\figurenum{7}
\hskip -22mm
\includegraphics[scale=0.85,angle=90]{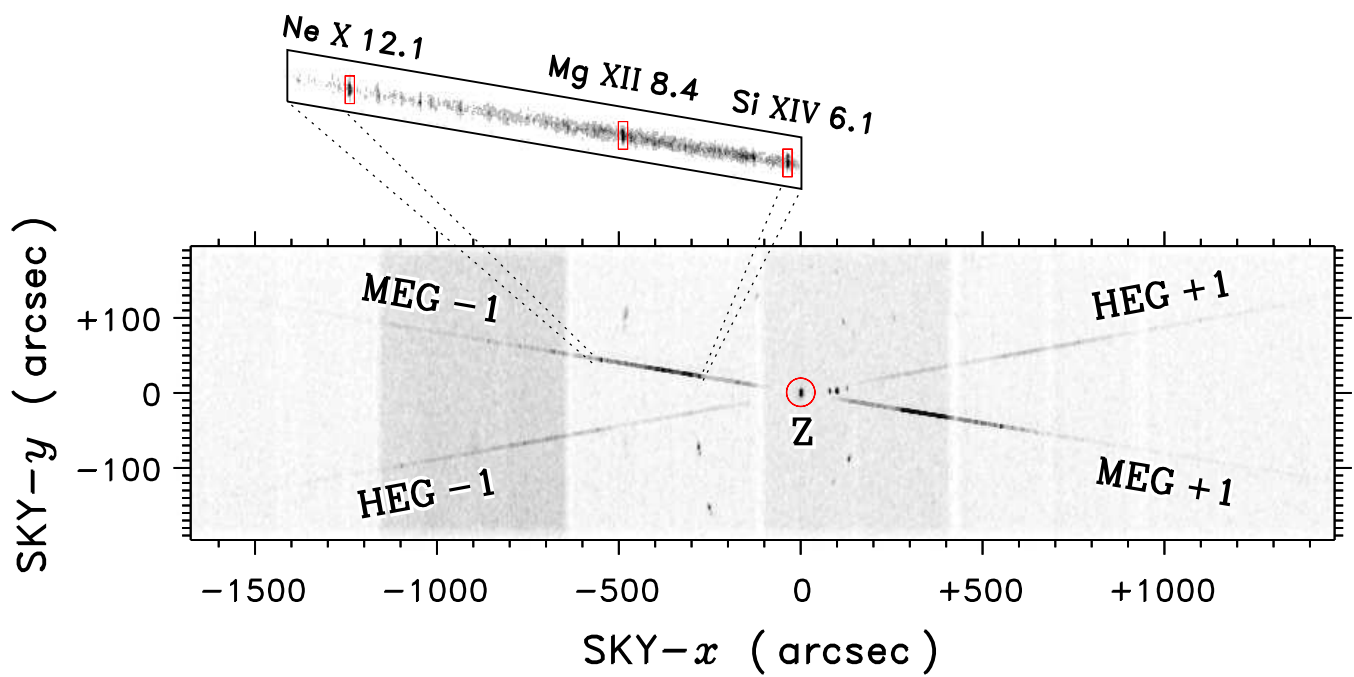} 
\vskip -10mm
\figcaption[]{\small Schematic HETGS spectral image of FK~Com, in negative, from the 2011 main campaign.  The plus and minus first orders of the HEG and MEG arms form a cross pattern centered on the zeroth-order image (marked ``Z'').  A blow-up of part of the MEG\,$-$1 order is depicted above the main image: the bright Ly$\alpha$ features of the hydrogen-like ionization stages of neon, magnesium, and silicon are highlighted (appended numbers are wavelengths in \AA).  In the main panel, there are several fainter zeroth-order images of point-like X-ray sources in the field, which appear multiple owing to the slightly different roll angles of the individual observations.  Note that the $y$-axis scale is magnified relative to the $x$-axis.
}
\end{figure}

\clearpage
\subsubsection{Ground-Based High-Resolution Spectroscopy, Photometry, and Spectropolarimetry}

FK~Com has been the subject of numerous ground-based studies over the past quarter century (see, e.g., summary by Vida et al.\ 2015).  The unusual yellow giant is an appealing target because it is bright enough to observe with a variety of spectroscopic, photometric, and polarimetric techniques (including ZDI), its spot patterns evolve relatively slowly ($\sim$months), yet there are profound changes over long time scales ($\sim$years) suggestive of phenomena known from long-term records of solar activity (such as the so-called active longitudes and their periodic ``flip-flops'').

The supporting ground-based campaign for COCOA-PUFS operated over the period 2011 April 11 to June 15, using a variety of instruments. These included: The European Southern Observatory (ESO) UVES high-resolution spectrometer at the Very Large Telescope (VLT) on Paranal in Chile; the ESO FORS2 low-resolution spectropolarimeter also on the VLT; the SES high-resolution spectrograph of the STELLA facility at the Izana Observatory on Tenerife, Spain; the high-resolution SARG spectrograph on the Telescopio Nazionale Galileo (TNG) on La Palma, Spain; high-resolution spectropolarimetry (ZDI) from the NARVAL spectrograph on the Bernard Lyot Telescope at Pic du Midi, France; and broad-band photometry from the Vienna-Potsdam Automatic Photoelectric Telescopes at Fairborn Observatory in Arizona, USA. The STELLA/SES coverage was by far the densest (at high spectral resolution) with typically 2--4 observations at 2-hour intervals per night during the main COCOA-PUFS campaign.

A full description of the ground-based program is deferred to a subsequent paper of this series, including  detailed thermal and magnetic surface maps (see, e.g., Vida et al.\ 2015).  Here, just the visible broad-band photometry, including $(V-Ic)$ colors, obtained in the weeks surrounding the main campaign, will be presented as a context for the higher energy measurements from {\em HST}\/ and {\em Chandra}.                                                                             

\section{ANALYSIS}

\subsection{STIS FUV Spectrum of FK Com}

The single-orbit  STIS E140M-1425 ``safety-check'' exposure of FK~Com was passed through the {\sf CALSTIS}\footnote{See: http://www.stsci.edu/institute/org/ins/cos{\small\textunderscore}stis/projects/Pipeline} pipeline, then post-processed using protocols developed for the Advanced Spectral Library Project (ASTRAL)\footnote{see: http://casa.colorado.edu/$\sim$ayres/ASTRAL/}.  Important aspects include a correction for small wavelength distortions present in the pipeline dispersion solutions, and a procedure to align and merge sub-exposures to boost signal-to-noise (S/N).  Figure~8 is an overview of the STIS FUV spectrum of FK~Com, compared to the similar, but more slowly rotating, yellow giant 24~UMa from the STIS stellar catalog StarCAT\footnote{see: http://casa.colorado.edu/$\sim$ayres/StarCAT/}.  The comparison emphasizes the extraordinary widths of the chromospheric and hotter emission lines of FK~Com, noting that 24~UMa, itself, has much wider features than a solar-like G dwarf star.  In fact, even when the 24~UMa spectrum is rotationally broadened to the photospheric $\upsilon\sin{i}$ of FK~Com (as depicted in the figure), the FK~Com emissions still are noticeably wider.  

Such ``super-rotational broadening'' has been identified in FUV emissions of several other fast-spinning yellow giants (e.g., Ayres et al.\ 1998), although FK~Com represents an extreme case.  The enhanced broadening of the FK~Com hot lines, like \ion{C}{3} 977~\AA\ and \ion{O}{6} 1031~\AA\,+\,1037~\AA, was noted in the singular {\em FUSE}\/ spectrum taken in 2004.  A similar effect has been seen over many decades in ground-based H$\alpha$ spectra of FK~Com, especially as captured in diagrams that display spectral differences as a function of photometric phase (for recent examples see Vida et al.\ 2015).

Table~3 lists integrated fluxes of the STIS FUV features, mainly for comparison with the corresponding emissions from the COS FUV visits of the main program, to be described later (Table~5).

Although the primary motivation for the STIS FUV pointing on FK Com was for the Ly$\alpha$ safety check, a more subtle, but equally important, reason has to do with the presence of systematic errors in the COS FUV wavelength scales.  This effect has been documented in a COS time-domain study of the young solar analog EK~Draconis (Ayres 2015a): a (wavelength-precise) STIS E140M-1425 spectrum of the star was compared directly to a combined COS G130M-1291\,+\,G160M-1577 tracing (same instrumental setup as for the ``long-program'' visits on FK Com here) to characterize the slight ``stretching'' of the COS wavelength scales (which amounted to a differential shift from the shortwavelength side of a setting to the longwavelength end of about 10 km s$^{-1}$).  

\clearpage
\begin{deluxetable}{rcccl}
\tablenum{3}
\tablecaption{{\em HST}\/ STIS FUV Measurements}
\tablewidth{0pt}
\tablecolumns{5}
\tablehead{
\colhead{Transition} &  
\colhead{$\lambda$} & 
\colhead{$\Delta\lambda$} &  
\colhead{$f_{\rm INT}$} &
\colhead{Notes}\\[3pt]
\colhead{} &  
\multicolumn{2}{c}{(\AA)} &
($10^{-13}$ erg cm$^{-2}$ s$^{-1}$)\\[3pt]
\colhead{(1)} &  
\colhead{(2)} &
\colhead{(3)} &  
\colhead{(4)} &
\colhead{(5)}  
}                
\startdata   
\cutinhead{Integrated flux measurements}
\ion{Si}{3}~1206  &  1206.2  &   1.8 &   2.0  &  \\
\ion{H}{1}~1215   &  1216.5  &   6.5 &   61   & excluding $\oplus$ Ly$\alpha$ \\
\ion{N}{5}~1240   &  1240.7  &   4.8 &   2.1  &  \\
\ion{O}{1}~1305   &  1304.0  &   4.0 &   7.2  &  \\
\ion{C}{2}~1335   &  1335.5  &   4.5 &   2.5  &  \\
\ion{Si}{4}~1400  &  1397.5  &   9.0 &   3.5  &  \\
\ion{C}{4}~1550   &  1549.0  &   6.0 &   6.1  &  \\
\ion{He}{2}~1640  &  1640.0  &   5.0 &   3.3  &  \\
\cutinhead{Interstellar absorption velocities}
                      & $\upsilon$ &  $\Delta\upsilon_{\rm D}$ \\
                      & \multicolumn{2}{c}{(km s$^{-1}$)} \\[3pt]
\hline
\ion{O}{1}  1302.168  &  $-29$  &   32  &   \nodata &  sGau \\
\ion{Si}{2} 1304.370  &  $-27$  &   19  &   \nodata &   \\
\ion{C}{2}  1334.532  &  $-28$  &   33  &   \nodata &  sGau \\
\enddata
\vskip 0mm
\tablecomments{Col.~(1) wavelength designator, following species, in \AA; laboratory wavelengths (in vacuum) for the ISM absorptions were taken from the NIST Atomic Spectra Database (Kramida et al.\ 2013) ``Ritz'' values. Col.~(2) is the central wavelength of the integrated flux measurement, or the velocity of the absorption measurement.  Col.~(3) is the half width of the integration band, or the (e-folding) Doppler width of the ISM absorption measurement.  Col.~(4): integrated fluxes ($f_{\rm INT}$).  In the Col.~(5) Notes, ``sGau'' indicates that an $e^{-x^4}$ profile was fitted, in place of the normal $e^{-x^2}$ Gaussian, to more closely model the partially saturated absorption profile.  Uncertainties are not quoted for the integrated fluxes because the large number of spectral bins in the summation reduces the formal photometric error to negligible levels.  Monte Carlo uncertainties for the Gaussian and pseudo-Gaussian modeling of the interstellar absorptions average about $\pm$1.4~km s$^{-1}$, while the differences between the results of the two profile fitting approaches (a rough measure of systematic errors) are of similar size.
}
\end{deluxetable} 

\clearpage
Unfortunately, owing to the large separation in time between the initial STIS observation of FK~Com and the main COS program several months later, and the more velocity-variable stellar lines of the yellow giant, the direct spectrum comparison strategy could not be applied.  Instead, the cross-calibration role fell to the few sharp, presumably interstellar, absorptions that appear in the STIS spectrum: \ion{O}{1} 1302~\AA, \ion{Si}{2} 1304~\AA, and \ion{C}{2} 1335~\AA.

Table~3 summarizes Gaussian measurements of the three narrow absorption features.  The average velocity is $-28$~km s$^{-1}$, and the standard error of the mean (s.e.) over the three independent ISM features is less than 1~km s$^{-1}$.  The STIS CCD G230MB-2836 spectrum of the \ion{Mg}{2} hk region also was measured, using a multi-component constrained fit to the blended stellar chromospheric emission profiles, and the narrow, separated ISM absorptions (as was applied to the main program COS NUV spectra of \ion{Mg}{2}, and described in more detail later [\S{3.2.3}]).  Compared to the COS NUV exposures of similar duration ($\sim$~1 minute), the STIS CCD \ion{Mg}{2} spectrum is lower in resolution ($R\sim 6\times10^3$ versus $2\times10^4$ for COS NUV) and appears noisier even though the STIS Exposure Time Calculator (ETC), and the assigned photometric error, indicated otherwise.  An examination of the CCD spectral image showed what appeared to be streaks along the CCD columns, apparently due to Charge Transfer Inefficiency (CTI) trails of hot pixels and cosmic rays, which likely are responsible for the large point-to-point excursions in the extracted spectrum.  The measured velocity of the narrow absorption components of h and k was $-29{\pm}2$~km s$^{-1}$, in agreement with the contemporaneous FUV value, but the STIS CCD value should be treated with caution owing to the CTI sleeking effect, which potentially could influence the shapes of the h and k absorption dips.  In fact, the STIS CCD \ion{Mg}{2} velocity is about 10~km s$^{-1}$ blueward of the $-17$~km s$^{-1}$ reported by Ayres et al.\ (2006) based on 15 {\em IUE}\/ high-dispersion NUV spectra.  

\clearpage
\begin{figure}
\figurenum{8}
\vskip  0mm
\hskip -17mm
\includegraphics[scale=1.125]{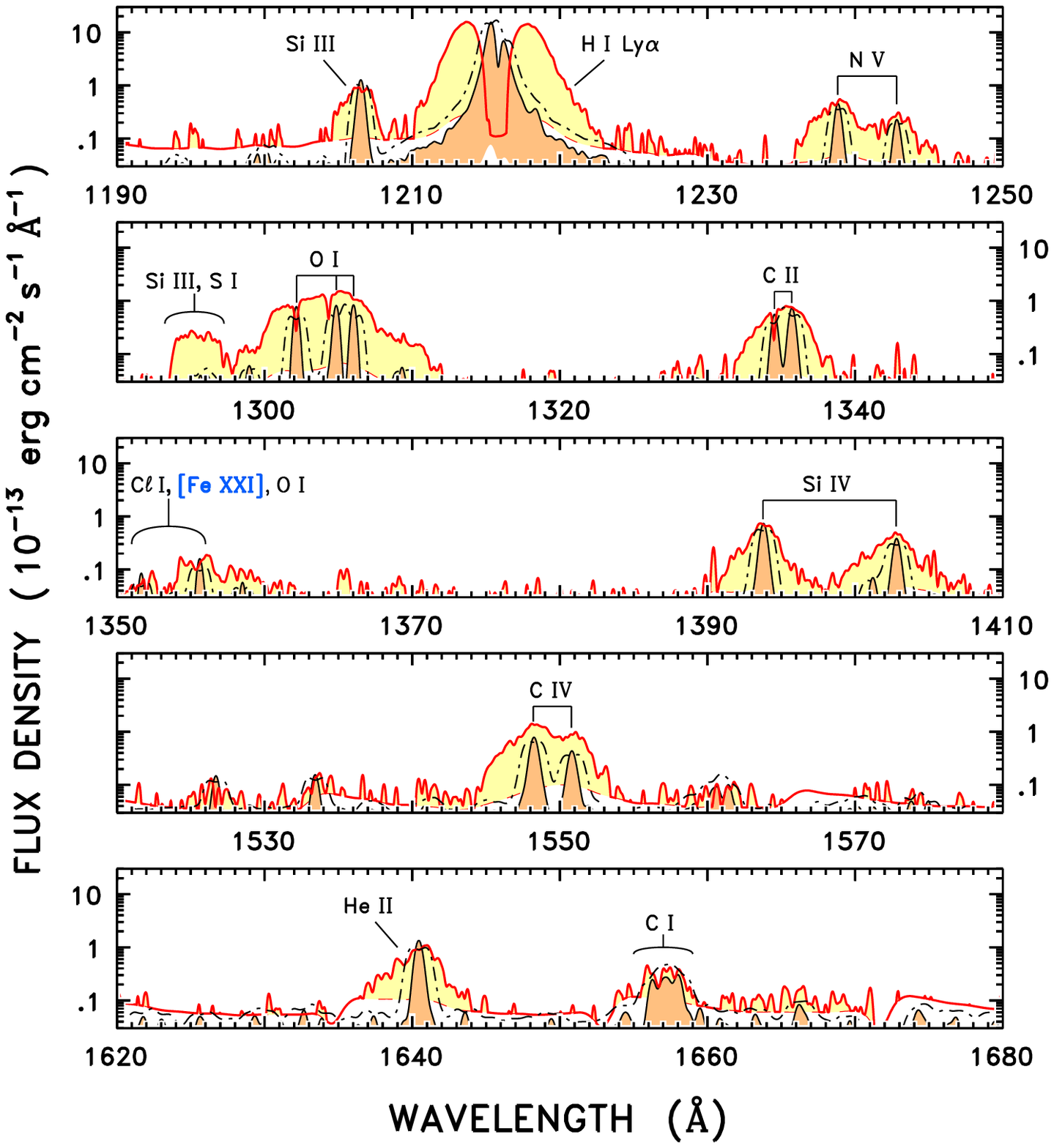}
\figcaption[]{
}
\end{figure}
 
\begin{figure}
\figurenum{8}
\vskip 0mm
\figcaption[]{\small STIS E140M-1425 spectrum of FK~Com (red outlined yellow shading) contrasted to a scaled FUV observation of the similar, but more slowly rotating, active yellow giant 24~UMa (black outlined orange shaded).  The (noisier) FK~Com spectrum was (Gaussian) smoothed by approximately 2 resolution elements (resels), which is of little consequence for the already extremely broad spectral features.  The bottom edge of the yellow shading (thin red dashed) follows the 1\,$\sigma$ photometric noise level for FK~Com.  The black dotted-dashed curve is the 24~UMa spectrum spun-up with a simple rotational profile representing $\upsilon\sin{i}= 160$ km s$^{-1}$, and re-normalized.  The extreme broadening of FK~Com's emission lines is conspicuous, well beyond the photospheric $\upsilon\sin{i}$.  Ground-state resonance absorptions, presumably due to the ISM, can be seen at \ion{H}{1} 1215~\AA, \ion{O}{1} 1302~\AA, \ion{Si}{2} 1304~\AA, and \ion{C}{2} 1334~\AA.
}
\end{figure}

\clearpage
\subsection{COS NUV \ion{Mg}{2} Profiles}

\subsubsection{Overview of the G285M-2676 \ion{Mg}{2} Spectra}

Although the main focus, and virtually all the clock time, of the FK~Com campaign was lavished on the FUV, every visit of the COS short and long programs began with a brief exposure of the \ion{Mg}{2} 2800~\AA\ region with COS NUV setting G285M-2676.  In fact, the \ion{Mg}{2} hk lines rival Ly$\alpha$ as the brightest features in the FK~Com UV spectrum, although the \ion{H}{1} emission is about three time wider in velocity (which should not be taken too literally in terms of Doppler displacements, because both \ion{H}{1} and \ion{Mg}{2} likely are extremely optically thick).

Figure~9 is a montage of the 13 \ion{Mg}{2} spectra taken during the FK~Com main program, in time order running from bottom to top, and offset in flux density for clarity.  The spectral traces were derived from the {\sf CALCOS}\footnote{See: http://www.stsci.edu/hst/cos/pipeline/} pipeline ``x1dsum'' datasets, and registered in velocity to the ISM \ion{Mg}{2} absorption dips (see below).  Photometric cycle numbers are listed on the right axis.  The original flux points were binned a factor of 5, roughly a resolution element (resel), and displayed as error bars representing 1~s.e.\ of the apparent variations of the constituent flux points, an empirical gauge of the local photometric uncertainties.  The two observations highlighted in blue are anomalous with respect to the neighboring time and/or phase points, and likely were affected by flares.  The dashed curves represent an epoch-average \ion{Mg}{2} profile, excluding the two suspected flares.  Darker shading highlights portions of the local profile that exceed the average, whereas lighter shading marks places where the local profile is lower.
 
\begin{figure}
\figurenum{9}
\vskip  0mm
\hskip  7mm
\includegraphics[scale=0.875]{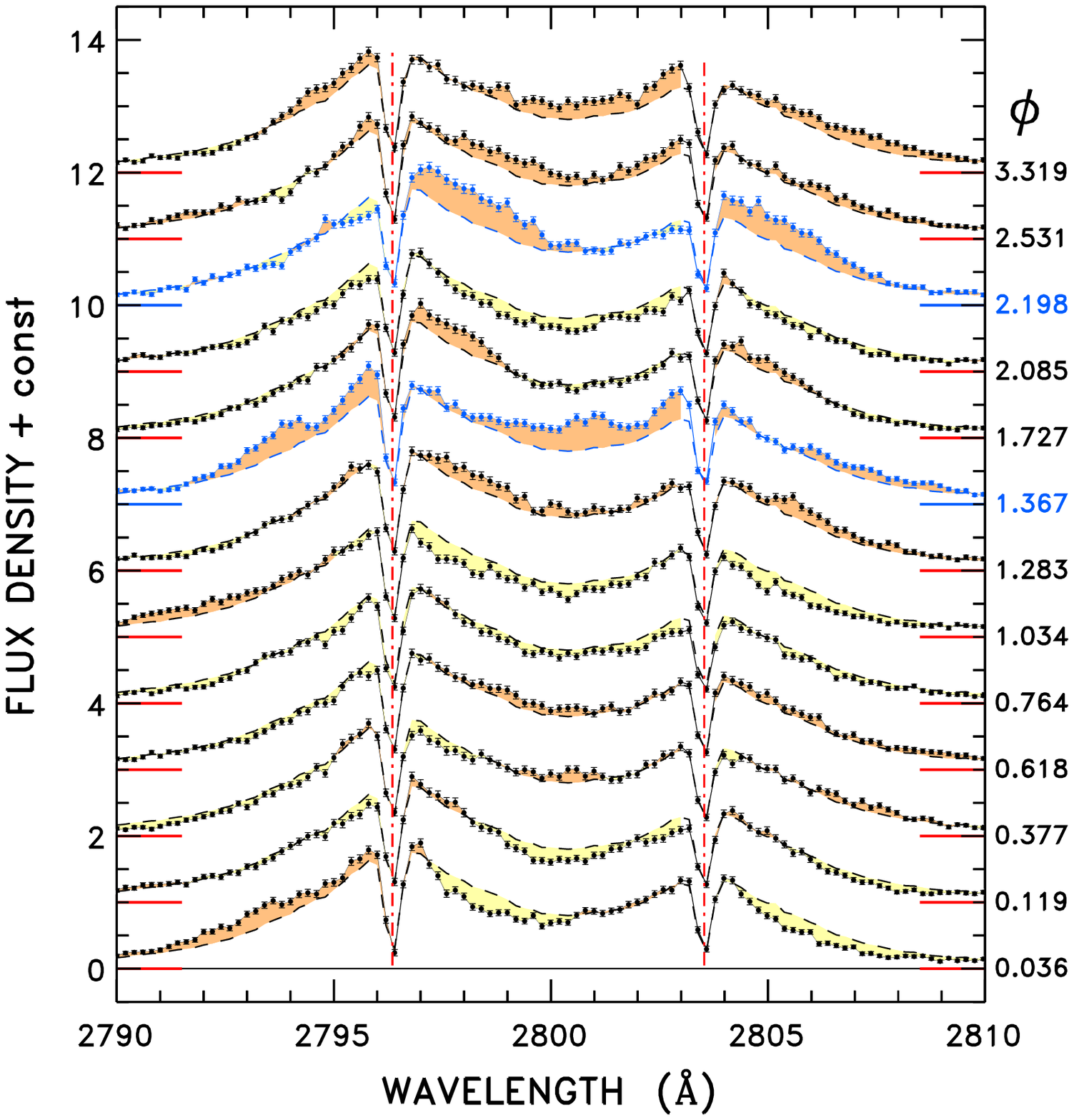}
\vskip -5mm
\figcaption[]{\small COS G285M-2676 spectra of the blended \ion{Mg}{2} doublet of FK~Com from the main program, registered to the \ion{Mg}{2} absorption velocity: k is left of center, h to the right.  Both have broad stellar chromospheric emissions punctuated by narrow absorptions (marked by vertical red dotted-dashed lines at the \ion{Mg}{2} laboratory wavelengths).  The profiles are arranged, from bottom to top, in increasing time, and are offset by one unit of the flux density axis ($10^{-12}$ erg cm$^{-2}$ s$^{-1}$ \AA$^{-1}$) for clarity (horizontal colored ticks refer to the zero levels of the individual spectra).  Photometric cycle numbers, relative to cycle 6841, are noted at the right.  The two profiles marked in blue are anomalous, probably affected by flares.  Thin black dashed curves represent the epoch-average profile, excluding the two suspected flares.  Darker (lighter) shading indicates whether the local profile is higher (lower) than the average.  Error bars represent a $\sim$1~resel rebinning of the original flux densities for display.}
\end{figure}

Cursory examination of the diagram reveals -- as noted previously by Ayres et al.\ (2006) based on a collection of {\em IUE}\/ NUV spectra spread over several years -- that the \ion{Mg}{2} lines are extraordinarily broad (about 470~km s$^{-1}$ full width at half maximum intensity [FWHM]), enough so to be partially blended, and are highly variable on time scales corresponding to a few tenths of a rotational cycle ($P= 2.40$~days).

\subsubsection{Integrated Fluxes of the COS NUV \ion{Mg}{2} Features}

Figure~10 depicts the integrated fluxes of the \ion{Mg}{2} hk doublet of FK~Com as derived from numerical integrations of the tracings illustrated in Fig.~9.  A baseline ``continuum'' level, determined from flux points well away from the blended emission feature, was subtracted from each profile prior to the measurement.  The specific integrated flux values are reported later (Table~4).  Note that the two suspected flares occurred at nearly the same phase, $\phi\sim$~0.3\,, although in different rotations.

\begin{figure}
\figurenum{10}
\vskip  0mm
\hskip  8mm
\includegraphics[scale=0.875]{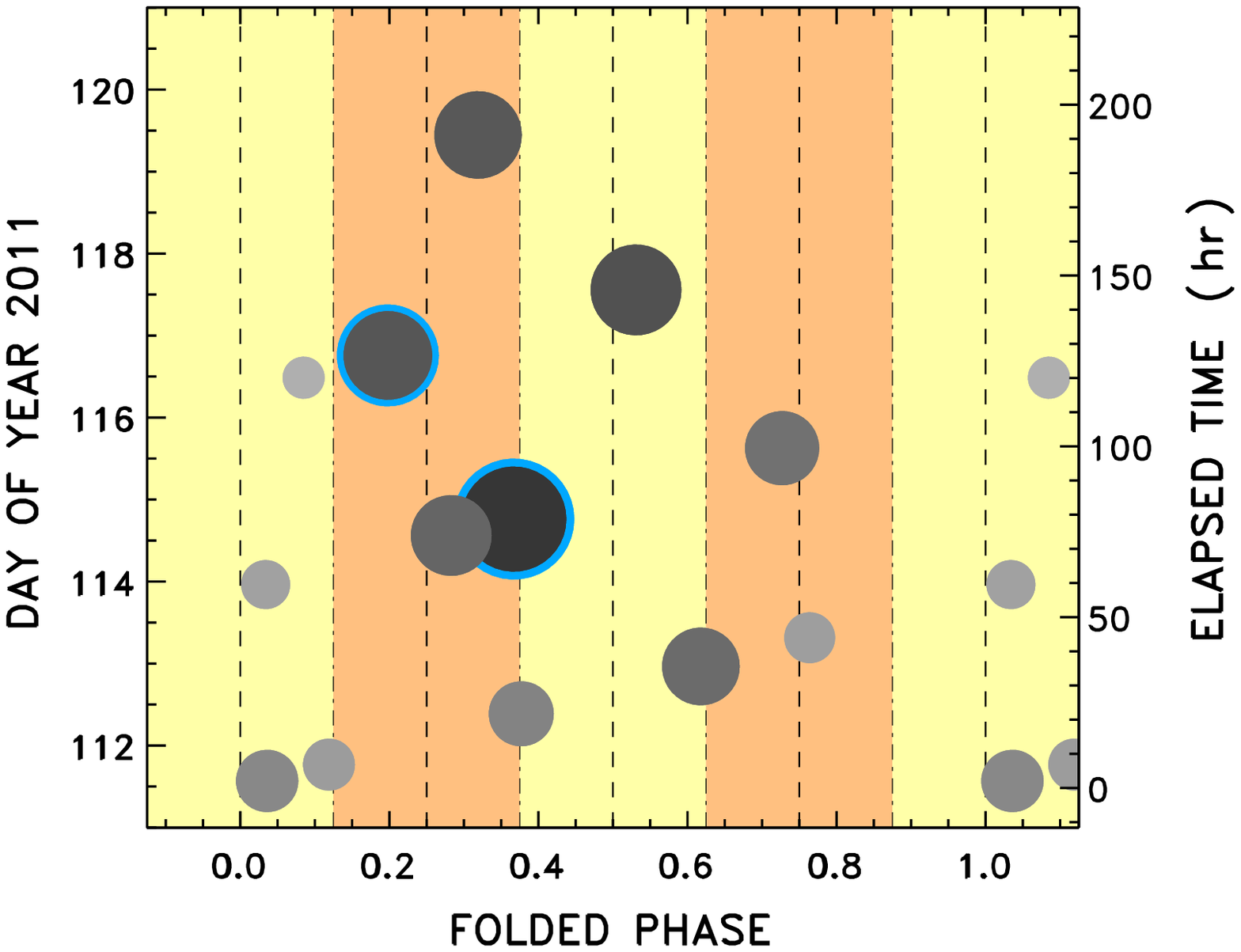}
\vskip -5mm
\figcaption[]{\small Schematic integrated fluxes of the \ion{Mg}{2} hk doublet of FK~Com over the period of the main campaign, in ascending time along the $y$-axis, and increasing (folded) phase along the $x$-axis.  Size of the symbol, and its shading, are related to the integrated flux: larger, darker circles represent higher fluxes.  The flux to symbol size scaling is exaggerated for display purposes.  Two suspected flares are outlined in blue.}
\end{figure}

\subsubsection{Profile Shapes of the COS NUV \ion{Mg}{2} Features}

A great deal of additional information, beyond the integrated fluxes, is encoded in the hk line shapes.  One can imagine, for example, that bright patches at discrete locations on the stellar surface would be mapped onto specific velocities in the blended line profiles according to the projected Doppler shifts across the visible hemisphere of the fast-spinning star.  This effect is apparent, for example, in the $\phi= 1.367$ \ion{Mg}{2} flare profile: distinct secondary emission bumps are present at $\sim 250$~km s$^{-1}$ blueward of the ISM reference velocity in both h and k.  This, of course, is the emission-line equivalent of the absorption ``Doppler imaging'' that has been applied so successfully for many decades to ground-based studies of FK~Com and its ilk.  In fact, the earlier {\em FUSE}\/ program recorded large {\em redshifts}\/ of the highly broadened FUV hot lines (\ion{C}{3} 977~\AA\ and \ion{O}{6} 1035~\AA\ doublet), but the few-hour pointing could not distinguish whether this was caused by an atmospheric dynamical effect -- say, strong downflows at $\sim$10$^{5}$~K as seen during a large FUV flare on EK~Dra (Ayres 2015a) -- or rather simply from an over-bright surface patch at the receding stellar limb (where the Doppler shift $\sim +\upsilon\sin{i}\sim 160$~km s$^{-1}$).  As mentioned earlier, resolving this issue was a major motivation for COCOA-PUFS.

In the earlier {\em FUSE}\/ study, an iterative deconvolution approach was applied to archival {\em IUE}\/ NUV echelle spectra to recover the intrinsic profile of the (stronger) \ion{Mg}{2} k line, plus the sharp interstellar absorption dip, in the face of the partial blending by the secondary h component.  Here, a slightly different approach was adopted; namely, a highly-constrained multiple-component numerical fit to the blended hk feature in each epoch.  The idea was to choose a reasonable analytic model for the stellar chromospheric emission profile(s), combine this with an assumed Gaussian line shape for the sharp absorption(s), but constrain the fitting process by assuming that the stellar h and k emission features have the same widths and velocity shifts (but possibly different fluxes), and similarly for the narrow central absorptions (but, of course, allowing different widths and velocities than the stellar counterparts).

Figure~11 illustrates the procedure for the best case: the epoch-average profile of \ion{Mg}{2} (ignoring the two spectra possibly compromised by flares).  The error bars are as before in Fig.~9: 5-pixel rebinning of the original fluxes with error flags set at $\pm$1~s.e.  The dotted curve well below the blended profile is the (heavily smoothed) average photometric error constructed by combining those of the individual spectra, in a way analogous to the standard error of the mean.  The horizontal orange dashed line is a baseline ``continuum'' level calculated from flux points (marked in red) on either side of, and well beyond, the emission wings of the hk blend; which was subtracted prior to the multi-component fitting (same as for the integrated fluxes mentioned earlier).  Green dots represent the original, un-binned epoch-average fluxes, which were the fundamental input to the constrained modeling technique.  Blue curves are the fitted k and h line stellar emission profiles, of the form:
\begin{equation}
(f_{\rm k})_{\lambda}= f_{\rm k}\, /\, (1 + (x^{\star}_{\rm k})^2)^\alpha\,\,.
\end{equation}
Here, $f_{\rm k}$ is the peak intensity of the k-line component; $x^{\star}_{\rm k}\equiv (\lambda - \lambda^{\star}_{\rm k}) / \Delta\lambda_{\rm L}$, where $\lambda^{\star}_{\rm k}$ is the observed wavelength of the stellar k emission and $\Delta\lambda_{\rm L}$ is a Lorentzian line width; and $\alpha$ describes any non-Lorentzian behavior of the empirical line shape ($\alpha\equiv 1$ is the normal Lorentzian profile).\footnote{This profile is related to the $\kappa$ distribution (e.g., Summers \& Thorne 1991).  It was exploited by Drake et al.\ (2008) to model bright X-ray lines in the earlier HETGS spectrum of FK~Com.  However, in that case the main influence was the well characterized HETGS instrumental profile, whereas here the stellar spectra are fully resolved by COS, and the generalized Lorentzian model simply is a way to capture non-Gaussian aspects of the FK~Com emission lines with a minimum of free parameters.}  The equivalent profile for the h line has $\lambda^{\star}_{\rm k}$ in the expression for $x^{\star}_{\rm k}$ replaced with $\lambda^{\star}_{\rm k}\,+\,\Delta\lambda_{\rm hk}$, where $\Delta\lambda_{\rm hk}= 7.179$~\AA\ is the laboratory separation between the doublet components.  In other words, the k and h lines were assumed to have the same velocity shift, as well as the same broadening parameter $\Delta\lambda_{\rm L}$, although the amplitude of the h-line profile, $f_{\rm h}$, was allowed to float. 

For the generalized Lorentzian profile, the key full width at half maximum intensity (FWHM) is given by:
\begin{equation}
{\rm FWHM}= 2\,\Delta\lambda_{\rm L}\,\sqrt{2^{(1/\alpha)}\,-1}\,.
\end{equation}

Superimposed on the sum of the two pseudo-Lorentzian emission profiles was a pair of Gaussian absorptions,
\begin{equation}
r_\lambda= 1 - r\,(\,\exp(-[x^{\rm abs}_{\rm k}]^2) + \exp(-[x^{\rm abs}_{\rm h}]^2)\,)\,\,.
\end{equation}
Here, $r$ is the maximum relative absorption depth; and $x^{\rm abs}_{\rm k}\equiv (\lambda - \lambda^{\rm abs}_{\rm k}) / \Delta\lambda_{\rm D}$, where $\lambda^{\rm abs}_{\rm k}$ is the observed wavelength of the k-line absorption, and $\Delta\lambda_{\rm D}$ is the Doppler line width.  The h-line absorption profile (last term, RHS) was assumed to be the same, aside from the $\Delta\lambda_{\rm hk}$ factor described above.  Note that $r$ could be slightly larger than unity owing to the continuum subtraction prior to the multi-component fitting.

\begin{figure}
\figurenum{11}
\vskip  0mm
\hskip -7mm
\includegraphics[scale=0.75,angle=90]{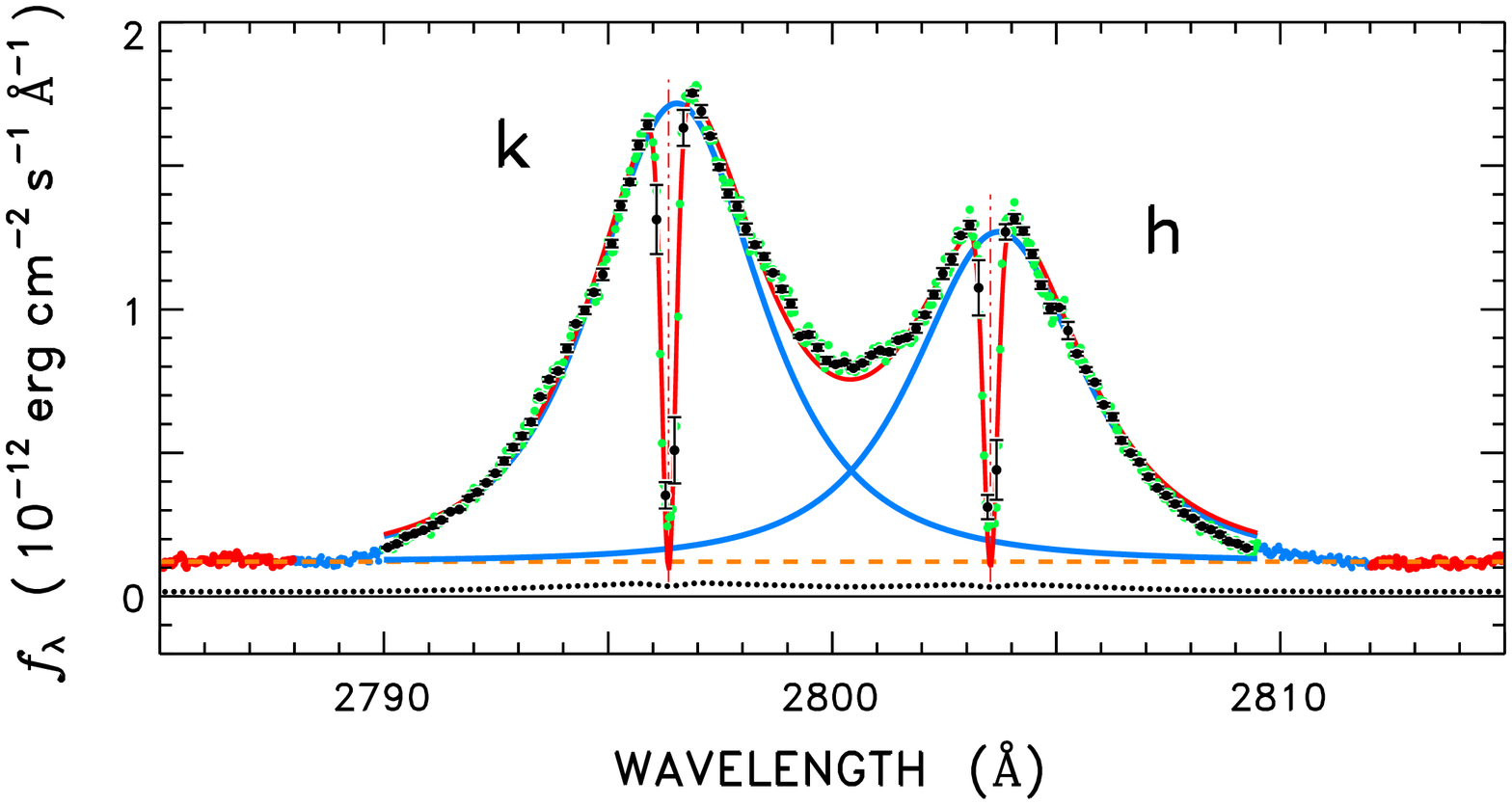}
\vskip 3mm
\figcaption[]{\small Multi-component constrained modeling of the blended stellar hk emission doublet of \ion{Mg}{2}, and the corresponding sharp absorptions, probably interstellar.  Red points in the extreme wings of the blended emission feature were used to define a continuum baseline (horizontal orange dashed line).  Green points represent the spectral fluxes upon which the modeling was based.  Blue points mark fluxes that were outside the fitted region but inside the continuum zone.  Error bars are 5-bin average fluxes (as in Fig.~9) for display.  The thick red curve is the combined emission plus absorption fit, while the blue curves illustrate the fitted stellar-only k and h components.  The vertical red dotted-dashed lines mark the laboratory wavelengths.  The observed profile is an average over the main program \ion{Mg}{2} observations, excluding two suspected flares, as registered to the apparent \ion{Mg}{2} absorption velocity.  The black dotted curve is the heavily smoothed 1\,$\sigma$ photometric error, per original wavelength point, calculated from the individual epoch values as a standard error of the mean.}
\end{figure}

Table~4 lists values of the various modeled parameters.  The stellar Doppler shifts are quoted relative to the \ion{Mg}{2} absorption velocity in each epoch.  Uncertainties were estimated by a Monte Carlo approach whereby the fitted line shape was perturbed by random realizations of the smoothed photometric error, then re-fitted.  Uncertainties were taken as the standard deviations of the re-modeled parameters over many trials.    The average \ion{Mg}{2} absorption velocity was $-19$~km s$^{-1}$ (1\,s.e.\ $< 1$~km s$^{-1}$), similar to what was derived originally from the older {\em IUE}\/ material, but significantly redward of the STIS FUV ISM velocity, or that inferred from the STIS G230MB-2836 spectrum of \ion{Mg}{2}, although with the caveats for the CCD observation mentioned earlier.  At the same time, the COS G285M-2676 setting is known to suffer systematic velocity errors, relative to the better calibrated STIS NUV medium-resolution echelle, of up to 2--4 pixels (10--20~km s$^{-1}$: Roman-Duval et al.\ 2014).  However, the specific calibration shifts referenced in that report go in the opposite direction of the discordant STIS--COS \ion{Mg}{2} absorption velocities, and would make the disparity even worse.

Note that the average \ion{Mg}{2} stellar emission profile appears to be redshifted with respect to the \ion{Mg}{2} absorptions by $+19$~km s$^{-1}$.  The inferred stellar radial velocity (i.e., RV$\sim$ 0~km s$^{-1}$ heliocentric) conflicts with the historical RV reported by Abt \& Biggs (1972: $-17$~km s$^{-1}$), or subsequent measurements by McCarthy \& Ramsey (1984: $-21{\pm}5$ km s$^{-1}$) and Huenemoerder et al.\ (1993: $-24{\pm}3$~km s$^{-1}$).  To be sure, as noted in Fig.~3, COS coverage is lacking for $\phi\sim 0.8-1.0$, so that potential contribution to the average profile is missing.  The average \ion{Mg}{2} profile appears to be more redshifted near $\phi\sim 0.75$, and more blueshifted near $\phi\sim 0$, so it is conceivable that the missing phase band would contribute a less redshifted profile to the average, and thus push the mean profile toward the stellar RV.  Because the apparent velocity excursions of the available \ion{Mg}{2} profiles are large ($1\,\sigma\sim 30$~km s$^{-1}$), the 20\% of the missing phases could make a non-negligible contribution to the average profile.

Despite the outward complexity of the \ion{Mg}{2} features of FK~Com, the generalized Lorentzian profile modeling does a respectable job reproducing the gross appearance of the empirical blends, except in some cases in the (less important) far line wings.  What results, then, is a simple representation of the stellar \ion{Mg}{2} profile that is built collectively from the two blended spectral components.  This provides a window into the mean behavior, consistent with the introductory nature of this report, although at the expense of glossing over some of the finer scale discrete emission structure carried in the individual line profiles.
  
Figure~12 illustrates the fitted stellar \ion{Mg}{2} k-line collectivized profiles alone (i.e., without the absorption components), displayed in the same way as Fig.~9.  This representation reveals the global profile distortions more readily.    

\clearpage
\begin{deluxetable}{lcccccc}
\tabletypesize{\small}
\tablenum{4}
\tablecaption{{\em HST}\/ COS NUV \ion{Mg}{2} Multi-Component Profile Parameters}
\tablewidth{0pt}
\tablecolumns{7}
\tablehead{
\colhead{dataset} & \colhead{$\lambda_{\rm 2796}$} & \colhead{${\rm FWHM}$} & 
\colhead{$f_{\rm 2796}$} & \colhead{$f_{\rm 2803}/f_{\rm 2796}$} & \colhead{$\alpha$} & 
\colhead{$f_{\rm INT}$} \\[3pt] 
\colhead{} & \multicolumn{2}{c}{(km s$^{-1}$)} & 
\colhead{($10^{-12}$ cgs \AA$^{-1}$) } & \colhead{} &\colhead{} & \colhead{($10^{-12}$ cgs) }  \\[3pt] 
\colhead{(1)} &  
\colhead{(2)} &
\colhead{(3)} &  
\colhead{(4)} &
\colhead{(5)} &  
\colhead{(6)} &             
\colhead{(7)}              
}
\startdata   
lbme02010  &  $-15$  &   340  &   1.31  &   0.72  &    2.2  &   7.5 \\
lbme03010  &  $-21$  &   425  &   1.70  &   0.67  &    1.4  &  12.9 \\ 
lbme04010$\star$  &  $-15$  &   530  &   1.69  &   0.78  &    1.6  &  16.6 \\
lbme55010  &  $+36$  &   370  &   1.68  &   0.70  &    1.2  &  11.6 \\
lbkm01010  &  $+31$  &   420  &   1.65  &   0.68  &    1.6  &  12.2 \\ 
lbkm02010  &  $+10$  &   500  &   1.47  &   0.79  &    1.8  &  13.2 \\
lbkm03010  &  $+40$  &   505  &   1.54  &   0.77  &    1.8  &  14.2 \\
lbkm04010  &  $+27$  &   440  &   1.59  &   0.67  &    1.6  &  12.2 \\
lbkm05010  &  $-20$  &   450  &   1.45  &   0.68  &    1.2  &  12.0 \\
lbkm06010  &  $+42$  &   535  &   1.62  &   0.73  &    3.8  &  14.4 \\
lbkm07010  &  $+46$  &   445  &   1.83  &   0.68  &    2.0  &  13.9 \\
lbkm08010  &  $+24$  &   390  &   1.58  &   0.71  &    1.2  &  11.6 \\
lbkm09010$\star$  &  $+84$  &   471  &   1.86  &   0.70  &    1.9  &  15.1 \\
lbkm10010  &  $+32$  &   515  &   1.66  &   0.73  &    1.8  &  15.3 \\
lbkm11010  &  $ +0$  &   520  &   1.60  &   0.81  &    2.0  &  15.0 \\
\cutinhead{Typical Monte-Carlo-based uncertainties}
  \nodata  &  ${\pm}3$ &   ${\pm}6$ & ${\pm}0.02$ & ${\pm}0.02$ & ${\pm}0.1$ & \nodata \\
\cutinhead{Main program average, non-flare epochs}
           & $+19$   &   $470$  &  1.60   &  0.72   &    1.7  & 13.4 \\
\cutinhead{Averages and standard deviations over all the epochs}
  \nodata  &  $+20{\pm}29$ &   $460{\pm}60$ & $1.62{\pm}0.14$ & 
$0.72{\pm}0.04$ & $1.8{\pm}0.6$ & $13{\pm}2$  \\
\enddata
\tablecomments{Col.~(1) datasets marked ``$\star$'' are suspected flares.  Cols.~(2) and (3) wavelength parameters expressed in equivalent velocity units; and in the case of Col.~(2), relative to the derived \ion{Mg}{2} absorption velocity in that spectrum.  Cols.~(4) and (7) cgs units are erg cm$^{-2}$ s$^{-1}$.  Col.~(5) is ratio of the peak intensities of the two stellar \ion{Mg}{2} emission components.  Col.~(6) is an exponent that describes the departure of the stellar emission profile from a pure Lorentzian line shape ($\alpha\equiv 1$).  The Col.~(7) fluxes were integrated between 2790--2810~\AA, above a continuum baseline, and have negligible formal uncertainties owing to the low noise per pixel and many points that contributed to the sum.  The continuum flux was derived from two flanking bands: $2786.5{\pm}1.5$~\AA\ and $2813.5{\pm}1.5$~\AA.  In addition to these more variable parameters, more constant values derived in the fits were: absorption (``ISM'') velocity-- $\upsilon= -19 {\pm} 2 ({\pm}1 [1\,{\rm s.e.}])$~km s$^{-1}$; absorption Doppler (e-folding) width-- $\Delta\lambda= 23 {\pm} 1$~km s$^{-1}$; relative absorption depth-- $r= 1.02 {\pm} 0.02$ (dimensionless); and continuum level-- $f_{\rm C}= 0.12 {\pm} 0.01$ (same units as Col.~[4]).  (In this, and subsequent Tables of this type, the Col.~(2) velocities have been rounded to the nearest 1~km s$^{-1}$, and the Col.~(3) widths to the nearest 5~km s$^{-1}$, consistent with the measurement errors.)
}
\end{deluxetable} 

\clearpage
\begin{figure}
\figurenum{12}
\vskip  0mm
\hskip  7mm
\includegraphics[scale=0.875]{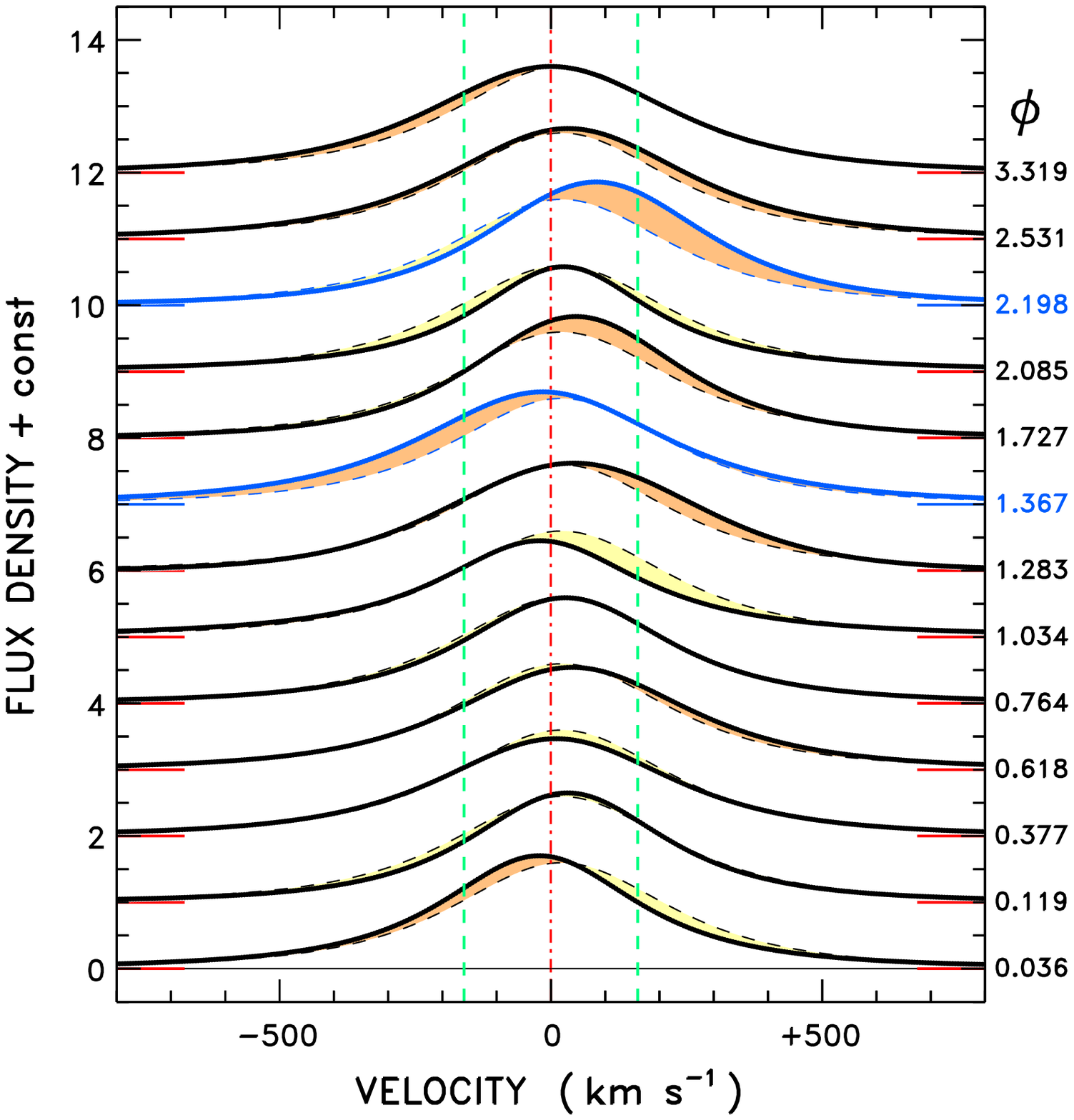}
\vskip 0mm
\figcaption[]{\small Similar to Fig.~9, but now just the fitted stellar-only ``collectivized'' profiles of \ion{Mg}{2} k (thick curves).  As in Fig.~9, thinner black dashed tracings represent the (fit to the) non-flare epoch-average profile, and shading highlights whether the local profile exceeds, or falls below, the average.  The vertical red dotted-dashed line marks the (absorption) velocity zero point, and the green dashed lines are ${\pm}\upsilon\sin{i}$.
}
\end{figure}

\clearpage
\subsection{Interstellar Velocities in the Direction of FK~Com}

As a prelude to a discussion of empirical corrections to the COS FUV wavelength scales for COCOA-PUFS, based on the velocities of sharp interstellar absorptions, it is worth considering what is known about the ISM toward FK~Com, 10$^{\circ}$ from the North Galactic Pole (NGP), at a distance of about 230~pc from the Sun.

There are two main local ($d< 100$~pc) interstellar clouds in that direction: the so-called Geminorum (Gem) and NGP structures, in the nomenclature of Redfield \& Linsky (2008).  FK~Com is firmly inside the angular boundaries of the NGP cloud, and possibly grazes the Gem feature.  The projected cloud motions in the direction of FK~Com, according to the velocity vectors of Redfield \& Linsky (2008), are $-8$~km s$^{-1}$ for Gem and $-11$~km s$^{-1}$ for NGP, both redward of either the STIS FUV \ion{O}{1}\,+\,\ion{Si}{2}\,+\,\ion{C}{2} absorption velocity ($-28$~km s$^{-1}$) or the COS NUV \ion{Mg}{2} absorption velocity ($-19$~km s$^{-1}$, at face value).  

On the other hand, the distant halo star vZ\,1128 (O8p: $d\sim 10$~kpc), 5$^{\circ}$ from FK~Com, has an even more blueshifted average absorption velocity, $-$39~km s$^{-1}$, based on archival STIS FUV echelle spectra (of \ion{O}{1} 1302~\AA, \ion{Si}{2} 1304~\AA, \ion{C}{2}~1334~\AA, and \ion{Si}{2}~1526~\AA).  These ISM absorptions of vZ\,1128 are so broad that it would be easy to hide a $-19$~km s$^{-1}$ component (e.g., FK~Com \ion{Mg}{2} absorption velocity) or a $-28$~km s$^{-1}$ component (FK~Com FUV absorption velocity).  However, vZ\,1128 does display a triplet of ground-state \ion{N}{1} absorptions (1199.5~\AA, 1200.2~\AA, and 1200.7~\AA) that appear to be less saturated than the features longward of 1300~\AA, and which are centered at the $-28$~km s$^{-1}$ velocity seen in the FK~Com STIS spectrum.

The hot white dwarf HZ\,43 (DA.9; $d= 39$~pc) lies in the direction of FK~Com, about 8$^{\circ}$ away, but in the foreground, rather than in the background like vZ\,1128 (the latter is far above the Galactic plane).  HZ\,43 has a single absorption system, at $-6$~km s$^{-1}$, associated with the dominant ISM cloud in that direction, the NGP (Redfield \& Linsky 2004).  Likewise, the yellow giant 31~Comae, about 9$^{\circ}$ from FK~Com at a distance of 89~pc, and the DA white dwarf GD\,153, 10$^{\circ}$ from FK~Com at 60~pc (intermediate to HZ\,43 and 31~Com), both have a single interstellar absorption system, at $-5$~km s$^{-1}$, again consistent with the NGP cloud.

Finally, another hot sdO, HD\,113001\,B, is in the direction of FK~Com, about 13$^{\circ}$ away and at almost the same distance ($\sim 220$~pc: see Ayres et al.\ 2006).  Like several of the other stars described earlier, HD\,113001\,B falls almost exclusively within the angular boundaries of the NGP cloud.  The predicted ISM velocity in that direction is only $-2$~km s$^{-1}$.  However, the {\em FUSE}\/ spectrum of HD\,113001\,B described by Ayres et al.\ (2006) suggests an absorption velocity of $-25$~km s$^{-1}$, and indeed the sharp lines of the subdwarf appear less saturated and less blueshifted than the corresponding {\em FUSE}\/ features of even more distant vZ\,1128 ($\upsilon_{\rm ISM}\sim -39$~km s$^{-1}$).  Nevertheless, the {\em FUSE}\/ wavelength scales are notoriously difficult to calibrate accurately (Bowen et al.\ 2008), and unfortunately there are no STIS FUV spectra of HD\,113001\,B to confirm the {\em FUSE}\/ velocities.

The tentative conclusion is that relatively unsaturated ISM absorptions in the FK~Com FUV spectrum (STIS\,+\,{\em FUSE}\,) are from an absorption system at $-28$~km s$^{-1}$, more blueshifted than the dominant interstellar cloud, NGP, in that direction.  At the same time, the likely more saturated \ion{Mg}{2} NUV features might have additional components at lower (negative) velocities contributed by the NGP cloud itself, or other substructures, including perhaps circumstellar gas around FK~Com itself.  Since the $-28$~km s$^{-1}$ system is not seen in other, less distant, stars in the FK~Com direction (e.g., HZ\,43, 31~Com, and GD\,153), the absorbing gas likely is beyond the 100~pc horizon of the Local ISM surveys.  It remains, then, to undertake a more direct comparison between the FK~Com FUV and NUV spectra to explore whether the apparent $\sim 10$~km s$^{-1}$ discrepancy between the STIS FUV absorption system and the COS \ion{Mg}{2} absorption velocity might be a result of saturation effects, pulling the NUV resonance lines redward; different absorbing structures (e.g., interstellar vs.\ circumstellar); or systematic errors in the COS NUV wavelengths.  This comparison will be described shortly.


\subsection{COS FUV Spectra}

In this introduction to COCOA-PUFS, the intent is to present top-level results from preliminary analyses of the multi-wavelength observations, to be followed by more detailed dissections of the diverse datasets in future publications.  The following section focuses on the correction of systematic errors in the COS G130M and G160M wavelength scales.  That will be followed by a description of the epoch-average FUV spectrum over the eight-day main campaign.  The presentation then moves to time-resolved COS profiles of \ion{C}{2} 1334~\AA\ and \ion{Si}{4} 1393~\AA\ during the main monitoring period; and higher time-resolution integrated fluxes of key diagnostic lines during that interval, as a prelude to a comparison with time-resolved soft X-ray fluxes from the contemporaneous {\em Chandra}\/ pointings, as well as the ground-based photometric context measurements in the visible.

\subsubsection{STIS Velocity Cross-Calibration of COS FUV Spectra}

The main objective of the COS part of COCOA-PUFS was to closely follow the dynamical evolution of the FUV emissions of FK~Com over several rotational cycles, as was done for EK~Dra, although in that instance only part of a rotation was captured (by design).  Because the narrower chromospheric and higher temperature line profiles of EK~Dra were much more stable in velocity than the FK~Com counterparts, it was possible to compare the STIS reference spectrum directly against the numerous COS FUV emissions in common, to evaluate deviations of the COS wavelengths against the better calibrated STIS scales.  The result was an empirical correction, in terms of equivalent velocities, of the form:
\begin{equation}
\lambda_{\rm new}= \lambda_{\rm old} \times (1\, +\, \Delta{\upsilon} / c)\,\,   ,
\end{equation} 
where,
\begin{equation}
\Delta{\upsilon}= \upsilon_{0} + a\, (\lambda - \lambda_{0})\,\,\,   .
\end{equation}
The same zero-point offset, $\upsilon_{0}= -3$~km s$^{-1}$, was derived for the three unique grating/segment combinations of the EK~Dra program (same settings as for COCOA-PUFS).  The $a$ coefficients ranged from $-$0.11~km s$^{-1}$ \AA$^{-1}$ to $-$0.05~km s$^{-1}$ \AA$^{-1}$, with $\lambda_{0}= 1340$~\AA\ for G130M/A, 1400~\AA\ for G160M/B, and 1640~\AA\ for G160M/A.  The uniformly negative $a$ coefficients signaled systematically increasing {\em redshifts}\/ of the (uncorrected) COS spectra, in both independent grating settings, relative to gold-standard STIS.

For the (larger) collection of more variable FK~Com spectra, the following strategy was followed.  First, all the independent COS FUV spectra taken during the main program (GO-12376 Visits 3\,+\,4 and   GO-12279 Visits 1--11 for G130M-1291; GO-12376 Visits 3\,+\,4 for G160M-1577) were fitted for the apparent absorption velocities of \ion{O}{1} 1302~\AA, \ion{Si}{2} 1304~\AA, and \ion{C}{2} 1334~\AA\ in G130M/A; \ion{Si}{2} 1526~\AA\ in G160M/B; and \ion{Al}{2} 1670~\AA\ in G160M/A.  (Unfortunately, the two absorptions in the G160M range were too weak to measure in the reference STIS spectrum.)  The model assumed a smooth distribution of the local background intensities with a superimposed sharp Gaussian absorption dip.  The basic datasets were the so-called ``x1dsum'' {\sf CALCOS} pipeline files, which co-add the individual FP-POS splits into a single spectrum, correcting for detector fixed patterns such as the grid wire shadows.  These observations necessarily average over some level of stellar variability during the partial orbit (GO-12279: G130M; GO-12376: G160M) or multi-orbit (GO-12376: G130M) exposures, but the intent was to measure the presumably stationary interstellar absorptions, so the varying background was of little concern.  Next, the individual datasets were corrected for the average local velocity shifts and combined to boost S/N.  This master spectrum was re-fitted for the set of reference absorption features, to determine the offsets relative to the assumed $-28$~km s$^{-1}$ average velocity of the three prominent ISM absorptions captured by STIS (1302~\AA, 1304~\AA, and 1334~\AA).  Finally, the coefficients in equation~6 were adjusted to achieve the best match to the individual segment velocities.  Because there were only two calibrators for G160M, one for each detector side, it was assumed that the $\upsilon_{0}$ and $a$ parameters were the same for both sides (this was the case for EK~Dra).  Figure~13 illustrates the fitting procedure schematically. 

With the available, though limited, calibration material, the best-fit parameters were: $\upsilon_{0}\sim -6$~km s$^{-1}$ and $a\sim +0.18$~km s$^{-1}$ \AA$^{-1}$ for G130M/A; $\upsilon_{0}\sim -1.5$~km s$^{-1}$ and $a\sim -0.08$~km s$^{-1}$ \AA$^{-1}$ for G160M/BA.  The values for G160M are similar to those found for EK~Dra, but the G130M/A parameters differ significantly.  Especially note that the key $a$ gradients are of opposite sign.  This suggests -- at least for G130M/A -- that the COS velocity distortions are not constant, but rather vary with an unknown secondary parameter.  Unfortunately, the primary G130M/A calibrators are located at the blue end of that segment, and there are no useful ISM absorptions at the red end.  Thus, it cannot be verified whether the linear velocity model derived from the blue end is valid 60~\AA\ away in the 1400~\AA\ part of G130M/A where the key \ion{Si}{4} doublet falls.  Accordingly, the \ion{Si}{4} velocities derived later should be viewed with that uncertainty in mind (but see end of {\S}3.4.2, below).

\begin{figure}
\figurenum{13}
\vskip  0mm
\hskip  6mm
\includegraphics[scale=0.875]{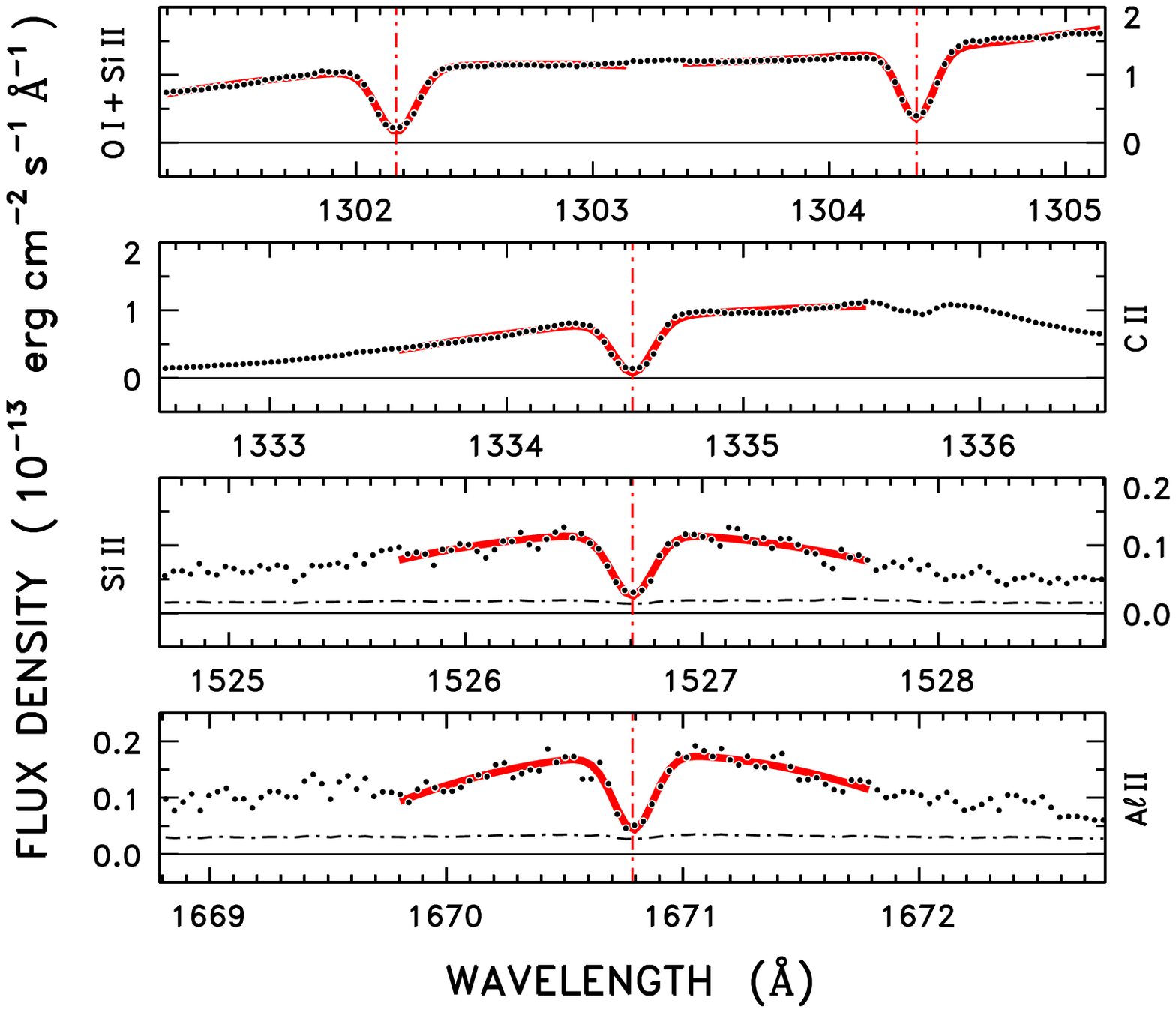}
\vskip 0mm
\figcaption[]{\small ISM absorption characterization for the two independent grating settings used in the COCOA-PUFS campaign: G130M-1291, upper two panels; G160M-1577, lower two panels (note large $y$-axis scale change).  The black dots are epoch-average tracings from the respective grating settings.  Black dotted-dashed curves in the lower two panels are smoothed 1\,$\sigma$ photometric errors.  For the upper two panels, the analogous curves would be indistinguishable from the zero level and are not displayed.  The fits (red solid curves) impose a narrow Gaussian absorption on a smooth quadratic background.  The spectra have been corrected for the systematic COS velocity errors, and registered to the absorption velocity frame.  The vertical red dotted-dashed lines represent the laboratory wavelengths of the ISM absorptions.
}
\end{figure}

\clearpage
\subsubsection{The Radial Velocity of FK Comae}

The cross-calibration described in the preceding section paves the way to correct the COS FUV spectra of FK~Com into the reference frame of the STIS absorption velocity ($-28$~km s$^{-1}$ heliocentric).  In the absence of evidence to the contrary, it will be assumed that the five FUV absorptions recorded by COS all are at the same reference velocity, so that the wavelength corrections to the COS G130M and G160M scales can be applied as stated.  Certainly, the assumption will be most appropriate for the heavily-used G130M/A segment, which has three of the five reference lines, all of which were recorded by STIS, with a minimal $\sim {\pm}1$~km s$^{-1}$ empirical velocity dispersion among them.  There is the possibility, however, that the strongest low-excitation absorptions -- NUV \ion{Mg}{2} hk -- might have added contributions at different velocities, say circumstellar material associated with FK~Com itself.  The $\sim +10$~km s$^{-1}$ redward displacement of the COS \ion{Mg}{2} absorption velocity with respect to the STIS FUV velocity might be an example of that effect (or, it might simply be due to a systematic velocity error in the COS NUV setting).

One lingering question is the RV of the stellar photosphere, for which there is some disagreement in the existing literature.  This bears on the discussion of Doppler shifts of chromospheric and higher temperature species at different phases.  To this end, a new estimate of the photospheric RV was derived from a series of ten UVES high-resolution spectra of FK~Com, taken over a period of five days (two rotations) just a few days prior to the COCOA-PUFS main program.  A cross-correlation was performed against an RV standard of similar type, rotationally broadened to the $\upsilon\sin{i}$ of the yellow giant, over the region 6050--6170~\AA, which has relatively isolated absorption features.  The derived RV's ranged from $-16$~km s$^{-1}$ to $-21$~km s$^{-1}$, with an average of $-18{\pm}2$~(0.5)~km s$^{-1}$ (parenthetical value is 1\,s.e.).  The dispersion of the velocity measurements is larger than the nominal accuracy of high-precision UVES, but not unexpected for a heavily spotted star whose absorption profiles can be distorted by the dark umbrae, and slighted shifted in apparent centroid depending on the disk locations of the spots (see, especially, the study by Huenemoerder et al.\ [1993], who documented systematic variations of their RV measurements against photometric phase with a semi-amplitude of $\sim$~3 km s$^{-1}$).  The UVES estimate of the RV of FK~Com is on the lower side of the range $-17$~km s$^{-1}$ to $-24$~km s$^{-1}$ reported in the historical work cited earlier.  The smaller velocity dispersion, nevertheless, puts even more stringent constraints on binary scenarios for FK~Com, as outlined by McCarthy \& Ramsey (1984).  The conclusion is that the apparent photospheric RV of FK~Com probably is sensitive to the distribution of spots in a given epoch, and thus depends on how the observations are averaged over phase.  In any event, the RV of FK~Com in the COCOA-PUFS epoch appears to be $\sim 10$~km s$^{-1}$ redward of the FUV absorption (``ISM'') velocity measured by STIS, but close to the nominal \ion{Mg}{2} absorption velocity.

Finally, low-excitation, optically-thin chromospheric features can be exploited as a possible check on the photospheric RV of FK~Com, and the COS velocity correction: such lines in stellar FUV spectra often are used as surrogates for the photospheric RV (Ayres et al.\ 2003).  Unfortunately, due to the heavy rotational blending, there are only two suitable candidates in the high-S/N G130M/A range: \ion{Cl}{1} 1351~\AA\ and \ion{N}{1} 1411~\AA.  Figure~14 illustrates these two features from the COS epoch-average FUV spectrum, shifted into the photospheric RV frame of FK~Com as indicated by UVES.  The agreement with a broadened version of the RV-corrected 24~UMa spectrum is good in both cases, which provides additional confidence in the UVES-derived RV, as well as the COS wavelength correction (especially for 1411~\AA, at the far red end of G130M/A).  The agreement also suggests that there are no large differential motions between the low chromosphere and photosphere, which bears on interpretations of the apparent \ion{Mg}{2} redshifts.

\subsubsection{The Epoch-Average FK Com FUV Spectrum}

Figure~15 is an overview of the epoch-average COS FUV spectrum of FK~Com together with comparison yellow giant 24~UMa, adjusted to the RV frames of both stars.  The FK~Com spectrum is noisier beyond 1415~\AA\ (where the G160M part was spliced in) owing to the lower exposure in the G160M segments compared with G130M/A.  Nevertheless, the epoch-average COS tracings are much higher in S/N overall than the partial-orbit STIS spectrum illustrated in Fig.~8.  COS reinforces the view earlier from STIS (and {\em FUSE}\,): the FK~Com lines are extraordinarily broad, so much so that features normally separated (e.g., \ion{C}{2} 1334~\AA\,+\,1335~\AA) now have become strongly blended, which (as with \ion{Mg}{2}) makes it challenging to isolate the individual profile shapes.  Table~5 lists integrated fluxes of several of the COS blended line complexes, analogous to the STIS measurements in Table~3.

\begin{figure}[hb]
\figurenum{14}
\vskip  0mm
\hskip  0mm
\includegraphics[scale=0.925]{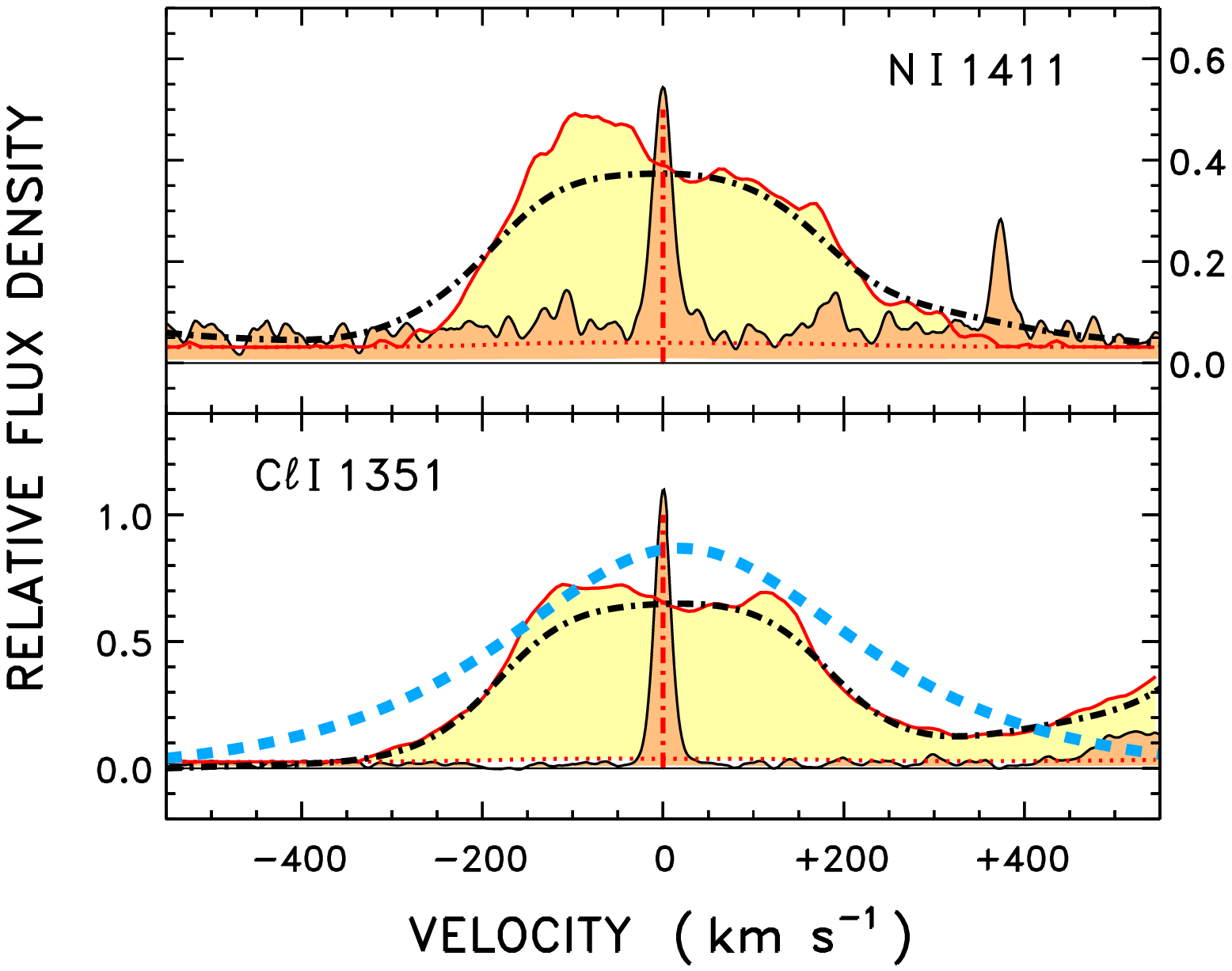}
\vskip 0mm
\figcaption[]{\small STIS FUV spectrum of 24~UMa (orange shaded, black bordered), at relatively isolated \ion{Cl}{1} 1351~\AA\ (lower panel) and \ion{N}{1} 1411~\AA\ (upper panel), contrasted with the FK~Com epoch-average wavelength-corrected COS G130M/A spectrum (yellow shaded, red bordered), shifted into the UVES-defined photospheric frame.  The red dotted horizontal lines indicate the smoothed 1\,$\sigma$ photometric errors for FK~Com.  Note the ``horns'' on the observed 1351~\AA\ and 1411~\AA\ profiles of FK~Com, which indicate a high degree of limb brightening (Saar et al.\ 2014; see also Huenemoerder et al.\ 1993, their Fig.~8 for limb-brightened \ion{He}{1} D$_{3}$).  The black dotted-dashed curve is the 24~UMa spectrum broadened with an ad hoc rotational profile and scaled to match the FK~Com counterpart.  (In some cases, features in the 24~UMa spectrum near the reference lines were modified in strength to better match the apparent FK~Com intensities beyond the reference line cores.)  The blue dashed curve in the lower panel is the (scaled) epoch-average \ion{Mg}{2}~k collective fit, for comparison, in its absorption reference frame taking the COS NUV wavelength scale at face value.  (If the \ion{Mg}{2} sharp absorptions instead are at the FUV ISM velocity, stellar \ion{Mg}{2}~k would be shifted $10$~km s$^{-1}$ to the blue, but still would be redshifted relative to the photospheric RV by $\sim$10 km s$^{-1}$.)  Note the extra high-velocity broadening of the \ion{Mg}{2} profile relative to the optically thinner \ion{Cl}{1} and \ion{N}{1} features.
}
\end{figure}

\clearpage
\begin{figure}
\figurenum{15}
\vskip  -10mm
\hskip  -18mm
\includegraphics[scale=1.125]{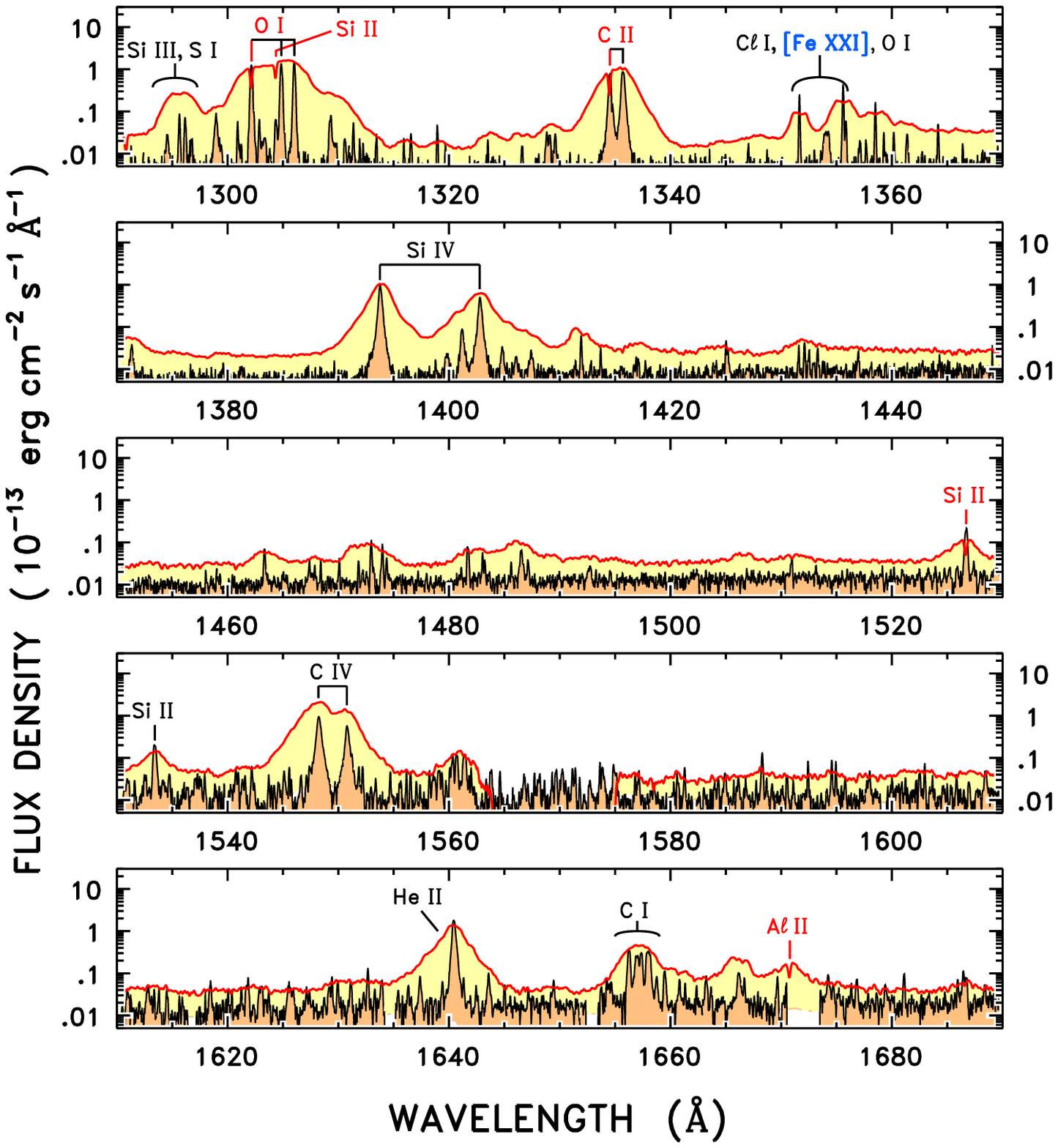}
\vskip 0mm
\figcaption[]{\small Epoch-average COS FUV spectrum of FK~Com (yellow shaded, red outlined) compared to 24~UMa (orange shaded, black outlined), now in the RV frames of both stars.  Prominent features are identified.  The five reference interstellar absorptions described earlier are highlighted in red.  
}
\end{figure}

\begin{deluxetable}{rcccc}
\tablenum{5}
\tablecaption{{\em HST}\/ COS FUV Measurements: Epoch-Average FK~Com Spectrum}
\tablewidth{0pt}
\tablecolumns{5}
\tablehead{
\colhead{Transition} &  
\colhead{$\lambda$} & 
\colhead{$\Delta\lambda$} &  
\colhead{$f_{\rm INT}$} &
\colhead{$f_{\rm C}$}\\[3pt]
\colhead{} &  
\multicolumn{2}{c}{(\AA)} &
($10^{-13}$ cgs) & ($10^{-13}$ cgs \AA$^{-1}$)\\[3pt]
\colhead{(1)} &  
\colhead{(2)} &
\colhead{(3)} &  
\colhead{(4)} &
\colhead{(5)}  
}                
\startdata   
\ion{Si}{3}~1295  &         1295.5  &   2.5 &   0.70 &  0.03\\
\ion{O}{1}~1305   &         1304.0  &   4.0 &   7.6  &  0.02\\
\ion{C}{2}~1335   &         1335.5  &   4.5 &   3.1  &  0.02 \\
\ion{Cl}{1}~1351   &  1351.5  &   1.5 &   0.12 &  0.02 \\
\ion{O}{1}~1355   &         1355.2  &   1.8 &   0.34 &  0.02\\
\ion{Si}{4}~1400  &         1397.5  &   9.0 &   4.3  &  0.02 \\
\ion{C}{4}~1550   &         1549.0  &   6.0 &   8.4  &  0.04\\
\ion{He}{2}~1640  &         1640.0  &   5.0 &   3.6  &  0.04 \\
\ion{C}{1}~1657   &         1656.5  &   2.5 &   1.3  &  0.04\\
\enddata
\vskip 0mm
\tablecomments{Col.~(1) wavelength designator, following species, in \AA.  Col.~(2) is the central wavelength of the integrated flux measurement.  Col.~(3) is the half width of the integration band.  The cgs units of the Col.~(4) integrated fluxes and Col.~(5) continuum flux densities are: erg cm$^{-2}$ s$^{-1}$.  These flux measurements have negligible formal photometric errors.}
\end{deluxetable} 

\clearpage
\subsubsection{Time-Resolved Profiles of \ion{C}{2} 1334~\AA}

Together with the short exposures of NUV \ion{Mg}{2} at the beginning of every pointing, the FUV G130M-1291 sequence -- either the short version of GO-12279 or the long program of GO-12376 -- formed the backbone of the multi-rotational-cycle monitoring effort.  The main spectral diagnostics in the G130M range are the \ion{C}{2} multiplet at 1335~\AA\ and the \ion{Si}{4} doublet at 1400~\AA.  The other dominant feature is the \ion{O}{1} triplet at 1305~\AA.  However, the blending of the multiple spectral components, including unrelated emissions, is more severe than for \ion{C}{2}, and the stellar \ion{O}{1} features are contaminated to an unknown degree in each visit by atomic oxygen airglow emission; so the \ion{O}{1} blend is not considered further.  The \ion{C}{2} triplet effectively is a close doublet of similar strength features, with a strong ISM absorption near the center of the shortward component, and is described first.

\begin{figure}
\figurenum{16}
\vskip  0mm
\hskip  7mm
\includegraphics[scale=0.875]{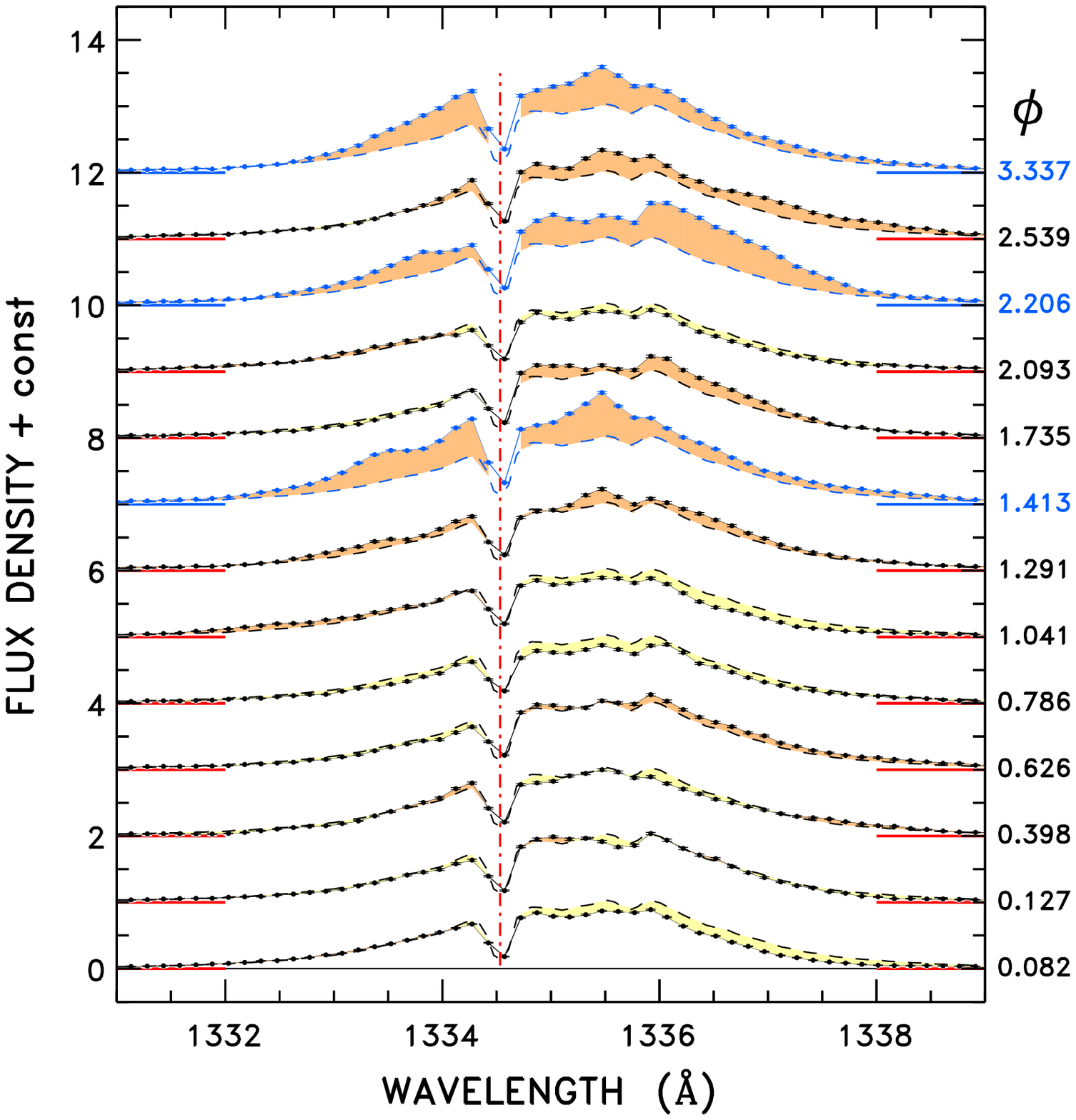}
\vskip -5mm
\figcaption[]{\small \ion{C}{2} profiles from the G130M-1291 exposures of the main program, in the velocity frame of the \ion{C}{2} 1334~\AA\ absorption feature.  The profiles are arranged, from bottom to top, in increasing time, and are offset by one unit of the flux density axis ($10^{-13}$ erg cm$^{-2}$ s$^{-1}$ \AA$^{-1}$) for clarity (horizontal colored ticks refer to the zero levels of the individual spectra).  Photometric cycle numbers are listed at the right.  The vertical dotted-dashed line marks the laboratory wavelength of the \ion{C}{2} 1334~\AA\ resonance transition.  The three profiles marked in blue are anomalous, probably affected by flares.  Thin black dashed curves represent the epoch-average profile, excluding the three suspected flares.  Darker/lighter shading indicates whether the local profile is higher/lower than the average.  Error bars represent a $\sim$2~resel re-binning of the (heavily oversampled) original flux densities.}
\end{figure}

Figure~16 depicts the \ion{C}{2} blend over time during the main program, including deviations from a mean profile constructed from the epochs identified as non-flaring.  As in Fig.~9, each visit-average profile is registered to its (now \ion{C}{2}) absorption velocity.  The average measured absorption value (from a detailed multi-component fit: see below) was $-28{\pm}2 ({\pm} 0.4)$~km s$^{-1}$.

Figure~17, like Fig.~10, illustrates integrated flux measurements of the \ion{C}{2} blend as a function of time and (folded) phase.  The values are reported later, in Table~6.  Note, again, the concentration of high intensities near $\phi\sim 0.3$, including the three suspected flares.

\begin{figure}
\figurenum{17}
\vskip  0mm
\hskip  8mm
\includegraphics[scale=0.875]{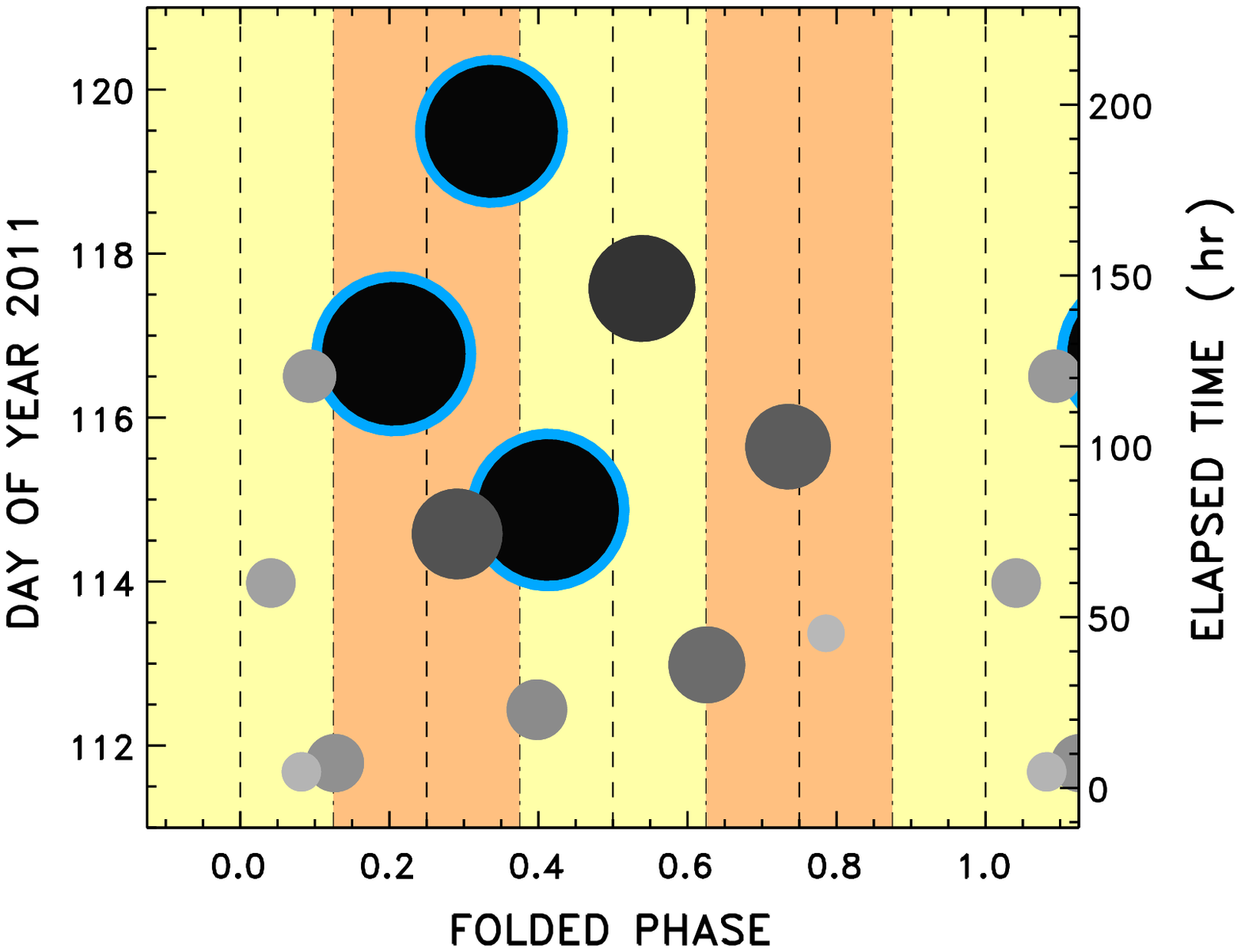}
\vskip -5mm
\figcaption[]{\small Schematic integrated fluxes of the \ion{C}{2} multiplet of FK~Com over the period of the main campaign, in ascending time along the $y$-axis, and increasing phase along the $x$-axis.  The flux to symbol size scaling is exaggerated for display.  Three suspected flares are outlined in blue.}
\end{figure}

As with the \ion{Mg}{2} blend described earlier, a multi-component emission/absorption model was applied to the \ion{C}{2} feature to extract global properties of the line shapes.  The analytical description was essentially identical to \ion{Mg}{2}, although here only a single absorption component, 1334~\AA, was fitted, and further the $f_{\rm 1335}/f_{\rm 1334}$ peak flux density ratio was fixed at the value 1.3 inferred from the 24~UMa spectrum.  The two emission components of \ion{C}{2} are close enough together that a floating peak flux density ratio sometimes would produce nonsensical ratios, so the additional constraint (in the Bayesian sense of utilizing prior information) was applied.  Figure~18 illustrates the fitting procedure; Table~6 summarizes derived values of the various parameters; and Figure~19 shows the fitted ``collectivized'' profiles.  Note in Fig.~18 that a weak secondary absorption component appears at 1335.7~\AA\ in the non-flare average profile.  This was not modeled in the individual profiles because the S/N was too low, and in any event the influence on the line shapes is minor.  However, the presence of the excited ground-level state \ion{C}{2} feature indicates that the absorbing gas is relatively warm (compared to normal LISM measurements where the 1335~\AA\ component is absent).

Comparing Tables~4 and 6, the centroid velocities, relative to the respective absorption systems, of the average non-flare emission profiles of \ion{Mg}{2} and \ion{C}{2} differ by 10~km s$^{-1}$, with \ion{C}{2} more redshifted relative to its absorption system than \ion{Mg}{2}.  At the same time, the profile shapes are closely congruent, with similar $\alpha$ parameters ($\sim 2$) and FHWMs (470~km s$^{-1}$ and 490~km~s$^{-1}$, for the non-flare, epoch-average profiles of \ion{Mg}{2} and \ion{C}{2}, respectively).  The similarity suggests that the two chromospheric features probably are similarly optically thick, and likely reflect the same kinematic environments.  If the respective reference absorption velocities ($-19$~km s$^{-1}$ and $-28$~km s$^{-1}$) are taken at face value, then the radial velocities of the two non-flare epoch-average chromospheric features would be nearly identical and close to 0~km s$^{-1}$ heliocentric, a roughly 20~km s$^{-1}$ {\em redshift}\/ beyond the stellar photospheric RV.  As noted earlier, the additional redshift might be due to the ``missing phases'' effect, or it might have a more subtle gas-dynamical origin (neither of which, apparently, are adversely affecting the weaker chromospheric \ion{Cl}{1} or \ion{N}{1} features modeled earlier).

\begin{figure}
\figurenum{18}
\vskip  0mm
\hskip -7mm
\includegraphics[scale=0.75,angle=90]{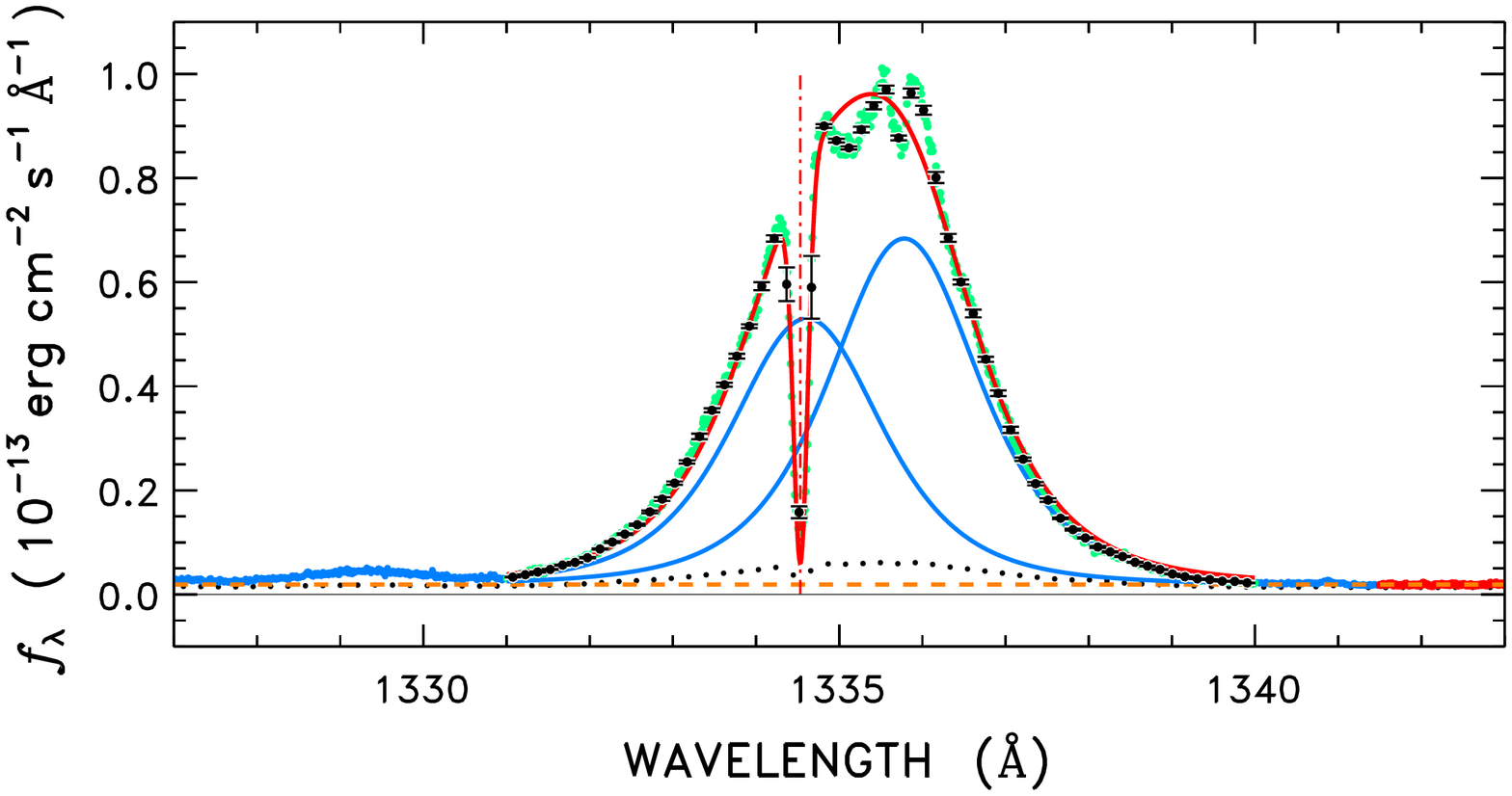}
\vskip 3mm
\figcaption[]{\small Multi-component constrained fitting of the two main blended stellar components of \ion{C}{2}, including the unrelated sharp, likely interstellar, \ion{C}{2} resonance absorption.  The observation is the epoch average from the main program, excluding three suspected flares, as registered to the apparent \ion{C}{2} absorption velocity.  Red points beyond the longwavelength wing of the blend were used to define a continuum background (horizontal orange dashed line).  Green points represent the fluxes upon which the modeling was based.  Blue points are fluxes that were not included in the fitting procedure or the background level.  Error bars are 15-bin average fluxes (about 2 resels), and local standard errors of the mean, for display.  The thick red curve is the combined fit; the blue curves illustrate the fitted emission sub-components.  The vertical red dotted-dashed line marks the laboratory wavelength of \ion{C}{2} 1334~\AA.  The black dotted curve is the heavily smoothed average photometric error, per original wavelength point, calculated from the individual epoch values as a standard error of the mean.  Note the weak absorption at 1335.7~\AA, which might be due to the excited ground state transition of \ion{C}{2}.}
\end{figure}

\begin{figure}
\figurenum{19}
\vskip  0mm
\hskip  7mm
\includegraphics[scale=0.875]{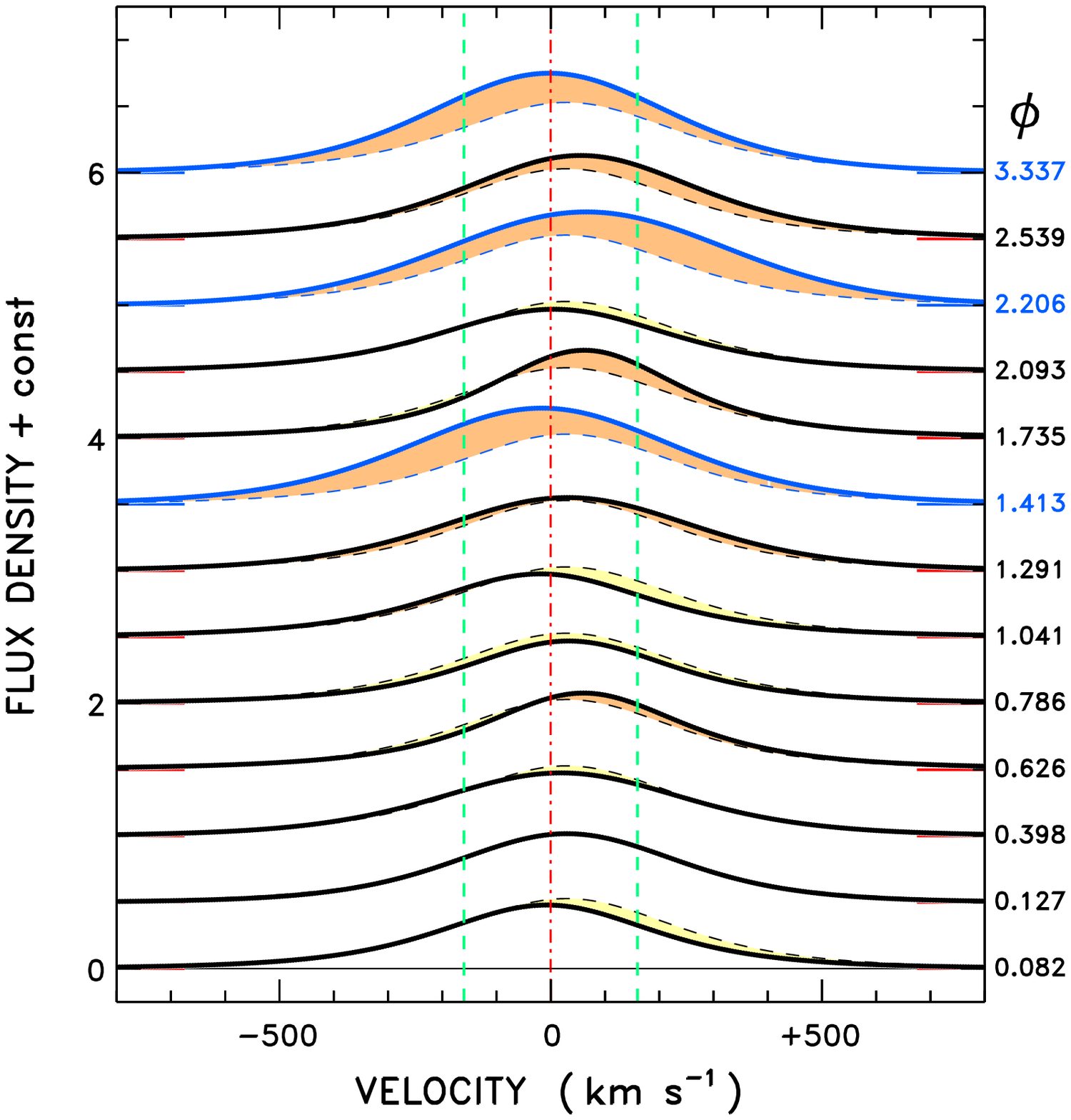}
\vskip 0mm
\figcaption[]{\small Similar to Fig.~16, but now just the fitted stellar-only emission profiles of \ion{C}{2} 1334~\AA\ (thick curves).  Thin black dashed tracings represent the (fit to the) epoch-average non-flare profile, and shading highlights whether the local profile exceeds, or falls below, the average.  The vertical red dotted-dashed line marks the velocity zero point of \ion{C}{2} 1334~\AA, in the \ion{C}{2} absorption frame, and the green dashed lines are ${\pm}\upsilon\sin{i}$.
}
\end{figure}

\clearpage
\begin{deluxetable}{lccccc}
\tabletypesize{\small}
\tablenum{6}
\tablecaption{{\em HST}\/ COS FUV \ion{C}{2} Multi-Component Profile Parameters}
\tablewidth{0pt}
\tablecolumns{6}
\tablehead{
\colhead{dataset} & \colhead{$\lambda_{\rm 1334}$} & \colhead{${\rm FWHM}$} & 
\colhead{$f_{\rm 1334}$} &  \colhead{$\alpha$} & 
\colhead{$f_{\rm INT}$} \\[3pt] 
\colhead{} & \multicolumn{2}{c}{(km s$^{-1}$)} & 
\colhead{($10^{-13}$ cgs \AA$^{-1}$)} & \colhead{} & \colhead{($10^{-13}$ cgs)}  \\[3pt] 
\colhead{(1)} &  
\colhead{(2)} &
\colhead{(3)} &  
\colhead{(4)} &
\colhead{(5)} &  
\colhead{(6)}               
}
\startdata   
lbme02030 &  $-$5  &  355  &   0.39  &     3.0  &   1.7  \\
lbme03040 &  $-$7  &  470  &   0.48  &     2.2  &   2.6  \\
lbme04040$\star$ & $-$19  &  580  &   0.72  &     3.8  &   4.6  \\
lbme55040 &   +37  &  515  &   0.47  &     3.6  &   2.7  \\
lbkm01020 &   +30  &  490  &   0.52  &     3.4  &   2.8  \\
lbkm02020 &   +19  &  540  &   0.48  &     3.4  &   2.9  \\
lbkm03020 &   +61  &  450  &   0.58  &     1.6  &   3.2  \\
lbkm04020 &   +32  &  450  &   0.47  &     2.0  &   2.5  \\
lbkm05020 & $-$18  &  485  &   0.48  &     2.0  &   2.7  \\
lbkm06020 &   +31  &  560  &   0.55  &     3.6  &   3.4  \\
lbkm07020 &   +62  &  420  &   0.66  &     1.8  &   3.3  \\
lbkm08020 &    +6  &  510  &   0.47  &     2.4  &   2.8  \\
lbkm09020$\star$ &   +67  & 620  &   0.70  &     9.0  &   4.6  \\
lbkm10020 &   +56  &  520  &   0.63  &     2.6  &   3.8  \\
lbkm11020$\star$ &  $-$3  &  530  &   0.75  &     3.8  &   4.3  \\
\cutinhead{Typical Monte-Carlo-based uncertainties}
 \nodata    & ${\pm}1$  &  ${\pm}4$ &   ${\pm}0.01$ & \nodata & \nodata \\
\cutinhead{Main program average, non-flare epochs}
 \nodata    & $+29$   &   490  &  0.53   &  2.2  & 2.80 \\
\cutinhead{Averages and standard deviations over all the epochs}
  \nodata  &  $+23{\pm}29$ &   $500{\pm}65$ & $0.56{\pm}0.11$ & 
$3.2{\pm}1.7$ & $3.2{\pm}0.8$  \\
\enddata
\tablecomments{Col.~(1) datasets marked ``$\star$'' are suspected flares.  Cols.~(2) and (3) wavelength parameters expressed in equivalent velocity units; and in the case of Col.~(2), relative to the derived \ion{C}{2} absorption velocity in that spectrum.  Cols.~(4) and (6) cgs units are erg cm$^{-2}$ s$^{-1}$.  The ratio of the peak intensities of the two stellar \ion{C}{2} emission components was fixed at $f_{\rm 1335}/f_{\rm 1334}= 1.3$\,.  Col.~(5) is an exponent that describes the departure of the observed stellar emission profile from a pure Lorentzian line shape ($\alpha\equiv 1$).  The Col.~(6) fluxes were integrated between 1331--1340~\AA, above a continuum level (based on the interval $1342.5{\pm}1.0$~\AA), and have negligible formal uncertainties.  In addition to these more variable parameters, more constant values derived in the fits were: absorption (``ISM'') velocity-- $\upsilon= -28 {\pm} 2 ({\pm}1 [1\,{\rm s.e.}])$~km s$^{-1}$; absorption Doppler (e-folding) width-- $\Delta\lambda= 27 {\pm} 2$~km s$^{-1}$; relative absorption depth-- $r= 0.95{\pm}0.03$ (dimensionless); and continuum level-- $f_{\rm C}= 0.02 {\pm} 0.01$ (same units as Col.~[4]).
}
\end{deluxetable} 

\clearpage
\subsubsection{Time-Resolved Profiles of \ion{Si}{4} 1400~\AA}

As mentioned earlier, a key spectral diagnostic for COCOA-PUFS is the \ion{Si}{4} doublet at 1400~\AA.  The 1393~\AA\ component is the strongest of the relatively isolated high-temperature ($T\sim 8{\times}10^4$~K) features in the FUV range.  The hotter \ion{N}{5} doublet ($T\sim 2{\times}10^5$~K) at 1240~\AA\ was excluded by the deactivation of G130M side B, and the bright \ion{C}{4} doublet at 1550~\AA\ ($T\sim 1{\times}10^5$~K) is badly blended (like \ion{C}{2}) and in any event was captured in only two observations of the main program (the G160M exposures of GO-12376 Visits~3 and 4; Visit~5 failed).  Figure~20 depicts the COS G130M spectra of the \ion{Si}{4} region from the 13 visits of the main program, in a similar way as the \ion{Mg}{2} and \ion{C}{2} profiles illustrated earlier. 

Referring back to Fig.~15, \ion{Si}{4} 1393~\AA\ appears to be relatively clean in the 24~UMa spectrum, but the 1402~\AA\ component is flanked by at least five distinct subsidiary emissions, from intersystem multiplets of \ion{O}{4} and \ion{S}{4}.  The brightest of these, \ion{O}{4}] 1401~\AA, is about 20\% the peak intensity of 1402~\AA, at least in more slowly spinning 24~UMa.

\begin{figure}
\figurenum{20}
\vskip -5mm
\hskip  7mm
\includegraphics[scale=0.875]{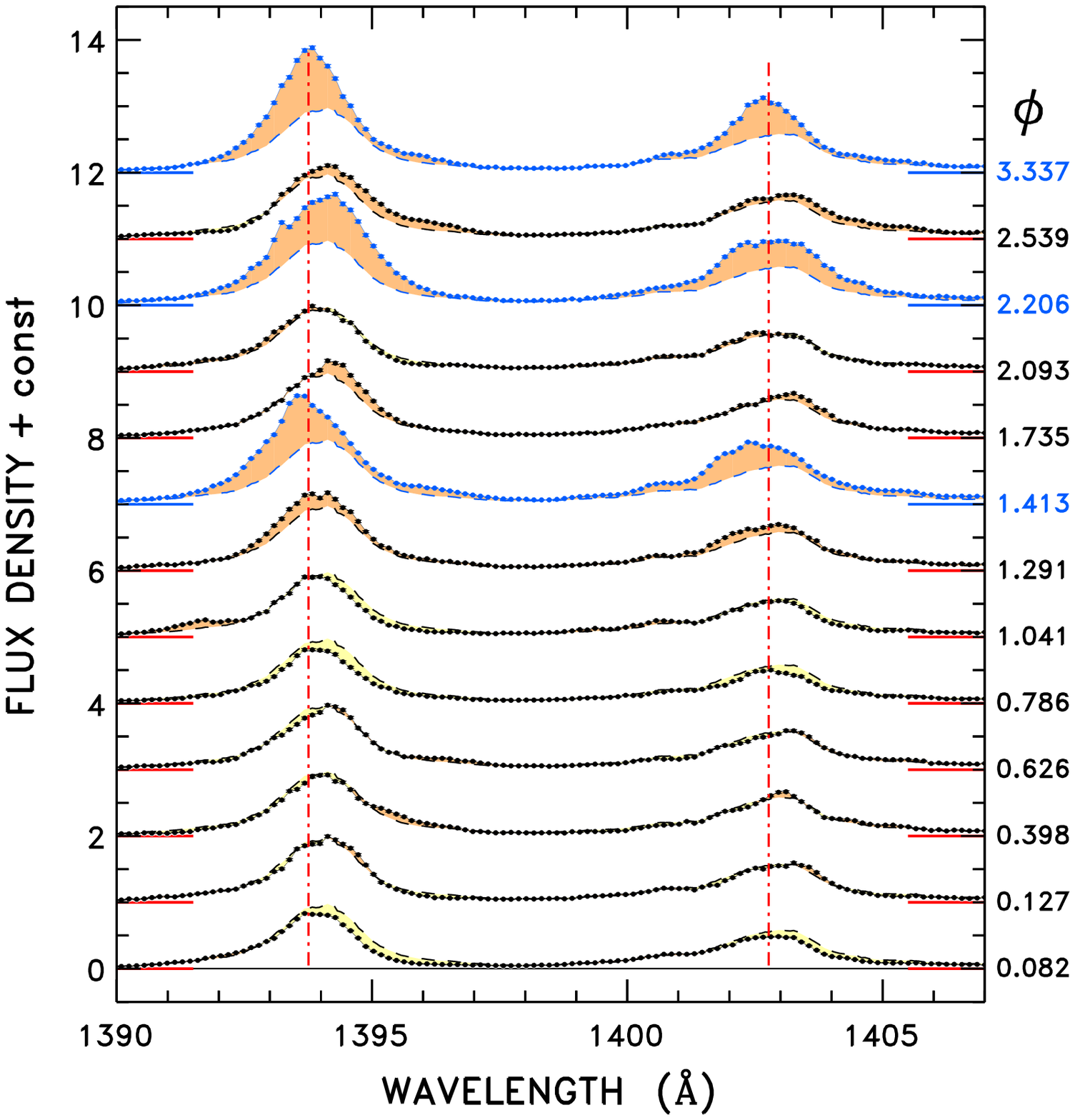}
\vskip -5mm
\figcaption[]{\small \ion{Si}{4} profiles from the G130M-1291 exposures of the main program, in the velocity frame established by the FUV interstellar features.  The \ion{Si}{4} profiles are arranged, from bottom to top, in increasing time, and are offset by one unit of the flux density axis ($10^{-13}$ erg cm$^{-2}$ s$^{-1}$ \AA$^{-1}$) for clarity (horizontal colored ticks refer to the zero levels of the individual spectra).  Photometric cycles are listed at the right.  The vertical dotted-dashed lines mark the laboratory wavelengths of the \ion{Si}{4} components.  Note that there is a weaker multiplet of \ion{O}{4}] (and \ion{S}{4}]) transitions surrounding \ion{Si}{4} 1402~\AA, especially \ion{O}{4}] 1401~\AA\ (see, e.g., narrow-line 24~UMa in Fig.~15).  The three profiles highlighted in blue are anomalous, probably affected by flares.  Thin black dashed curves represent the epoch-average profile, excluding the three suspected flares.  Darker/lighter shading indicates whether the local profile is higher/lower than the average.  Error bars represent a $\sim$2~resel rebinning of the original flux densities.}
\end{figure}

Figure~21 displays the integrated fluxes of the \ion{Si}{4} complex (including the blended \ion{O}{4}] and \ion{S}{4}] multiplets) as a function of time and (folded) phase (values reported in Table~7, below).  As in the previous diagrams of this type, subcoronal \ion{Si}{4} shows higher activity in the $\phi\sim 0.3$ phase band, mainly thanks to the several flares.

\begin{figure}
\figurenum{21}
\vskip  0mm
\hskip  8mm
\includegraphics[scale=0.875]{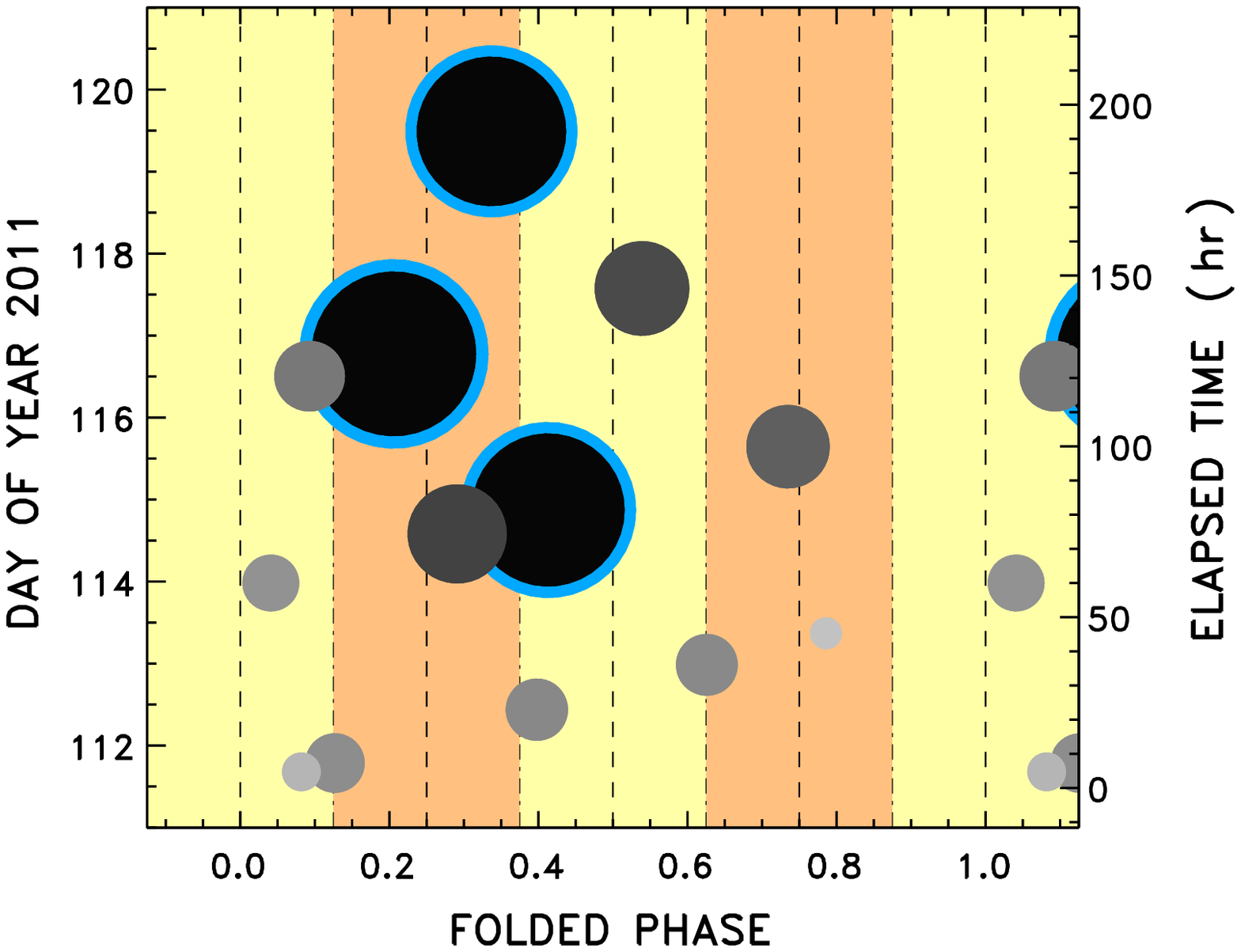}
\vskip -5mm
\figcaption[]{\small Schematic integrated fluxes of the \ion{Si}{4} complex (including \ion{O}{4}] and \ion{S}{4}]) of FK~Com over the period of the main campaign, in ascending time along the $y$-axis, and increasing (folded) phase along the $x$-axis.  The flux to symbol size scaling is exaggerated for display.  Three suspected flares are outlined in blue.}
\end{figure}
 
Figure~22 illustrates the multi-component constrained fit applied to the G130M/A \ion{Si}{4} profiles.  Each COS spectrum was registered to the FUV absorption velocity frame derived from that observation.  The two components of \ion{Si}{4} were represented by the same pseudo-Lorentzian line shape exploited earlier, taking the same width, $\Delta\lambda_{\rm L}$, and $\alpha$ value for each, and assuming that the wavelength separation was the laboratory value 9.013~\AA\ (see Ayres 2015b for a discussion of the wavelengths of \ion{Si}{4}, \ion{O}{4}], and \ion{S}{4}]).  The peak intensities of the well-separated \ion{Si}{4} components were allowed to float.  The five main components of the weaker \ion{O}{4}] plus \ion{S}{4}] complex were assigned the same line shape as for the \ion{Si}{4} features, and the velocities of the features were slaved to that of \ion{Si}{4} 1393~\AA\ by the laboratory wavelength differences.  The peak intensity of the strongest feature, \ion{O}{4}] 1401~\AA, was allowed to float, but the four subsidiary weaker components were assigned fixed peak intensity ratios relative to 1401~\AA.  The specific ratios in the 24~UMa STIS FUV spectrum range from 0.3--0.4.  However, trial fitting suggested that multiplying the 24~UMa ratios by 0.6 produced a better match to the far red side of the blended \ion{Si}{4} feature, where the influence of the weaker \ion{O}{4}] and \ion{S}{4}] emissions is greatest.  The intersystem lines are density-sensitive, so the apparent weakening of the subsidiary components of the \ion{O}{4}] multiplet, relative to \ion{O}{4}] 1401~\AA, in FK~Com compared to 24~UMa could be a pressure effect.  A baseline continuum flux density was subtracted prior to the fitting procedure.  It was determined from a 2~\AA\ window in the far blue wing of \ion{Si}{4}, well outside the region affected by the stellar emission blend.  Table~7 summarizes the resulting fit parameters.  Figure~23 displays the derived 1393~\AA\ component profiles, again highlighting any deviations with respect to an epoch-average line shape fitted to the non-flare observations of the main program.

With respect to the FUV absorption system, the average \ion{Si}{4} profile is redshifted by +50 km s$^{-1}$, or about +22 km s$^{-1}$ heliocentric, a further $\sim 20$~km s$^{-1}$ beyond \ion{Mg}{2} and \ion{C}{2} (the latter recorded simultaneously with \ion{Si}{4}).  The extra hot-line redshift at first glance might seem surprising, but is reminiscent of those seen in other active yellow giants (Ayres et al.\ 1998).

\begin{figure}
\figurenum{22}
\vskip  0mm
\hskip -4mm
\includegraphics[scale=0.75,angle=90]{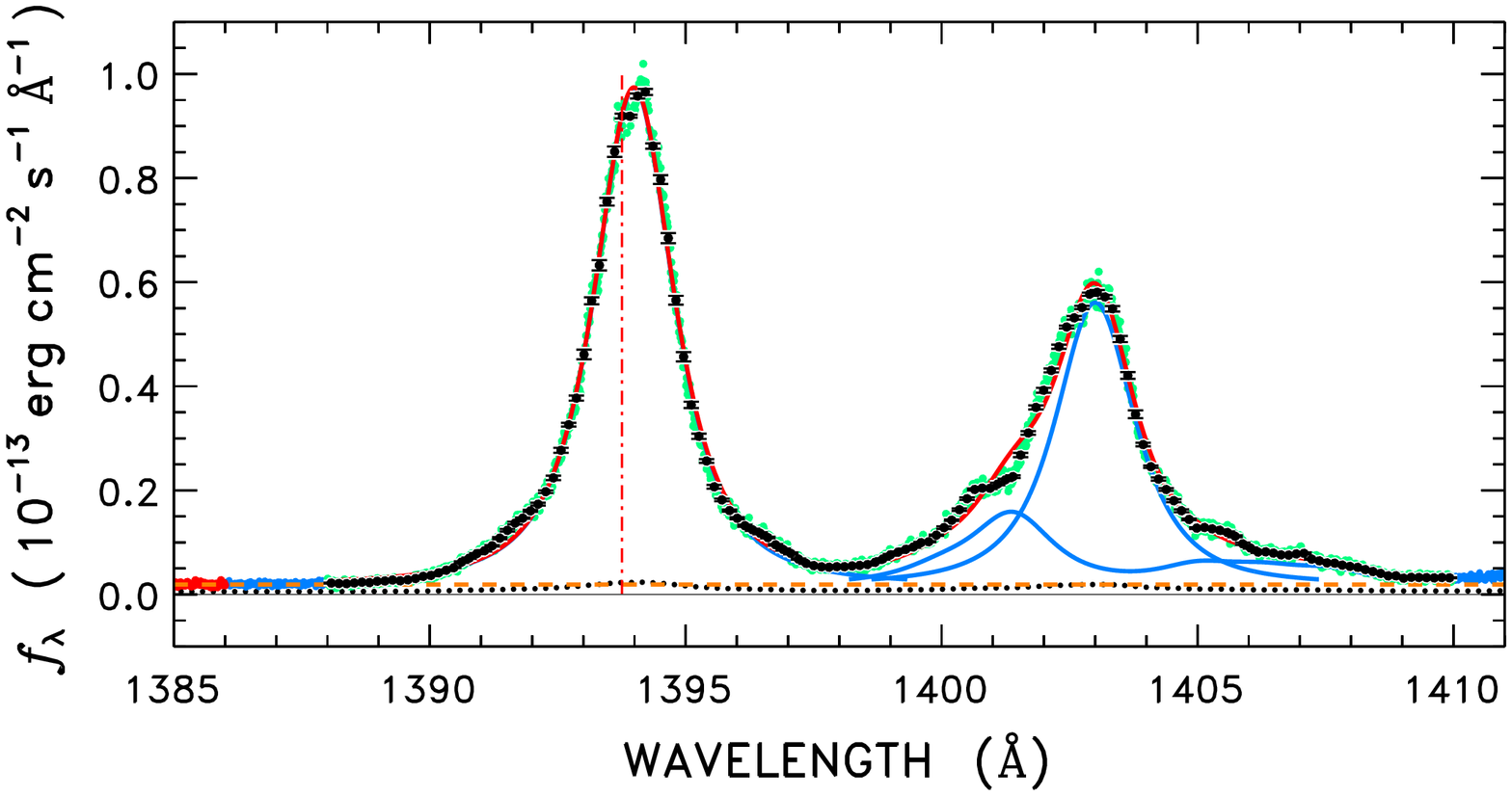}
\vskip 3mm
\figcaption[]{\small Schematic depiction of the multi-component constrained fit to the blended stellar emission doublet of \ion{Si}{4}, including subsidiary weaker features, mainly \ion{O}{4}] but with a minor contribution by \ion{S}{4}].  The observation is the epoch average from the main program, excluding flare intervals, registered to the apparent FUV absorption system velocity.  Red points in the extreme blue wing of the overall feature were used to define a continuum background (orange dashed line).  Green points represent the flux densities upon which the modeling was based.  Blue points are fluxes that were not incorporated in the fitting procedure or the background correction.  Error bars are 2-resel average fluxes, for display.  The red curve is the combined fit; blue curves illustrate the fitted sub-components.  (The red curve obscures the blue one for the more isolated 1393~\AA\ component.)  The vertical red dotted dashed line marks the laboratory wavelength of \ion{Si}{4} 1393~\AA.  The black dotted curve is the heavily smoothed average photometric error, per original wavelength point, calculated from the individual epoch values as a standard error of the mean.}
\end{figure}

\begin{figure}
\figurenum{23}
\vskip  0mm
\hskip  7mm
\includegraphics[scale=0.875]{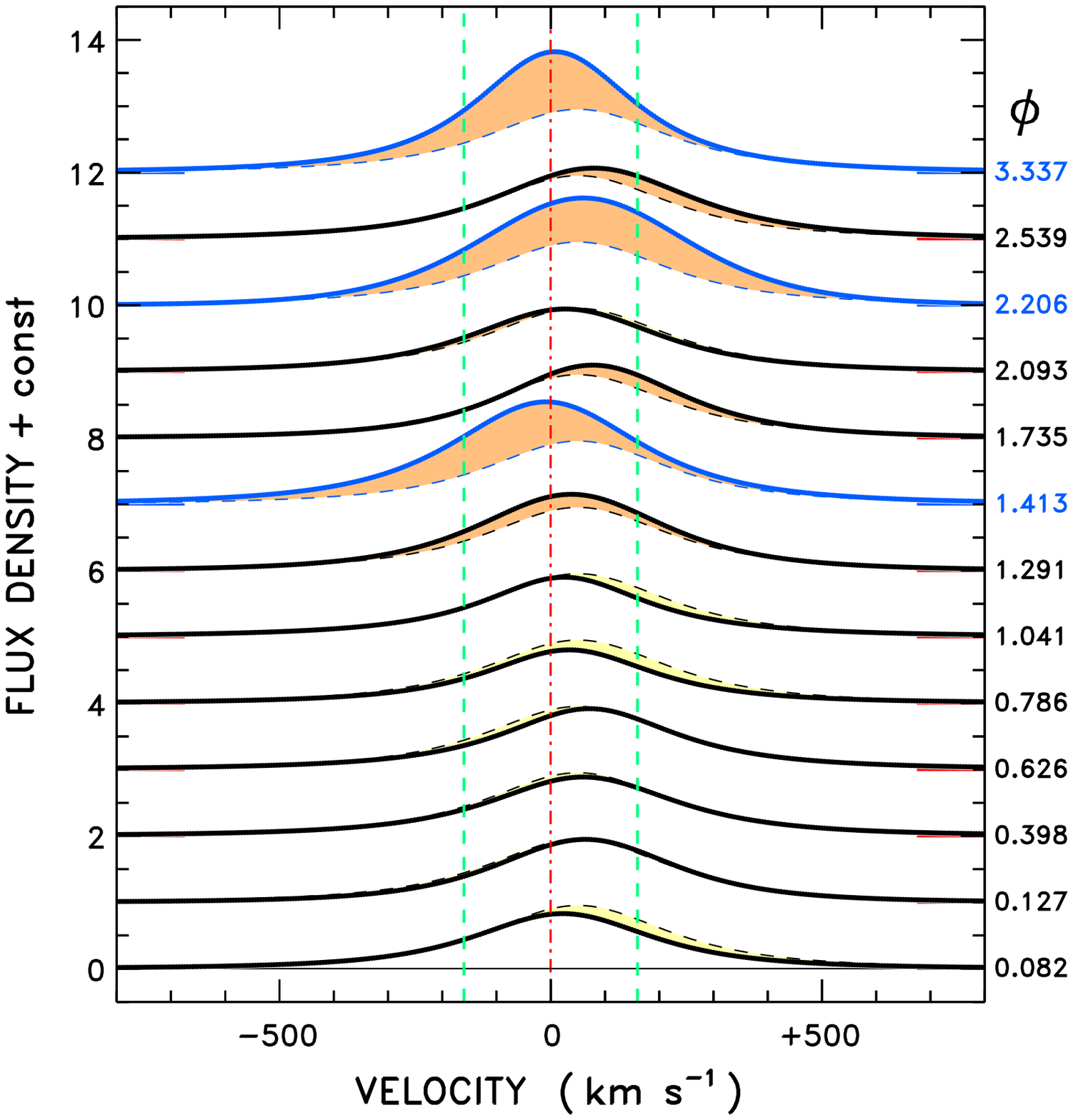}
\vskip 0mm
\figcaption[]{\small Similar to Fig.~20, but now just the fitted profiles of \ion{Si}{4} 1393~\AA\ (thick curves).  Thin black dashed tracings represent the (fit to the) epoch-average non-flare profile, and shading highlights whether the local profile exceeds, or falls below, the average profile.  The vertical red dotted-dashed line marks the velocity zero point of \ion{Si}{4} 1393~\AA\ in the FUV absorption frame, and the green dashed lines are ${\pm}\upsilon\sin{i}$.  The large redshift of the epoch-average \ion{Si}{4} profile in that frame is evident.
}
\end{figure}

\clearpage
\begin{deluxetable}{lccccc}
\tabletypesize{\small}
\tablenum{7}
\tablecaption{{\em HST}\/ COS FUV \ion{Si}{4} Multi-Component Profile Parameters}
\tablewidth{0pt}
\tablecolumns{6}
\tablehead{
\colhead{dataset} & \colhead{$\lambda_{\rm 1393}$} & \colhead{${\rm FWHM}$} & 
\colhead{$f_{\rm 1393}$} &  \colhead{$\alpha$} & 
\colhead{$f_{\rm INT}$} \\[3pt] 
\colhead{} & \multicolumn{2}{c}{(km s$^{-1}$)} & 
\colhead{($10^{-13}$ cgs \AA$^{-1}$)} & \colhead{} & \colhead{($10^{-13}$ cgs)}  \\[3pt] 
\colhead{(1)} &  
\colhead{(2)} &
\colhead{(3)} &  
\colhead{(4)} &
\colhead{(5)} &  
\colhead{(6)}               
}
\startdata   
lbme02030  &     +7  &  345  &  0.65  &  4.0  & 2.4 \\
lbme03040  &    +22  &  385  &  0.83  &  1.6  & 3.4  \\
lbme04040$\star$  &   $-$8  &  410  &  1.54  &  1.6  & 6.5  \\
lbme55040  &    +63  &  405  &  0.95  &  1.8  & 3.9  \\
lbkm01020  &    +64  &  385  &  0.95  &  1.9  & 3.8 \\
lbkm02020  &    +60  &  405  &  0.89  &  1.5  & 3.9  \\
lbkm03020  &    +71  &  380  &  0.92  &  1.3  & 3.9  \\
lbkm04020  &    +34  &  360  &  0.80  &  1.4  & 3.3  \\
lbkm05020  &    +25  &  370  &  0.90  &  1.3  & 3.8   \\
lbkm06020  &    +38  &  405  &  1.15  &  1.9  & 4.8 \\
lbkm07020  &    +78  &  385  &  1.10  &  1.8  & 4.4   \\
lbkm08020  &    +27  &  405  &  0.94  &  1.7  & 4.1 \\
lbkm09020$\star$  &    +60  &  450  &  1.62  &  3.4  & 7.0  \\
lbkm10020  &    +79  &  425  &  1.07  &  1.7  & 4.7  \\
lbkm11020$\star$  &     +8  &  345  &  1.82  &  1.5  & 6.4   \\
\cutinhead{Typical Monte-Carlo-based uncertainties}
 \nodata    & ${\pm}1$  &  ${\pm}2$ &   ${\pm}0.01$ & \nodata & \nodata \\
\cutinhead{Main program average, non-flare epochs}
 \nodata    & $+50$   &   390  &  0.95   &  1.6  & 3.9 \\
\cutinhead{Averages and standard deviations over all the epochs}
  \nodata  & $+42{\pm}27$ & $390{\pm}30$ &  $1.08{\pm}0.32$ & $1.9{\pm}0.7$  &  $4.41{\pm}1.25$\\ 
\enddata
\tablecomments{Col.~(1) datasets marked ``$\star$'' are suspected flares.  Cols.~(2) and (3) wavelength parameters expressed in equivalent velocity units; and in the case of Col.~(2), relative to the derived FUV absorption velocity in that G130M/A spectrum (average was $-28{\pm}2$~km s$^{-1}$).  Cols.~(4) and (6) cgs units are erg cm$^{-2}$ s$^{-1}$.  Col.~(5) is an exponent that describes the departure of the observed stellar emission profile from a pure Lorentzian line shape ($\alpha\equiv 1$).  The Col.~(6) fluxes were integrated between 1388.5--1406.5~\AA, above a continuum level (based on the interval $1385.0{\pm}1.0$~\AA), and have negligible formal uncertainties.  In addition to these more variable parameters, more constant values derived in the fits were: doublet peak intensity ratio-- $f_{1402}/f_{1393}= 0.57{\pm}0.02$; \ion{O}{4}]/\ion{Si}{4} flux ratio-- $f_{1401}/f_{1393}= 0.13{\pm}0.02$; and continuum level-- $f_{\rm C}= 0.03{\pm}0.01$ (same units as Col.~[4]).}
\end{deluxetable}

\clearpage
\subsubsection{Dynamic Spectra of \ion{Mg}{2}, \ion{C}{2} and \ion{Si}{4}}

Figures~24a--c are phase-velocity maps (``dynamic spectra'' in the nomenclature of Vida et al.\ 2015), for \ion{Mg}{2}, \ion{C}{2}, and \ion{Si}{4}, respectively, depicting the average profile distortions on a uniform phase grid.  The maps were constructed by reordering in (folded) phase the collectivized line shapes (i.e., Fig.~12 for \ion{Mg}{2}) from all the non-flare epochs, after subtracting the epoch-average non-flare profile; then interpolating in phase at each velocity bin; and finally smoothing the result to a resolution of roughly 0.1 in phase.  The red dotted-dashed line indicates the reference absorption velocity zero point, while the green dashed lines are $\pm\upsilon\sin{i}$, roughly marking the velocities of the advancing ($-$) and retreating ($+$) limbs of the star.  Note that the ``zero points'' might differ between the NUV and FUV maps, but the nominal absorption velocities both are within an essentially negligible $<10$~km s$^{-1}$ of the stellar RV, relative to line widths of several hundred km s$^{-1}$.  Red hatched zones in each map highlight the $\phi\sim$~0.8--1.0 band where the behavior is less well defined.

The dynamic spectra bear superficial similarities to each other, and qualitatively are like the H$\alpha$ maps presented by Vida et al.\ (2015), from more than two decades of ground-based measurements.  There are phases where the spectrum is weaker than average ($\phi\sim 0.4$), stronger than average ($\phi\sim 0.5$), double-lobed (\ion{Mg}{2}: $\phi\sim 0.3$), more redshifted ($\phi\sim 0.0$), and more blueshifted ($\phi\sim 0.7$).  As with the H$\alpha$ maps, the ultraviolet emissions frequently extend well beyond the photospheric limb ($\phi\sim 0.3$), although one must be cautious in such comparisons because at least the \ion{Mg}{2} and \ion{C}{2} lines probably are extremely optically thick, and formed primarily by collisional excitation, whereas the H$\alpha$ emission likely is thinner and formed mainly by recombination.  Of all these diagnostics, \ion{Si}{4} is most likely to be collisionally excited, and least likely to be very optically thick.

\clearpage
\begin{figure}
\figurenum{24}
\vskip  0mm
\hskip  -21mm
\includegraphics[scale=0.575]{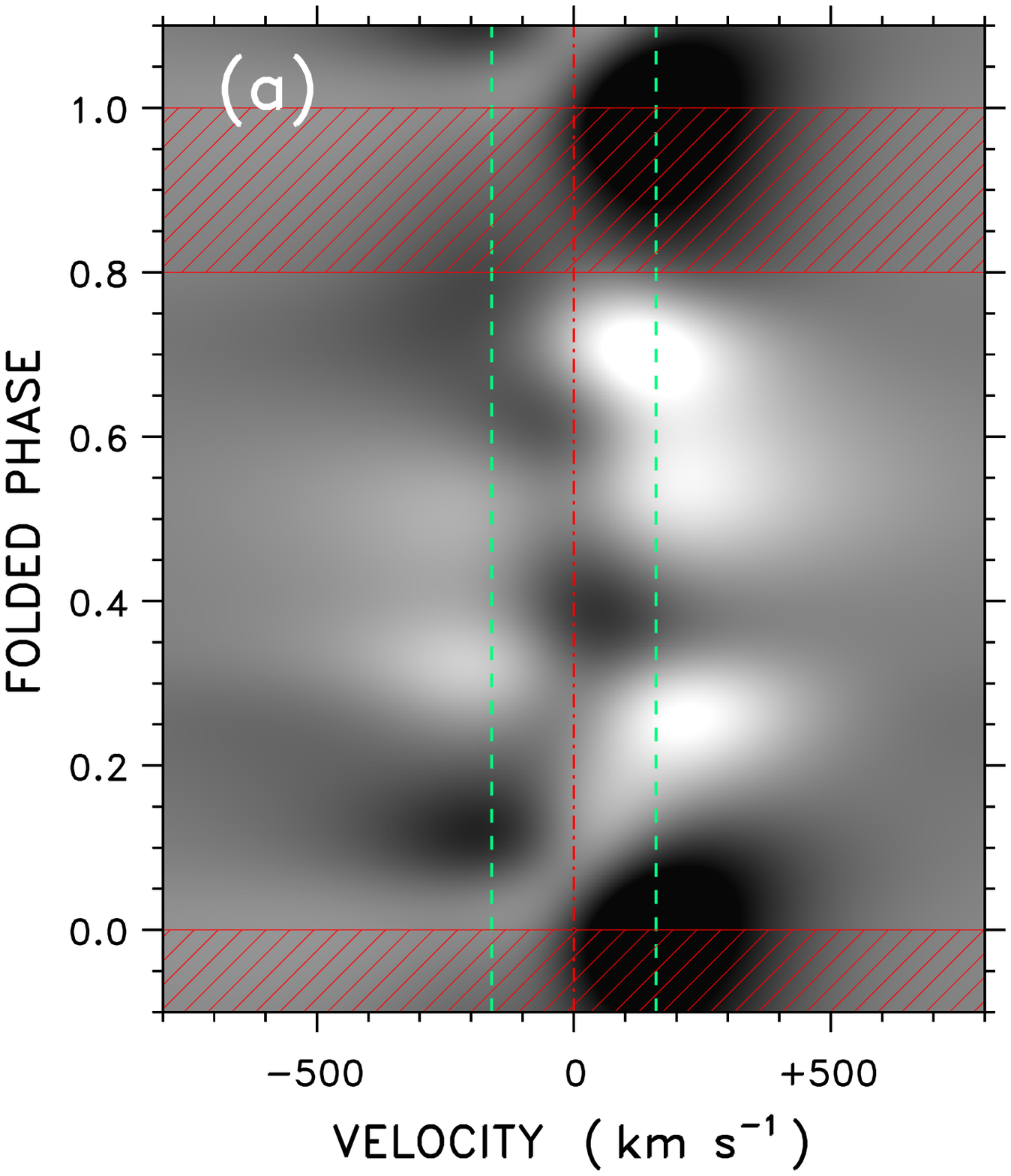}
\vskip -98mm
\hskip 74mm
\includegraphics[scale=0.575]{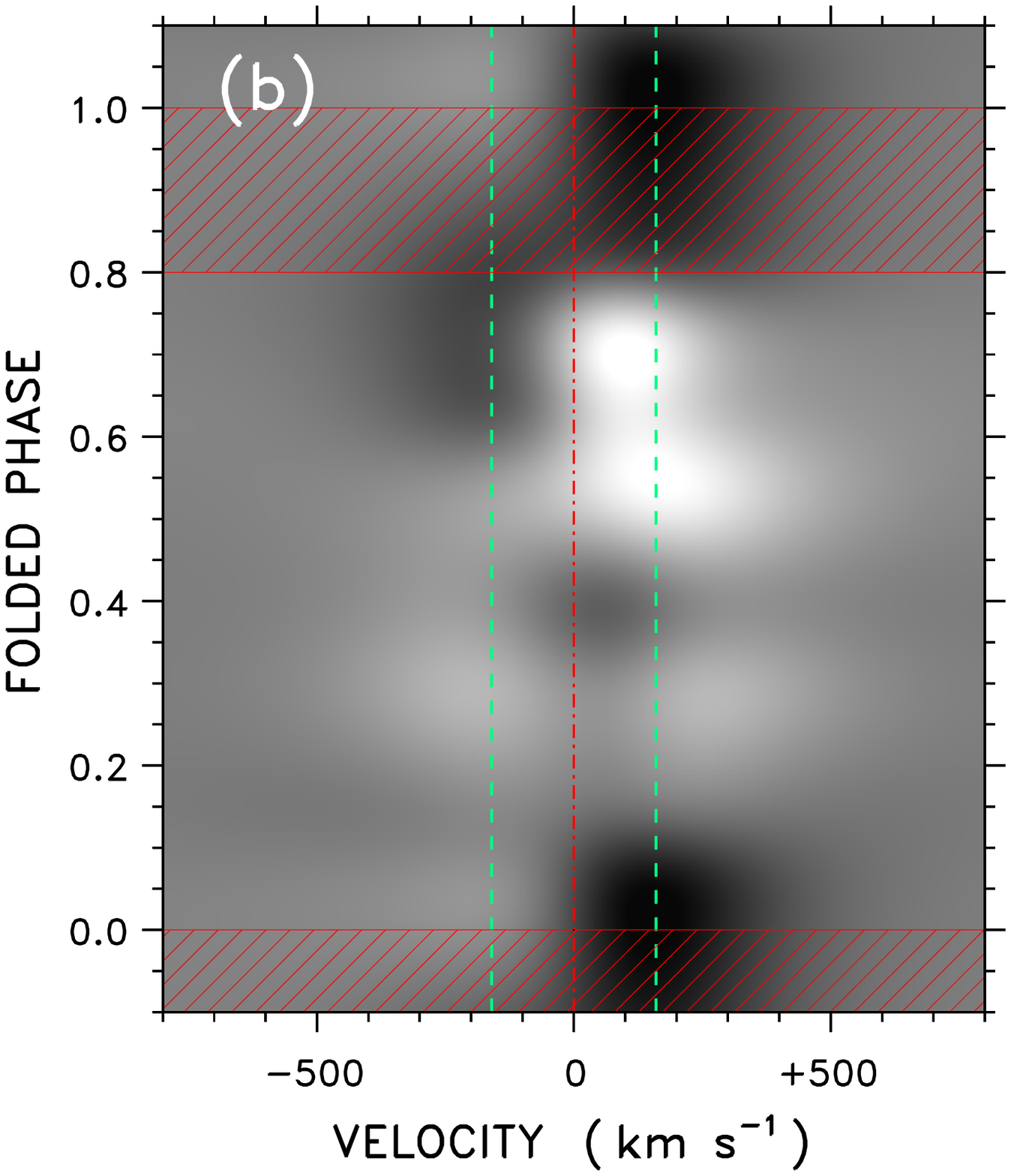}
\vskip 8mm
\hskip 27.0mm
\includegraphics[scale=0.575]{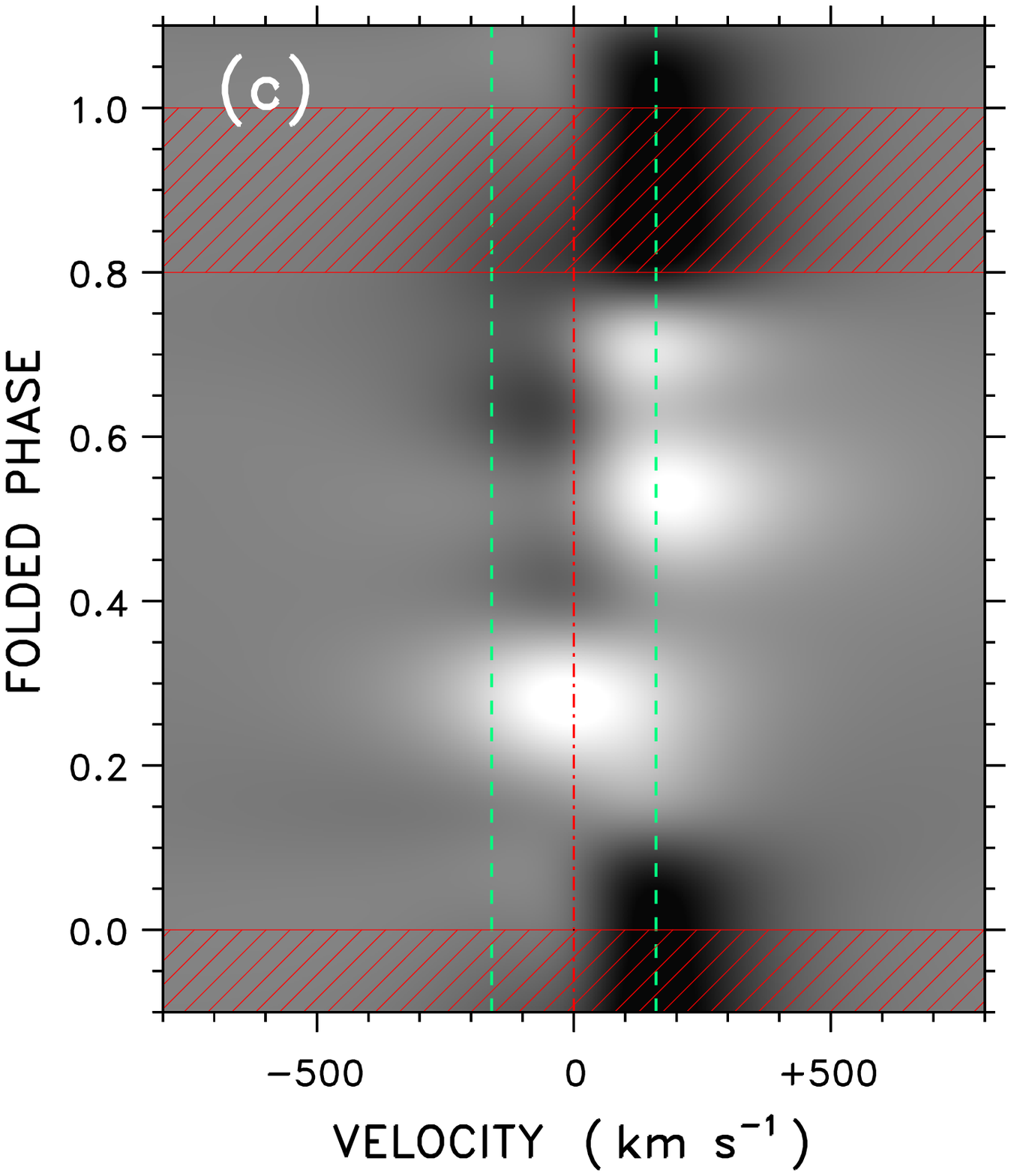}
\vskip 0mm
\figcaption[]{
}
\end{figure}

\begin{figure}
\figurenum{24}
\figcaption[]{\small ``Dynamic spectra'' derived from the collectivized profiles illustrated earlier: \ion{Mg}{2} (upper left: {\em a}\,), \ion{C}{2} (upper right: {\em b}\,), and \ion{Si}{4} (lower middle: {\em c}\,).  Lighter areas indicate where the local profile is higher than the epoch average; darker areas are where the local profile falls below.  The red shaded bands are cautionary: no observations were available for $0.8 \le \phi \le 1.0$, so the profile behavior in that zone is less well defined (essentially interpolated).  Vertical green dashed lines are at $\pm$160~km s$^{-1}$ ($\pm\upsilon\sin{i}$).  The image gray scale saturates at $\pm 0.15{\times}10^{-12}$ erg cm$^{-2}$ s$^{-1}$ \AA$^{-1}$ for \ion{Mg}{2}, $\pm 0.10{\times}10^{-13}$ for \ion{C}{2}, and $\pm 0.15{\times}10^{-13}$ for \ion{Si}{4}, about 10\% of the peak intensity of the respective epoch-average profiles.
}
\end{figure}

\clearpage
\subsubsection{Time-Resolved FUV Photometric Variations}

Figure~25a illustrates time-resolved line fluxes and profiles from the GO-12279 Visit~6 G130M/A exposures.  The top panel displays the total spectral image (the two FP-POS event lists combined), with specific integration bandpasses marked for the \ion{C}{2} multiplet (red), the \ion{Cl}{1}\,+\,[\ion{Fe}{21}]\,+\,\ion{O}{1} multi-line complex (blue), and the \ion{Si}{4} doublet (green); as well as two separated ostensibly line-free continuum bands (yellow/black dashed); and a pair of flanking off-spectrum background swaths (black dashed).  Table~8 lists the parameters defining these measurements, and the subsequently described spectral profile extractions.

The middle panel of Fig.~25a compares the time-resolved line and continuum count rates (CR: s$^{-1}$), with the same color-coding, including the two continuum bands (combined: black outlined yellow diamonds).  The individual points represent counts accumulated in fixed time bins ($\Delta{t}\sim$~150~s) to achieve high signal-to-noise ($<{\rm S/N}>$: see Table~8).  The background count rates (black dots; taken from each full FP-POS time interval) were scaled upward by 100 times relative to the combined continuum bands (taking into account the continuum integration wavelengths and cross-dispersion width: see Table~8) to illustrate the possible influence of the background for the worst-case scenario (broad, faint wavelength bands).  The background corrections for the combined continuum count rates in this instance are minor, but still were compensated.  The background corrections for the more compact line spectral footprints are entirely negligible.

The lower panels contain spectrally resolved profiles of three bright features -- \ion{C}{2} 1335~\AA\ (blended with weaker \ion{C}{2} 1334~\AA), the 1354~\AA\ complex, and \ion{Si}{4} 1393~\AA\ -- each accumulated in the same time intervals as for the corresponding integrated fluxes and representing a significant total number of counts per profile (for which $<{\rm S/N}_{\rm peak}>$ in Table~8 is a proxy).  The specific accumulation intervals are highlighted by small ticks in the middle panel time sequences.  In each of the bottom spectral panels, the time-resolved profiles are the thin colored curves.  The black dotted curve is the same region from the 24~UMa STIS FUV spectrum, filtered to COS resolution, and scaled to the FK~Com tracings. 

Figure~25b is similar, for the three G160M/B FP-split exposures of GO-12376 Visit~4, which followed GO-12279 Visit~6 by only a few hours.  Here, the integrated count rates are for the \ion{Si}{4} complex and the \ion{C}{4} blend, and extracted profiles are for \ion{Si}{4} 1393~\AA\ and \ion{C}{4} 1548~\AA\ (blended with \ion{C}{4} 1550~\AA).  There is only a single continuum band.  The background count rates, adjusted to the total $(\Delta{x},\,\Delta{y})$ continuum bandpass, were scaled upward a factor of 100 relative to the displayed continuum values.  \ion{Si}{4} is elevated compared to GO-12279 Visit~6, just a few hours earlier, although this partly is due to the higher sensitivity of G160M at 1400~\AA.

Figure~25c illustrates the two-orbit G130M/A sequence of the second part of GO-12376 Visit~4.  The line, continuum, and background count rates were scaled as in Fig.~25a, and the ordinate is the same.  The \ion{C}{2} and \ion{Si}{4} fluxes are higher and rising during the first four FP-POS splits, but are lower and more constant during the second sequence, about 50~minutes later following the Earth occultation.  Note, however, that the 1354~\AA\ complex appears to be steady during the apparent flare rise, despite the presence of the highly volatile [\ion{Fe}{21}] coronal forbidden line ($T\sim 10$~MK), which is expected to be strongly enhanced during such events.  The \ion{Si}{4} 1393~\AA\ profiles show a distinct trend of larger blueshifts with increasing line intensity during the flare rise; the \ion{C}{2} behavior is less clear.

\clearpage
\begin{deluxetable}{lccccc}
\tabletypesize{\small}
\tablenum{8}
\tablecaption{Measurement Parameters for the COS Event Lists}
\tablecolumns{6}
\tablewidth{0pt} 
\tablehead{\multicolumn{3}{c}{\underline{~~~~Integrated Count Rates~~~~}}  &  \multicolumn{3}{c}{\underline{~~~~~~~Spectral Profiles~~~~~~~}} \\
\colhead{Type} & \colhead{Band~(\AA)} & \colhead{$<{\rm S/N}>$} & \colhead{$\lambda_{0}$~(\AA)} & \colhead{$\Delta{\lambda}$~(\AA)}  &  \colhead{$<{\rm S/N}_{\rm peak}>$}\\
\colhead{(1)} & \colhead{(2)} & \colhead{(3)} & \colhead{(4)} & \colhead{(5)} & \colhead{(6)} 
} 
\startdata
\cutinhead{G130M-1291}
LINE-1 & $1335.5{\pm}4.5$  &  100  &  1335.50   &  0.25  &  50 \\[3pt]
LINE-2 & $1355.0{\pm}5.0$  &   45  &  1355.00   &  0.25  &  20 \\[3pt]
LINE-3 & $1397.5{\pm}9.0$  &  110  &  1393.75  &  0.25  &  50 \\[3pt]
CON-A & $1320.0{\pm}5.0$, $1380.0{\pm}5.0$   &  25  &  \nodata  &  \nodata & \nodata \\[3pt]
BKG-A & $1360{\pm}70$   &  \nodata   &  \nodata  &  \nodata & \nodata \\[3pt]
\cutinhead{G160M-1577}
LINE-1 & $1397.5{\pm}9.0$  &  100  &  1393.75 &  0.25  &  60 \\[3pt]
LINE-2 & $1549.0{\pm}6.0$  &   95  &  1548.20  &  0.25  &  60 \\[3pt]
CON-B & $1440{\pm}20$   &  45  &  \nodata  &  \nodata & \nodata \\[3pt]
BKG-B & $1475{\pm}85$   &  \nodata  &  \nodata  &  \nodata & \nodata \\[3pt]
\enddata
\tablecomments{Col.~(1): ``LINE'' for emission line; ``CON'' for continuum band; ``BKG'' for off-spectrum background.  Suffixes ``A'' and ``B'' refer to the respective detector segments.  For the integrated count rates, Col.~(2) is the integration bandpass, and Col.~(3) is the average signal-to-noise of each measurement (for the $\Delta{t}{\sim}150$~s bin).  For the line shape extractions, Col.~(4) is the central wavelength, Col.~(5) is the spectral bin size, and Col.~(6) is the approximate average peak S/N of each profile (integrated over the $\Delta{t}$ interval).  For the fluxes and profiles, the cross-dispersion ($y$) extraction window was ${\pm}11$~pixels for G130M and ${\pm}9$~pixels for G160M.  Background swaths were positioned $\Delta{y}= {\pm}40$ pixels from the spectrum center, with 20 pixel widths.  Background count rates were determined from the full duration of each FP-POS sub-exposure, regardless of S/N.}
\end{deluxetable}

\clearpage
\begin{figure}
\figurenum{25a}
\vskip  0mm
\hskip  -11mm
\includegraphics[scale=0.925,angle=90]{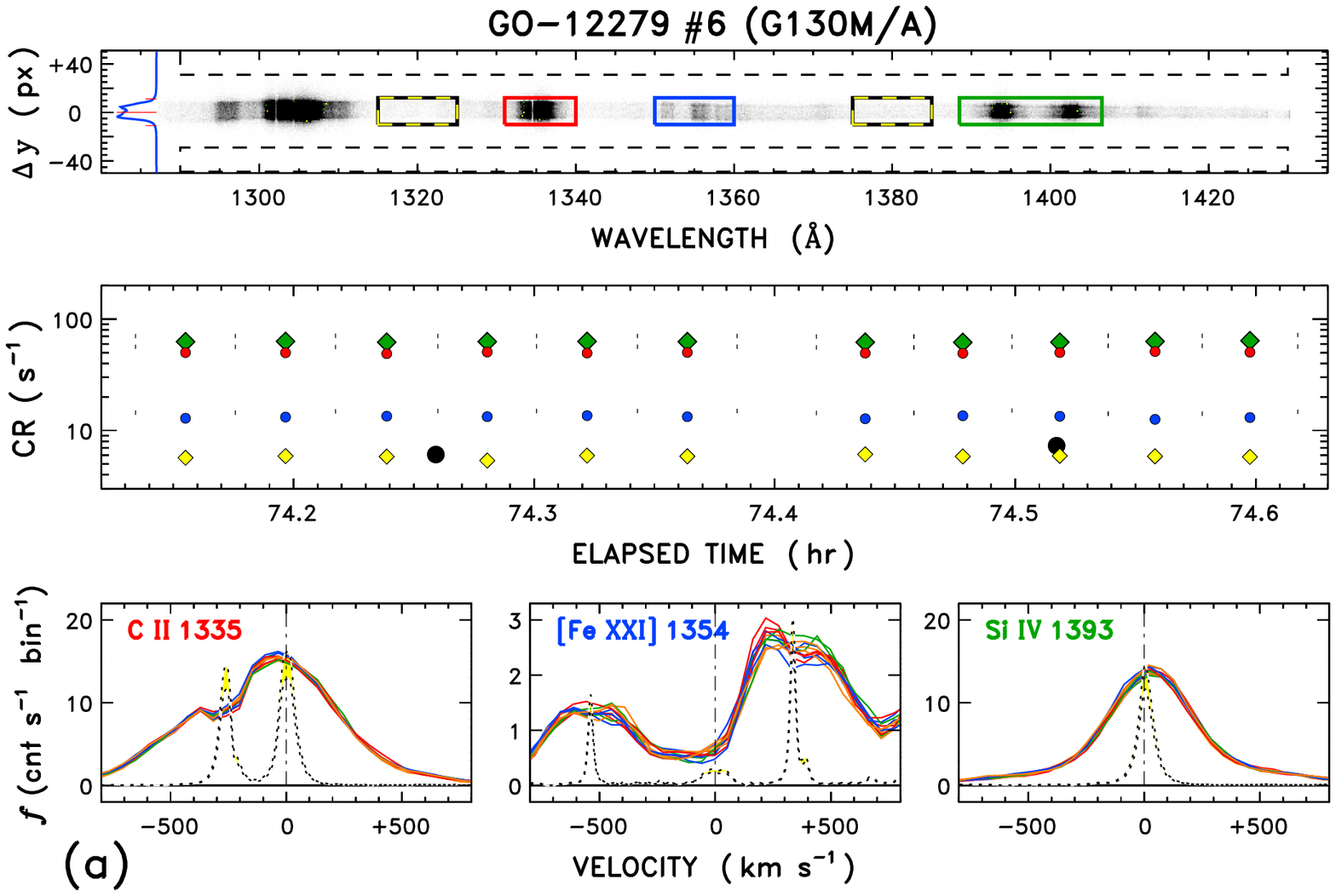} 
\vskip 10mm
\figcaption[]{
}
\end{figure}

\clearpage
\begin{figure}
\figurenum{25b}
\vskip  0mm
\hskip  -11mm
\includegraphics[scale=0.925,angle=90]{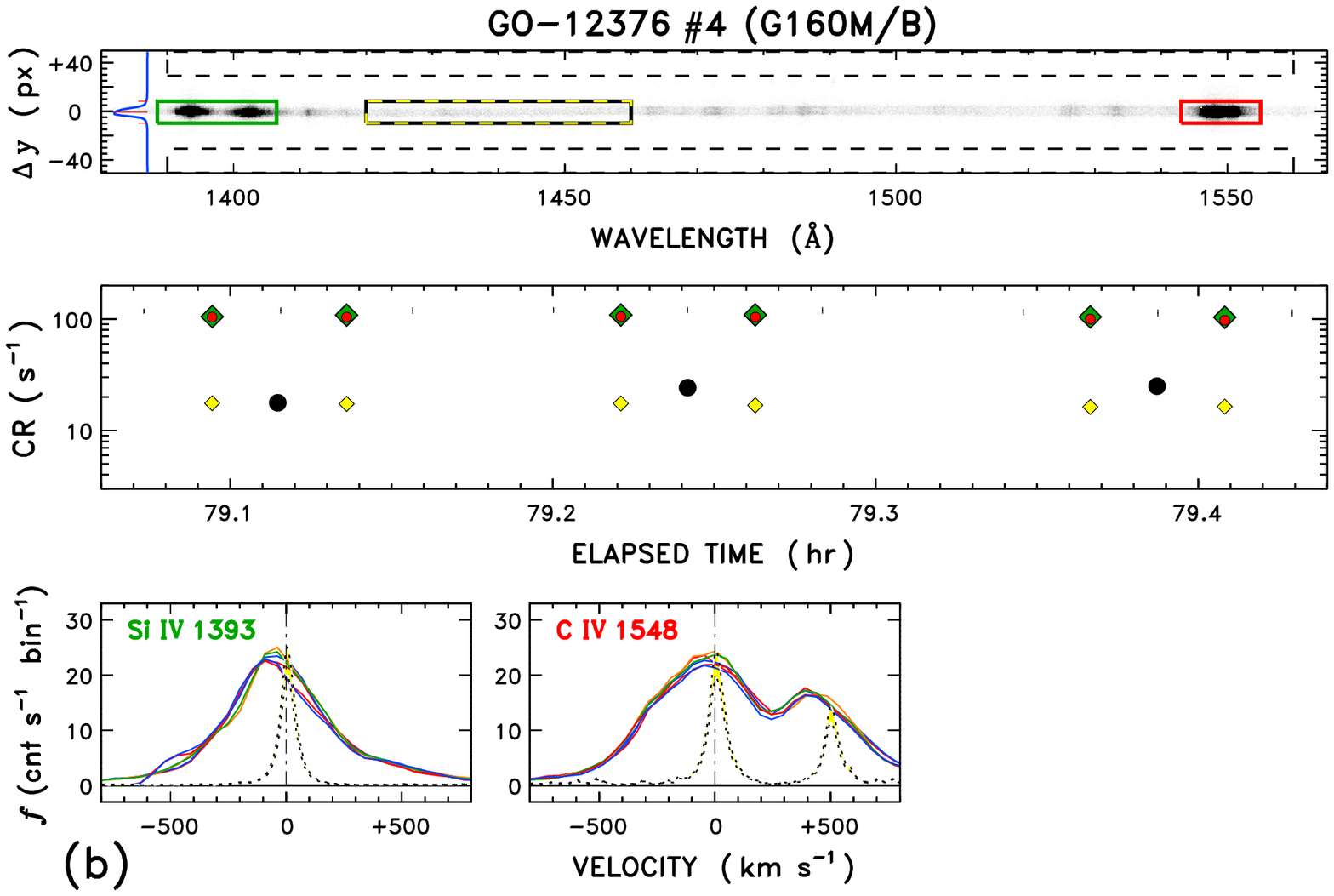} 
\vskip 10mm
\figcaption[]{ 
}
\end{figure}

\clearpage
\begin{figure}
\figurenum{25c}
\vskip  0mm
\hskip  -11mm
\includegraphics[scale=0.925,angle=90]{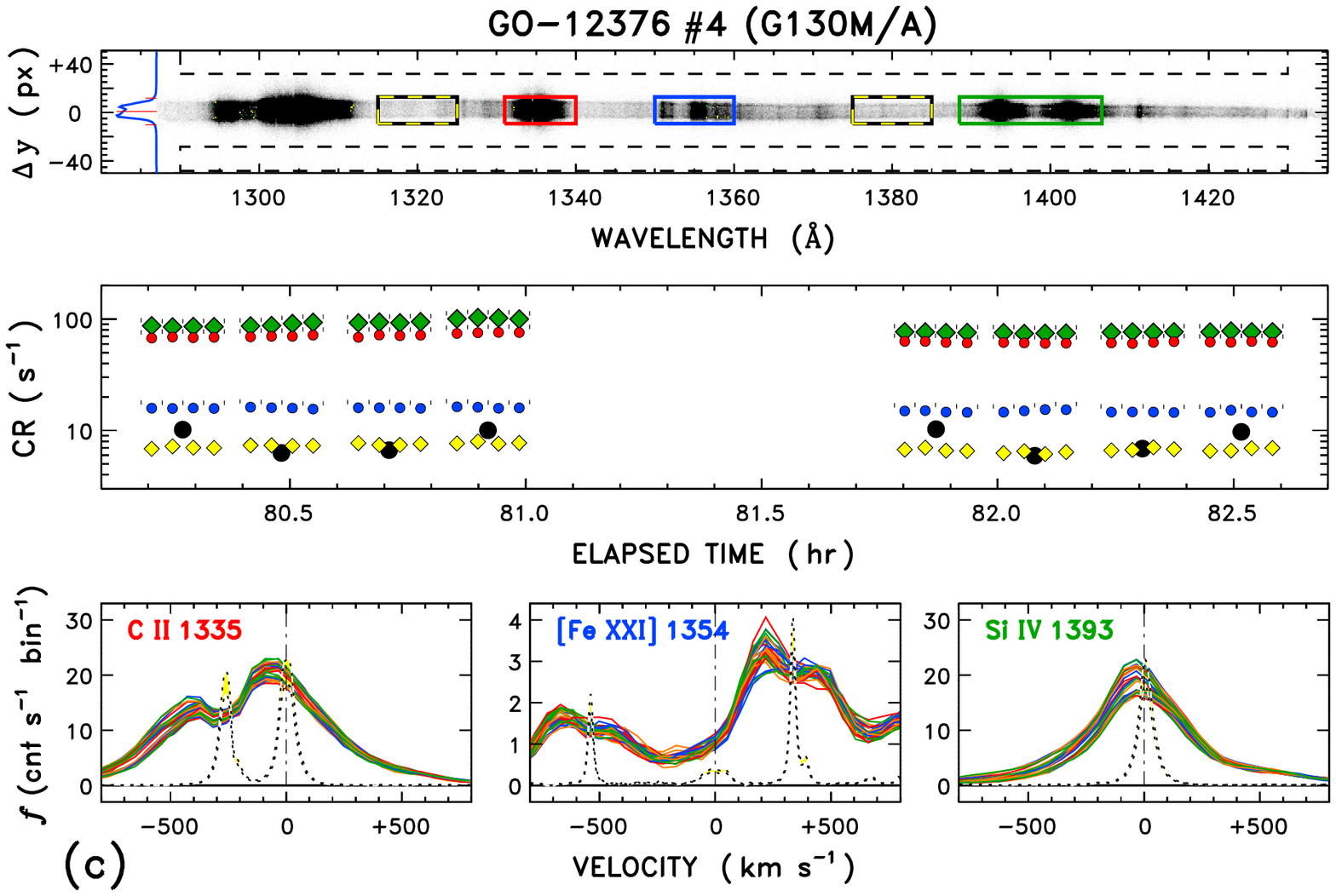} 
\vskip 10mm
\figcaption[]{
}
\end{figure}

\begin{figure}
\figurenum{25}
\vskip -10mm
\figcaption[]{\small Time-resolved COS FUV line fluxes and profiles: see text for detailed description.  (a) From the G130M/A exposures of GO-12279 Visit~6, prior to a large flare.  The blue curve at the left hand side of the diagram is the average cross-dispersion spatial profile; small red ticks delimit the spectral extraction window.  Green diamonds are for \ion{Si}{4}; small red circles for \ion{C}{2}; small blue circles for the 1354~\AA\ multi-line complex; yellow diamonds for the sum of the twin continuum bands; and large black circles for the (scaled) off-spectrum background level.  (b) For the G160M/B exposures during first part of GO-12376 Visit~4, a few hours after GO-12279 Visit~6.  The symbols are the same as in (a), except the small red circles now are for the \ion{C}{4} doublet.  (c) For the two-orbit G130M/A sequence of GO-12376 Visit~4 during the FUV flare.  The symbols are the same as in Fig.~25a.  In panel (c), the \ion{Si}{4} and \ion{C}{2} fluxes are rising during the first four FP-POS splits, but are lower and steadier during the second four.  Curiously, there is little evidence for enhanced emission at [\ion{Fe}{21}] 1354~\AA, which is expected to strengthen during a flare.
}
\end{figure}

\clearpage
Figure~26 summarizes high time resolution count rate measurements for the G130M/A visits of the main program (thirteen in all), including the lead-off G285M exposures (\ion{Mg}{2}).  The three intervals noted earlier as anomalous in the \ion{C}{2} and \ion{Si}{4} visit-average profiles now are seen to be clear enhancements over the average levels.  A conspicuous flare rise and fall, especially in \ion{Si}{4}, is visible in the middle of the diagram, from GO-12376 Visit~4 (e.g., Fig.~25c).  In this broad view, FK~Com is seen to be highly variable from hour to hour as well as from day to day, with the highest amplitudes in the hottest features, although the FUV continuum band also displays a surprisingly large degree of variability.  A similarly enhanced continuum variability was noted in the large FUV flare on EK~Dra mentioned earlier, and in that situation there even were two episodes when the continuum showed strong bursts without corresponding changes in the FUV hot lines (or coronal-proxy [\ion{Fe}{21}], which is comparatively strong in EK~Dra).

\begin{figure}[hb]
\figurenum{26}
\vskip  0mm
\hskip   0mm
\includegraphics[scale=0.675,angle=90]{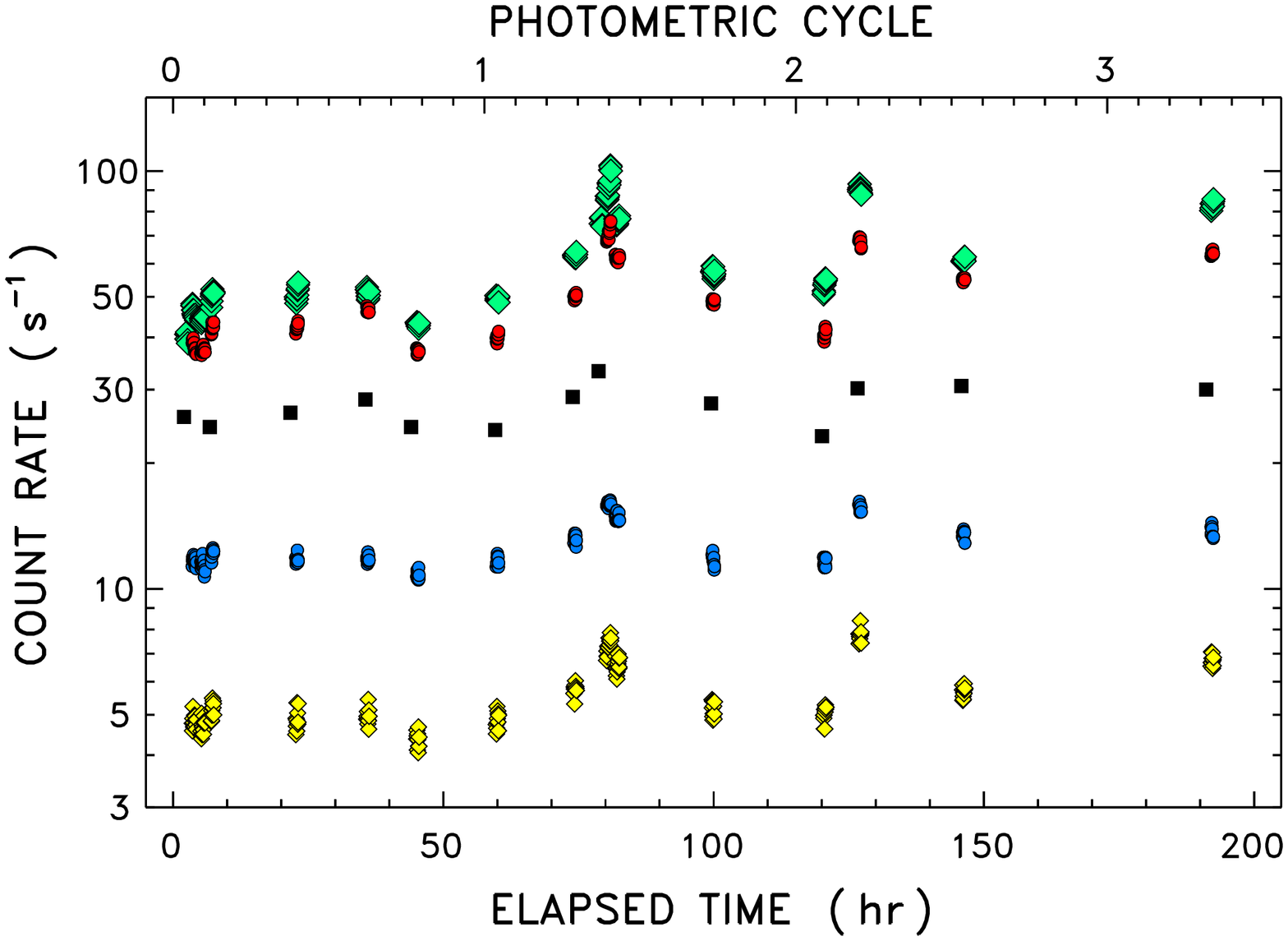} 
\vskip -3mm
\figcaption[]{\small Time series of COS UV count rates during the main program from the G130M and G285M exposures.  \ion{Si}{4} is represented by green diamonds, \ion{C}{2} by (smaller) red dots, \ion{Mg}{2} fluxes  (scaled) by black squares, the 1354~\AA\ complex as blue circles, and the FUV continuum as yellow diamonds.  Several flare enhancements are visible, especially in the hotter features, although all the species tend to vary in step.
}
\end{figure}

\clearpage
\subsection{{\em Chandra}\/ HETGS X-ray Photometry}

For this introductory report, the {\em Chandra}\/ HETGS pointings on FK~Com will be described mainly in terms of integrated spectral count rates.  This global description glosses over the (considerable) additional information available from the pulse-height distribution of the ACIS-S zeroth-order image, which allows changes in the coronal temperature to be inferred; as well as the dispersed MEG and HEG spectra, for which the resolved X-ray emission lines can provide refined knowledge of the coronal temperature structure, including compositional changes, as well as some velocity information in exceptional cases (see, e.g., Ayres et al.\ 2001; Drake et al.\ 2008).  As a preview, Figure~27 illustrates an X-ray spectrum derived from the MEG plus and minus first orders averaged over the main program.  Table~9 lists fluxes of selected bright X-ray lines.  Details of the spectral processing and line measurements will be described elsewhere. 

Figure~28 summarizes HETGS count rates for the several {\em Chandra}\/ visits of the main campaign, accumulated in 2~ks bins.  The S/N per bin ranges from about 15 to 20 (outside flares).  The integrated count rates were summed over both the zeroth-order grating image and the plus and minus arms of the first-order spectra.  The full COS FUV \ion{Si}{4} time series (including the G160M values, divided by 1.4 to compensate for the sensitivity difference relative to G130M at 1400~\AA) is reproduced in the figure for comparison.  Note the several occasions, especially at the end of the {\em Chandra}\/ sequence, of elevated X-ray count rates likely due to flare events.  However, note also that the rapid rise and fall of \ion{Si}{4} at $T\sim$~80~hours occurs on the leading flank of the X-ray rise at that time, rather than, say, at or after the X-ray peak about five hours later.  The delayed soft X-rays probably explain the lack of a strong coronal [\ion{Fe}{21}] response during GO-12376 Visit~4.  Likewise, the \ion{Si}{4} intensity in the middle of the large X-ray event at $T\sim$~190~hours is higher than normal, but not particularly exceptional, whereas the HETGS enhancement exceeds the other X-ray peaks by a considerable margin.  The issue of the association of the FUV hot lines and the coronal X-ray emissions will be addressed in \S{3.7}.  

\begin{figure}
\figurenum{27}
\vskip  0mm
\hskip  -3mm
\includegraphics[scale=0.75,angle=90]{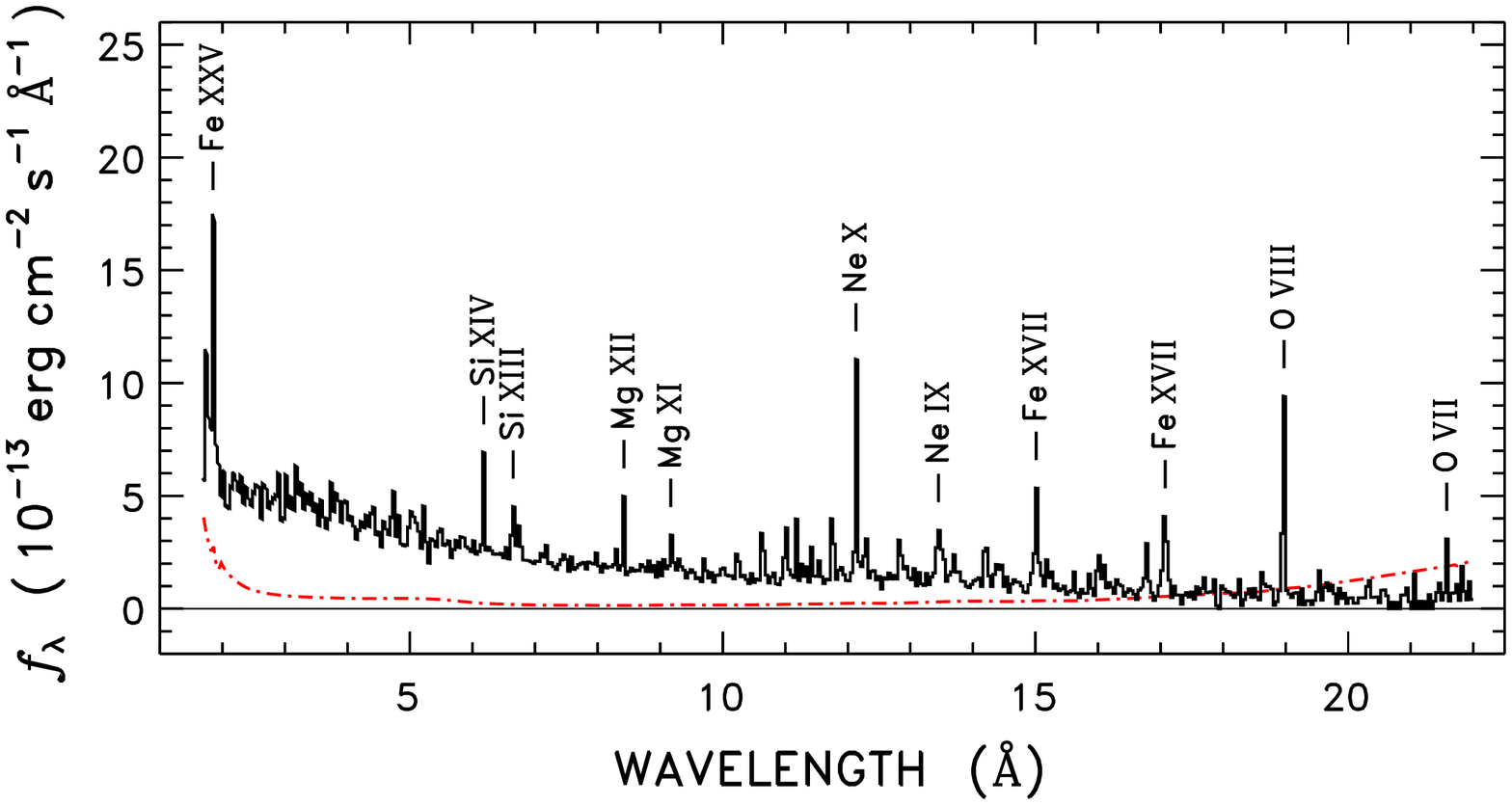} 
\vskip 0mm
\figcaption[]{\small HETGS MEG\,$\pm$1 spectrum of FK~Com, averaged over the COCOA-PUFS main program.  The spectrum was accumulated in 40~m\AA\ bins, and is presented in the same flux density units as for the previous FUV and NUV spectra.  The red dotted-dashed curve is the 1\,$\sigma$ photometric noise level.  Selected bright lines are identified.  These mostly are resonance transitions of hydrogen-like and helium-like ionization stages of abundant elements, although a pair of strong iron L-shell features is marked as well.  Note, also, the hot bremsstrahlung continuum rising toward shorter wavelengths.
}
\end{figure}

\clearpage
\begin{deluxetable}{rcl}
\tablenum{9}
\tablecaption{Selected {\em Chandra}\/ HETGS Measurements}
\tablewidth{0pt}
\tablecolumns{3}
\tablehead{
\colhead{Transition} &  
\colhead{$f_{\rm INT}$} &
\colhead{Notes}\\[3pt]
\colhead{} &  
\colhead{($10^{-13}$ erg cm$^{-2}$ s$^{-1}$)} & \colhead{}\\[3pt]
\colhead{(1)} &  
\colhead{(2)} &
\colhead{(3)}\\[3pt] 
}                
\startdata   
  \ion{S}{16}~4.73    &  $ 0.11{\pm}0.02 $  & Ly$\alpha$ \\
  \ion{Si}{14}~6.18   &  $ 0.20{\pm}0.02 $  & Ly$\alpha$ \\
  \ion{Si}{13}~6.65   &  $ 0.12{\pm}0.01 $  & He\,{\em r}\\
  \ion{Si}{13}~6.74   &  $ 0.06{\pm}0.01 $  & He\,{\em f}\\
  \ion{Mg}{12}~8.42   &  $ 0.15{\pm}0.01 $  & Ly$\alpha$ \\
  \ion{Mg}{11}~9.17   &  $ 0.03{\pm}0.01 $  & He\,{\em r}\\
  \ion{Ne}{10}~10.24  &  $ 0.06{\pm}0.01 $  & Ly$\beta$  \\
  \ion{Fe}{24}~10.62  &  $ 0.08{\pm}0.01 $  & \\
  \ion{Fe}{24}~10.66  &  $ 0.05{\pm}0.01 $  & \\
  \ion{Fe}{23}~10.98  &  $ 0.04{\pm}0.01 $  & \\
  \ion{Fe}{24}~11.18  &  $ 0.11{\pm}0.01 $  & \\
  \ion{Fe}{24}~11.43  &  $ 0.05{\pm}0.01 $  & \\
  \ion{Fe}{23}~11.74  &  $ 0.11{\pm}0.01 $  & \\
  \ion{Fe}{22}~11.77  &  $ 0.07{\pm}0.01 $  & \\
  \ion{Ne}{10}~12.14  &  $ 0.33{\pm}0.03$  & Ly$\alpha$ \\
  \ion{Fe}{23}~12.16  &  $ 0.09{\pm}0.02 $  & \\
  \ion{Fe}{17}~15.01  &  $ 0.19{\pm}0.02 $  & \\
  \ion{O}{8}~18.97    &  $ 0.47{\pm}0.06 $ & Ly$\alpha$ \\
\enddata
\vskip -6mm
\tablecomments{Col.~(1): wavelength designator, following species, in \AA.  Col.~(3) Notes: ``Ly'' indicates a hydrogenic resonance transition ($\alpha= 1$--2; $\beta= 1$--3); ``He'' denotes a helium-like transition (``{\em r}\,'' for the resonance line, ``{\em f}\,'' for the forbidden line).
}
\end{deluxetable}

\begin{figure}
\figurenum{28}
\vskip  0mm
\hskip  -3mm
\includegraphics[scale=0.675,angle=90]{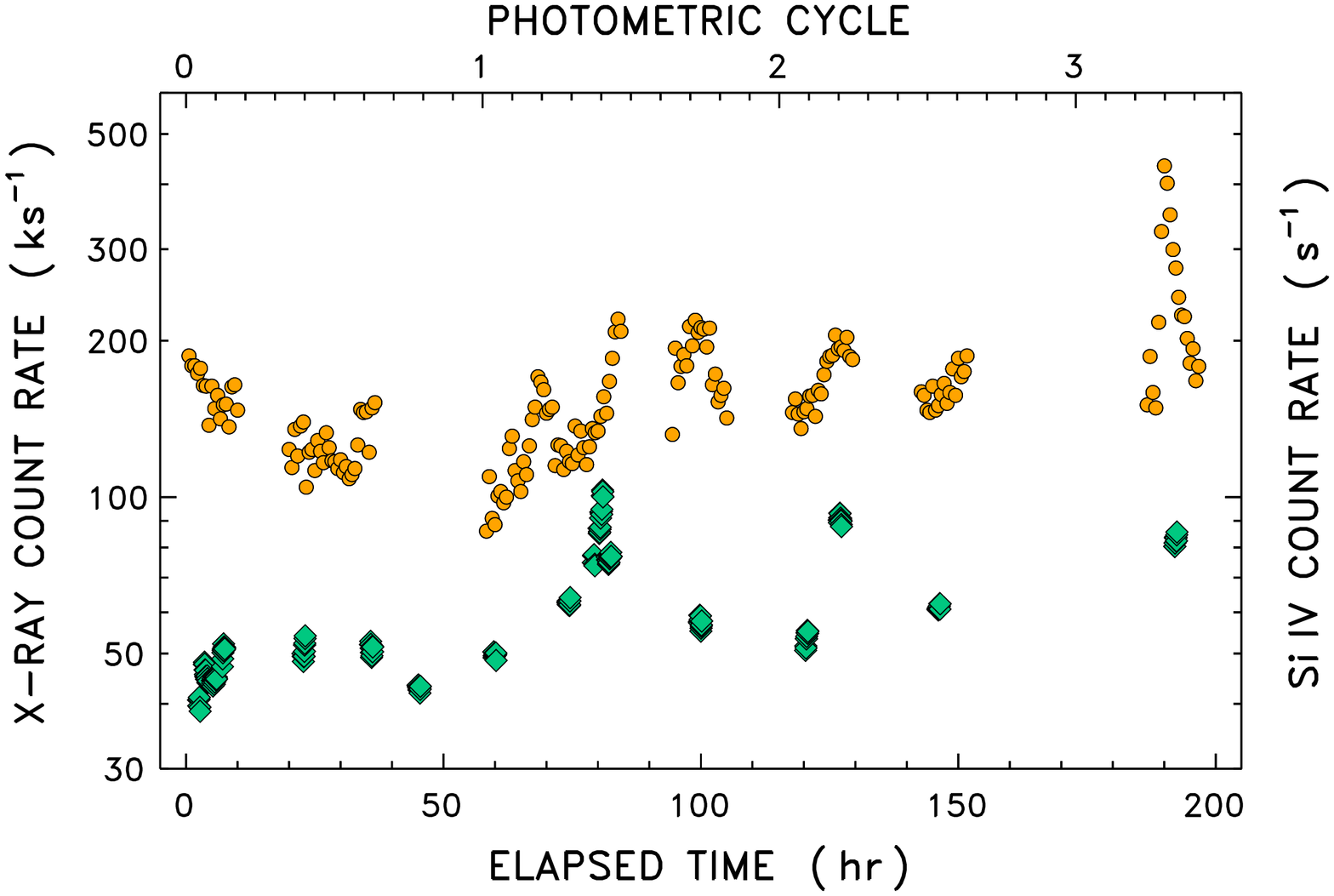} 
\vskip 5mm
\figcaption[]{\small Orange dots are soft X-ray count rates (left hand scale) derived from the HETGS event lists of the zeroth-order image and the combined $\pm$1 grating orders, from the {\em Chandra}\/ pointings of the main campaign.  Green diamonds are the contemporaneous COS FUV \ion{Si}{4} count rates for comparison (right hand scale: same ordinate values as left hand scale, but now s$^{-1}$).
}
\end{figure}

\clearpage
\subsection{Ground-Based Photometry}

Figure~29 summarizes the $V$ and $(V-Ic)$ photometry of FK~Com recorded during the main campaign, and for the week preceding and following it.  The photometry was obtained on a near-nightly basis, and appears noisy during a given observing session, but in reality the photospheric brightness levels, and colors, change in a very systematic way during each interval (as will be clear later, when a phase-folded version of the photometry is illustrated).  The amplitude of the $V$ modulations is about a tenth of a magnitude, which represents a very significant optical brightness change of the star, considering that the equivalent variations for the (albeit low-activity) Sun would be measured in millimagnitudes.  Note that the $(V-Ic)$ color increases when the star becomes fainter in $V$, and vice versa.  This is the well-known property of hyper-active stars that the darker hemispheres are redder owing to the presence of large cool starspot umbrae.  Conversely, however, the upper atmosphere in the vicinity of the starspot groups can be significantly hotter than normal, owing to the impact of enhanced magnetic activity on the energization of the chromosphere and overlying corona, and the propensity of the ``active regions'' to flare.

\begin{figure}
\figurenum{29}
\vskip  0mm
\hskip  -8mm
\includegraphics[scale=0.675,angle=90]{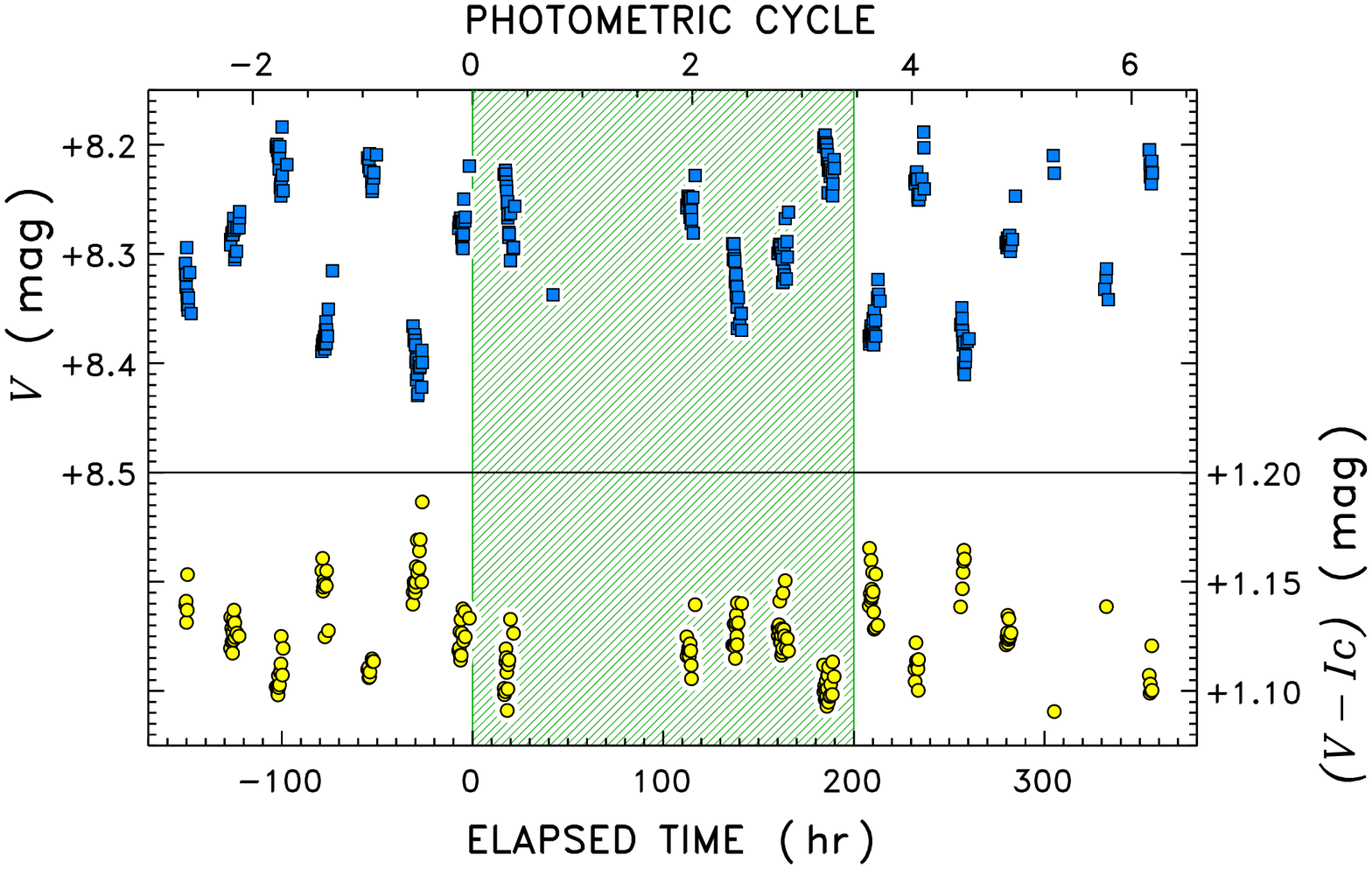} 
\vskip 0mm
\figcaption[]{\small Optical photometry of FK~Com before, during, and after the main campaign period.  Upper part of the figure depicts the $V$ magnitudes (blue squares; note reversed $y$-axis scale: brighter visual luminosities toward top of diagram), while the lower part illustrates the $(V-Ic)$ color (yellow circles; higher values indicate redder colors).  The apparent (nightly) changes are real, and highlight the large and rapid photometric variability of the fast-rotating, heavily spotted G giant.
}
\end{figure}

\clearpage
\subsection{Correlations between Emissions during Quiescent and Flaring Periods}

Figure~30 collects together the main activity diagnostics illustrated in previous figures -- \ion{Si}{4}, soft X-rays, and optical fluxes (${\rm const}{\times}10^{-V/2.5}$)-- into a single diagram, phase-folded according to the photometric period.  The values were normalized to an approximate minimum of each time series, and refer to the left hand scale; except for the optical fluxes which are displayed relative to their maximum, on the right hand scale.  Note the large X-ray flare near $\phi= 0.3$, in the phase band where enhanced activity was seen in chromospheric \ion{Mg}{2} and \ion{C}{2} and subcoronal \ion{Si}{4}.  This active phase band lies between the photometric maximum at $\phi\sim 0.2$ and photometric minimum $\phi\sim 0.6$ (when the dark spots are more numerous on the visible hemisphere).  Finally, discounting the flare enhancements, the phased X-ray light curve appears to be bimodal, with enhancements prior to and following the photometric minimum; reminiscent of the extensive bright plage regions that surround photospheric spots on the Sun.

\begin{figure}
\figurenum{30}
\vskip  0mm
\hskip  -5mm
\includegraphics[scale=0.675,angle=90]{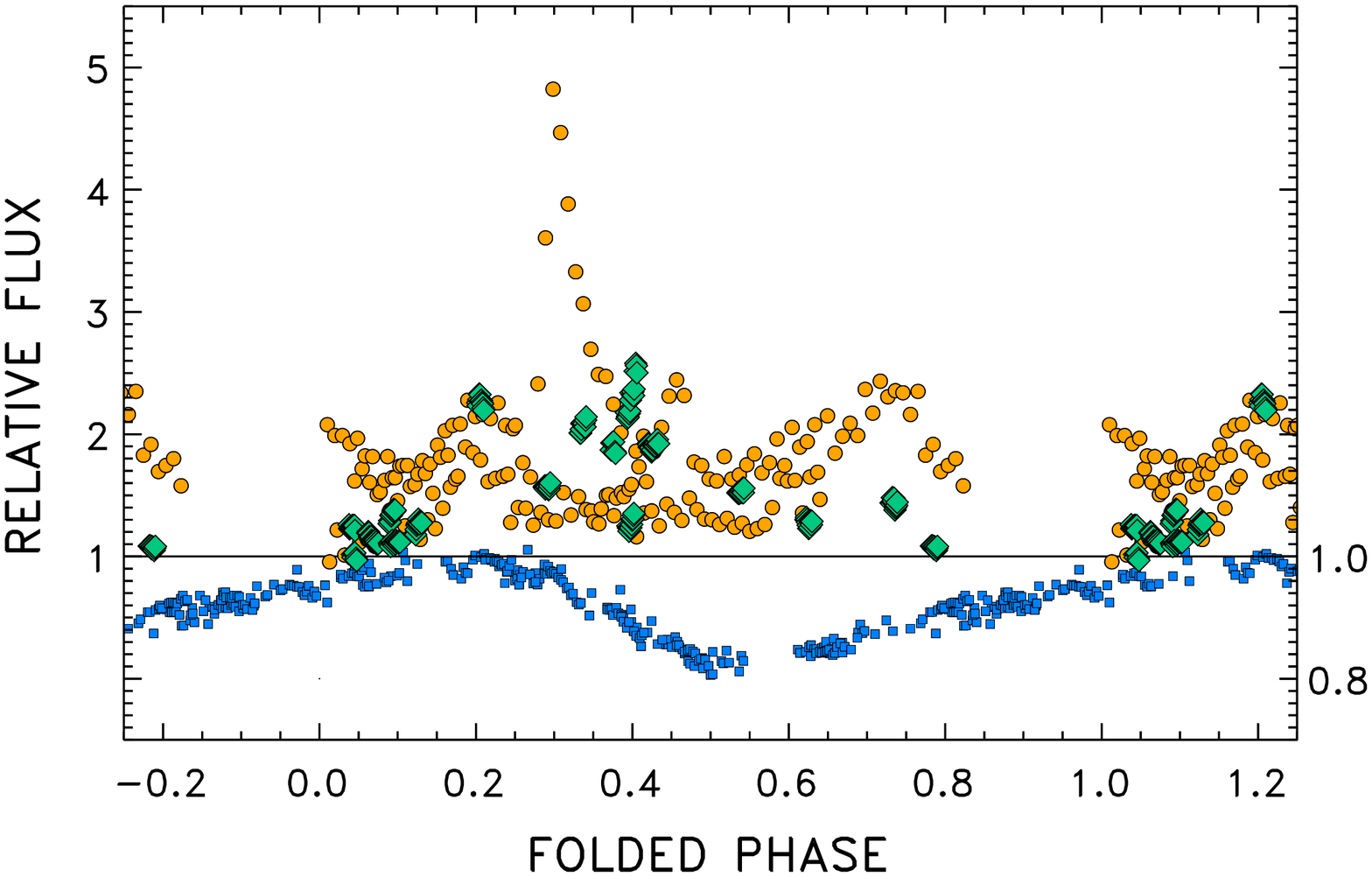} 
\vskip 0mm
\figcaption[]{\small Phase-folded diagram, for the main campaign period, of normalized FUV \ion{Si}{4} (green diamonds) and X-rays (orange dots), on the left hand relative flux scale; and broad-band optical fluxes (smaller blue squares), on the right hand scale.  
}
\end{figure}

Figure~31 compares chromospheric \ion{C}{2} and coronal X-ray fluxes against contemporaneous subcoronal \ion{Si}{4}, all normalized to an average flux obtained from the lower half of each intensity distribution (to exclude flares).  The \ion{C}{2} and \ion{Si}{4} fluxes were collected simultaneously (in $\sim 150$~s bins in the G130M exposures), but the HETGS soft X-rays were accumulated in coarser time intervals to achieve similar S/N.  To allow a fair comparison, the \ion{Si}{4} values (G130M/A or G160M/B) within a distinct {\em HST}\/ orbit were averaged, then an average X-ray flux was derived from the three HETGS bins closest to the mid-time of the \ion{Si}{4} values.  Error bars in the figure are (${\pm}$1\,$\sigma$) standard deviations of the average fluxes, to highlight the temporal variability within the \ion{Si}{4}/X-ray super-bin. The systematic trend of \ion{C}{2} toward the upper right represents the influence of several FUV flares, while the dispersions of the clusters of points near the origin (1,1) reflect the typical variability during quieter times.  The dashed line approximates an $\alpha= 0.9$ power law, as was found for \ion{C}{2} vs.\ \ion{Si}{4} in a large FUV flare captured on hyper-active EK~Dra (Ayres 2015a).  Although \ion{C}{2} is well-correlated with \ion{Si}{4}, surprisingly the soft X-rays appear to be much less so.

\begin{figure}[hb]
\figurenum{31}
\vskip  -5mm
\hskip  21mm
\includegraphics[scale=0.675]{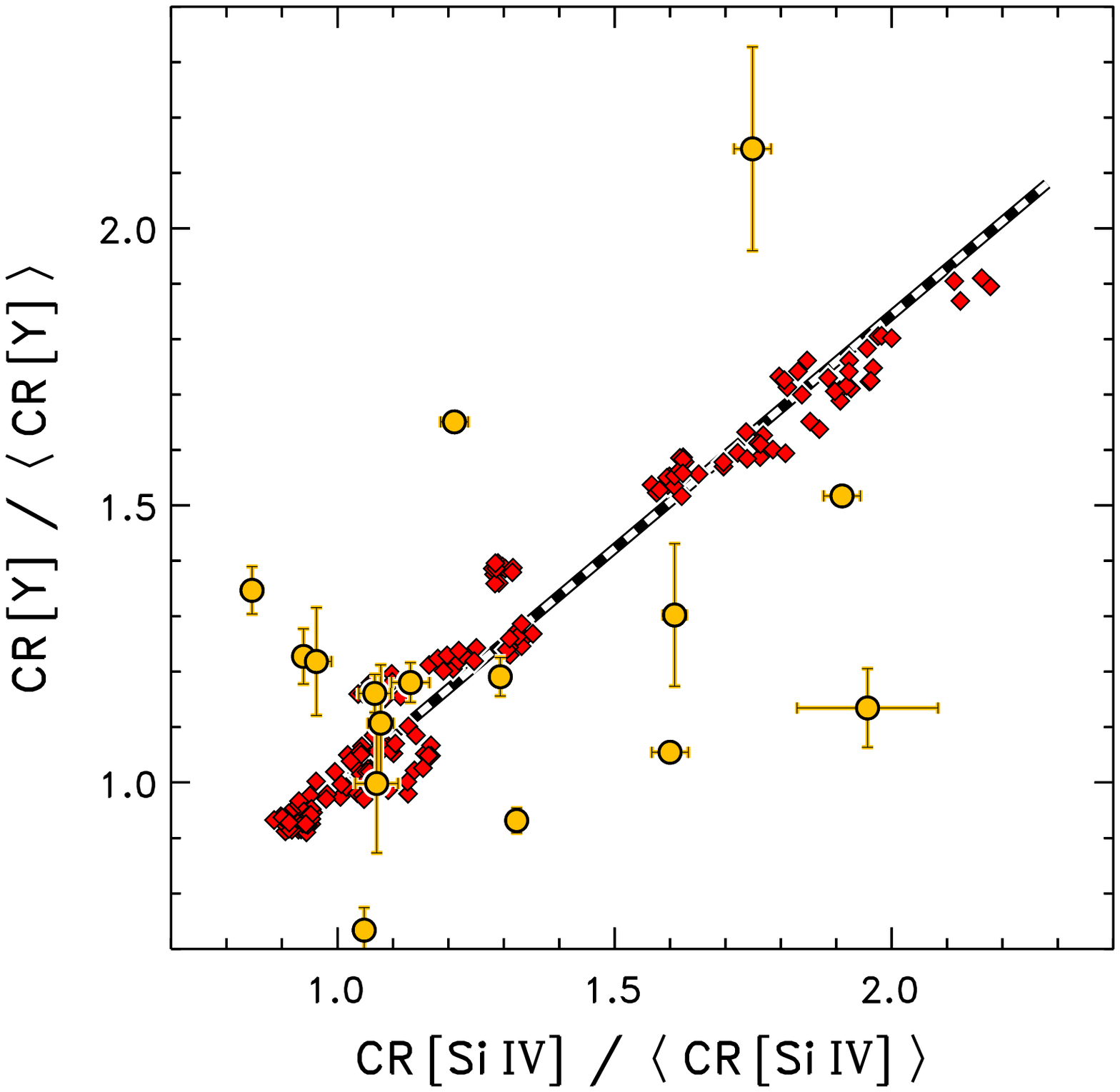} 
\vskip -3mm
\figcaption[]{\small Correlation diagram pitting 0.5-10~keV soft X-rays (orange circles and error bars) and \ion{C}{2} 1335~\AA\ (red diamonds), normalized to flare-excised averages, against the similarly normalized \ion{Si}{4} doublet count rates.  Chromospheric \ion{C}{2} appears to be well-correlated with subcoronal \ion{Si}{4}: the dashed line approximates an $\alpha= 0.9$ power law.  Conversely, the coronal soft X-rays appear much less correlated.
}
\end{figure}

\clearpage
\section{CONCLUSIONS}

This introduction to the COCOA-PUFS Project lays a foundation for future, more focused, studies that will consider the visible, FUV, and X-ray observations of FK~Com in more detail.  These future efforts will include, for example: Doppler imaging and ZDI maps of the distribution of spotted areas on the stellar surface; modeling the chromospheric properties necessary to reproduce the apparent rotational profiles of optically thick features like \ion{Mg}{2}, as well as optically thinner transitions like \ion{Cl}{1}; time-resolved profiles of chromospheric \ion{C}{2} and subcoronal \ion{Si}{4} during the flaring intervals; a careful search for so-far elusive [\ion{Fe}{21}]; and the HETGS spectra and ACIS-S zeroth-order pulse-height distributions, including emission-measure modeling, compositional changes during flares, and possible Doppler effects in the strongest X-ray lines (individually or collectively: see Drake et al.\ 2008).  Layered on top of the primarily observational studies, is envisioned a theoretical effort to derive, from the ZDI maps and rotational behavior of the various FUV and X-ray diagnostics, a magnetospheric model to capture the extreme coronal conditions on the exceptional yellow giant rotating near breakup (see, e.g., Cohen et al.\ 2010).

In the meantime, the initial foray into COCOA-PUFS has several conclusions of its own:

\begin{itemize}

\item[(1)] The chromospheric and subcoronal features of FK~Com are exceptionally broad, super-rotationally so; and exhibit large Doppler shifts, especially \ion{Si}{4}, which appears to be significantly redshifted in the epoch-average non-flare profile.  The chromospheric features of \ion{C}{2} and \ion{Mg}{2} also appear to be redshifted in the respective average profiles, although less so than \ion{Si}{4}.  There are, however, issues with the appropriate RV to apply to these spectra; and the influence of the ``missing phases'' (no observations in phase band $\phi= 0.8-1.0$) on the average profiles.  Nevertheless, these results help to clarify the singular {\em FUSE}\/ observation, which also had recorded large redshifts of, especially, subcoronal \ion{O}{6}.  Such shifts now are seen to be mainly due to rotational modulation effects, but likely partly also to an underlying systemic redshift of the hot lines, perhaps analogous to the steady radial downflows recorded in solar counterparts (e.g., Peter 2006, and references therein).  An early interpretation of the solar redshifts attributed the mass flows to spicular material injected into the corona and heated up, then cooling and draining back to the chromosphere below (Pneuman \& Kopp 1978).  More recent explanations, motivated by 3D MHD simulations, have suggested that episodic energy releases within the chromosphere itself impulsively heat the local gas to coronal temperatures, then the back-reaction to the explosive expansion pushes on the lower atmosphere causing the redshifted emissions at intermediate temperatures (Guerreiro et al.\ 2013, and references to previous work therein).  The latter scenario places the redshifted sources at relatively low altitudes, in compact magnetic loops.  This mechanism might be less relevant to FK~Com, where the super-rotational broadening of the hot lines suggests that the redshifts are occurring much further out.  Another solar phenomenon associated with redshifted subcoronal lines, which perhaps is more relevant to flare-dominated FK~Com, is the observed ballistic splashdowns associated with back-falling flare material (Reale et al.\ 2014).  
 
\item[(2)] There was no obvious sign of the [\ion{Fe}{21}] 1354~\AA\ coronal forbidden line in the COS FUV spectra, even during flaring periods.  One of the key objectives of the originally conceived COCOA-PUFS was to record Doppler widths of this $T{\sim}10$~MK feature at COS resolution ($\sim 20$~km s$^{-1}$ versus the $\sim$500~km s$^{-1}$ achieved, at best, with HETGS for coronal resonance lines) to gauge whether the hot coronal plasma is as extended as apparently is the cooler chromospheric gas traced by optical H$\alpha$.  It was possible, for example, that the cooler gas is organized in extended solar-like ``prominences,'' confined by high-altitude magnetic loops (see, e.g., Strassmeier [1994] for the case of HU~Virginis, which exhibited a large-scale chromospheric temperature siphon flow likely confined by a giant flux rope), while the multi-million degree coronal plasma is localized closer to the stellar surface in compact high-pressure loop systems.  However, the pilot synthetic spectra that indicated [\ion{Fe}{21}] should be easily detected had not taken into account what now is seen as the over-broadening, well beyond ${\pm}\upsilon\sin{i}$, of the flanking \ion{Cl}{1} 1351~\AA\ and \ion{O}{1} 1355~\AA\ emissions.  These blend together in their red and blue wings, respectively, making detection of intrinsically faint [\ion{Fe}{21}] (see Fig.~15 for 24~UMa) more challenging.  There is hope, nevertheless, that constraints can be placed on the coronal forbidden line width by carefully modeling the chromospheric blending (e.g., Saar et al.\ 2014).  At the same time, the apparent super-rotational broadening of \ion{Si}{4} indicates that gas at least as hot as $8{\times}10^4$~K exists far from the star.  It is not unreasonable to imagine that the subcoronal gas was derived from the cooling of even hotter, i.e., coronal material.  In particular, the systemic redshifts of \ion{Si}{4} in the epoch-average profile strongly suggest that the radiating gas is falling back to the surface, which in turn suggests that the gas is well above its normal equilibrium scale height.  In order to have achieved the (higher) altitude in the first place, the original material must have been much hotter.  This also is true for cool prominences in the solar corona, whose apparent stability belies a continual condensation of hot coronal material into the magnetic saddle region (see, e.g., Schmit et al.\ 2013, and references therein).  If the hot corona of FK~Com is as extended, or more extended, than indicated by the cooler species, super-rotationally broadened [\ion{Fe}{21}] likely will be difficult to detect.  However, lack of a detection (based on predicted visibilities derived from HETGS X-ray lines of similar excitation) also might prove the point.\\[3mm]

\item[(3)]  Flares are a common occurrence on FK~Com, both ultraviolet and X-ray.  An X-ray enhancement of five times was captured during one of the HETGS pointings, a significant outburst for a star whose normal coronal $L_{\rm X}$ already is $10^4$ times solar, but not remarkable given the similar scales of high-energy variability seen in previous X-ray observations (including the earlier HETGS pointing on FK~Com in 2000, described by Drake et al.\ [2008], which captured the peak and decay of a similar-size event).  The FUV outbursts were smaller, about two times the epoch-average level.  In both cases there is the issue of a relatively limited sampling interval, albeit much more extensive than previous FUV (e.g., {\em IUE}\,) and X-ray efforts ({\em ROSAT, ASCA, XMM-Newton, Chandra}\,).  Nevertheless, the specific X-ray and FUV enhancements during the FK~Com campaign did not show a strong correlation.  This is somewhat surprising in light of solar experience, where flares tend to affect all the atmospheric temperature regimes together; or the example of the large FUV flare decay on EK~Dra where [\ion{Fe}{21}] (proxy for soft X-rays) varied in step with the (simultaneously observed) \ion{Si}{4} emission.  Ayres \& France (2010) suggested that isolated ``flares'' at \ion{Si}{4} temperatures might represent catastrophic cooling events in previously hot coronal gas, producing cooling (down)flows, an ``anti-flare'' if you will.  Note, however, that the best subcoronal flare example in the FK~Com campaign appears to show significant extra blueshifts in the evolving \ion{Si}{4} profiles (Fig.~25c), although that still could represent a downflow if the event was above the limb on the far side of the star, but still in the line of sight.
 
\end{itemize}

The summary conclusion is that ultra-fast-rotating FK~Com is surrounded by an extended region filled by material at least as hot as $\sim 10^5$~K, and probably much hotter, although pervaded by high-altitude structures at chromospheric ($\sim 10^4$~K) temperatures, something like cool prominences on the Sun embedded in the hot corona.  This reinforces the conclusion that the previously well observed highly variable optical H$\alpha$ emission of FK~Com arises in a complex stellar magnetosphere, rooted in the strong fields of the pervasive surface starspots; rather than in a temporary ``excretion disk'' ejected during the binary merger thought to have created the fast-spinning yellow giant in the first place.  In this view, FK~Com is not a unique object, but rather just an extreme example of the more slowly rotating (but still quite fast) Hertzsprung gap yellow giants (see Ayres et al.\ 1998 and Fig.~1 here), which also display super-rotational broadening of their subcoronal lines, as well as [\ion{Fe}{21}] in a few favorable cases (Ayres et al.\ 2003).  At the same time, the prevalence of redshifts among the chromospheric and subcoronal species raises the possibility that the strongly confining magnetosphere might inhibit mass outflows and associated angular momentum loss.  For example, the magnetosphere could stall the breakout of coronal mass ejections (CME), which normally might carry away significant mass (and angular momentum) from the frequently flaring giant.  This notion has support in the theoretical magnetospheric models of a simulated FK~Com-like corona described by Cohen et al.\ (2010), which show the development of a strong toroidal field configuration that could suppress equatorial mass loss.  The inhibition of angular momentum loss could postpone what might have been a rather short-lived catastrophic spindown phase, allowing FK-Com-like binary mergers to persist longer as fast rotators, and thus be recognized observationally as unique objects.  It also is possible that similar magnetospheric trapping occurs on other hype-ractive stars, and perhaps is the cause of the surprisingly long-lived rapid rotator sequence seen in G and K dwarfs of young clusters (e.g., the ``C-branch'' of Barnes 2003).

Finally, although the initial episode of COCOA-PUFS is concluding, the saga is far from over.  The planned more focused examinations of the ground-based, {\em HST,} and {\em Chandra}\/ material, and theoretical magnetospheric modeling in parallel, hold great promise for unraveling the details of late-type activity at its outer limits.  As they say, ``Stay tuned.''

\acknowledgments
This work was supported by grants GO-12279, GO-12376, and GO-13938 from the Space Telescope Science Institute, based on observations from {\em Hubble Space Telescope}\/ collected at STScI, operated by the Associated Universities for Research in Astronomy, under contract to NASA; and grants GO1-12021X and GO1-12037X from the Smithsonian Astrophysical Observatory, based on observations from the {\em Chandra}\/ X-ray Observatory, collected and processed at the {\em Chandra}\/ X-ray Center, operated by SAO under contract to NASA.  VK and JJD also were supported by NASA contract NAS8-03060 to the CXC.  DPH was supported by NASA through SAO contract SV3-73016 to MIT for the CXC and Science Instruments.  The research leading to these results has received funding from the European Community's Seventh Framework Program (FP7/2007-2013) under grant agreement number RG226604 (OPTICON).  Special thanks to the schedulers of {\em HST}\/ and {\em Chandra}\/ who were able to successfully carry out the challenging coordination of the two space observatories during the main campaign period in 2011 April.  This project made use of public databases hosted by {SIMBAD}, maintained by {CDS}, Strasbourg, France; the Mikulski Archive for Space Telescopes at STScI in Baltimore, Maryland; and the High Energy Astrophysics Science and Research Center at the NASA Goddard Space Flight Center, in Greenbelt, Maryland.  We also thank the various ground-based facilities for their support of the FK~Com campaign, including especially the ESO Telescopes at the La Silla Paranal Observatory under program 087.D-0294.    

\clearpage
\appendix
\section{Glossary of Selected Abbreviations}
\small
\begin{tabular}[ht]{ll}
24~UMa     &                       24~Ursa Majoris (G-type giant star)\\
3D         &                         Three-Dimensional\\
ACIS-S     &                            {\em Chandra}\/ Advanced CCD Imaging Spectrometer, Spectroscopy Array\\
ACQ        &                             Target Acquisition\\
ACQ/SEARCH  &                      Type of COS Target Acquisition\\
{\em ASCA}       &         Advanced Satellite for Cosmology and Astrophysics (Japanese  X-ray Mission)\\
ASTRAL     &              Advanced Spectral Library Project (STIS spectra of selected bright stars)\\
BKM         &                  {\em HST}\/ program GO-12279 \\
BME         &                  {\em HST}\/ program GO-12376 \\
BOA        &                            COS Bright Object Aperture \\
CALCOS     &                        COS Calibration Pipeline \\
CALSTIS    &                        STIS Calibration Pipeline \\
CCD        &                              Charged-Coupled Device (Camera)\\
{\em Chandra}\/  &        Chandra X-ray Observatory (NASA Great Observatory)\\
CME         &                       Coronal Mass Ejection\\
COCOA-PUFS  & Coordinated Campaign of Observations and Analysis, Photosphere to \\
            & ~~~~Upper Atmosphere, of a Fast-rotating Star\\
COS          &                           Cosmic Origins Spectrograph ({\em HST}\/ instrument)\\
CR           &                       Count Rate\\
CTI          &                   Charge Transfer Inefficiency\\ 
CXO          &        Chandra X-ray Observatory\\
D            &                        (In-)Dispersion (COS peak-up type)\\
D$_3$        &                   {He}~{\footnotesize I} 5876 \AA\ \\
DA           &                     Hot White Dwarf \\
EK~Dra       &                      EK Draconis (G-type dwarf star)\\
ESO          &                              The European Southern Observatory \\
ETC          &                  Exposure Time Calculator \\
FGS          &                            Fine Guidance Sensor ({\em HST}\/ instrument: part of pointing control loop)\\
FK~Com       &                   FK Comae Berenices (star of COCOA-PUFS Project) \\
FORS2        &             Focal Reducer/Low Dispersion Spectrograph 2 (VLT instrument) \\  
FP-POS       &                         COS Grating Offset Position \\
FUSE         &                       Far-Ultraviolet Spectroscopic Explorer (NASA Mission, 912-1180~\AA) \\
FUV          &                       Far-Ultraviolet (approx.\ 1100--1700~\AA) \\
FWHM~~~~~~~~~~         &              Full Width at Half Maximum Intensity \\
Gem          &                  Geminorum Interstellar Cloud \\
H$\alpha$    &                  Balmer $\alpha$ line of Hydrogen ({H}~{\footnotesize I} 6563~\AA) \\
\end{tabular}

\begin{tabular}[ht]{ll}
HEG          &                              High Energy Grating Arm of HETGS \\
HETGS        &                    High-Energy Transmission Grating Spectrometer ({\em Chandra}\/ instrument) \\
HJD          &                    Heliocentric Julian Date \\
hk           &                       {Mg}~{\footnotesize II} h \& k lines \\
{\em HST}\/          &                       Hubble Space Telescope (NASA Great Observatory)\\
$Ic$  &                  Cousins Near-Infrared Broad-Band Filter\\
ISM          &                  Interstellar Medium \\
{\em IUE}\/          &                         International Ultraviolet Explorer  (NASA UV Mission) \\
ks           &                 Kiloseconds ($10^3$~s)\\
LASP         &                Laboratory for Atmospheric and Space Physics, University of Colorado \\
LHS          &                  Left Hand Side \\
LISIRD       &               LASP Interactive Solar Irradiance Data Center \\
LISM         &                    Local Interstellar Medium \\
LP           &                            Large Project \\
Ly{$\alpha$} &                 Lyman $\alpha$ ({H}~{\footnotesize I} 1215~\AA) \\
MEG          &                  Medium Energy Grating Arm of HETGS \\
MHD          &                    Magneto-Hydrodynamic \\
MIRROR-A     &               COS Acquisition Mirror\\
MIRROR-B     &               COS Acquisition Mirror\\
MIRVIS       &                         STIS Optical Element (mirror, often used in target acquisitions) \\
MJD          &                            Modified Julian Date \\
MK           &                               Million Kelvin ($10^6$~K)\\
NARVAL       &           High-Resolution Spectropolarimeter on Bernard Lyot Telescope, Pic du Midi \\
ND3          &              STIS Neutral Density Filter (factor of $10^3$ attenuation)\\
NGP          &                   North Galactic Pole \\
NIST~~~~~~~~~~         &                 National Institute of Standards and Technology \\
NUV          &                       Near-Ultraviolet (approx.\ 1700--3200~\AA) \\
ObsID        &                          {\em Chandra}\/ Observation Identification Number \\
PSA          &                           COS Primary Science Aperture \\ 
resel        &                  Resolution Element \\
RHS          &                  Right Hand Side \\
{\em ROSAT}        &                     {\em R{\"o}ntgensatellit}\/ (German X-ray Mission) \\
RS CVn       &                    RS Canum Venaticorum (active late-type binary systems) \\
RV           &                    Radial Velocity \\
S/N          &                   Signal-to-Noise Ratio \\
SARG         &               TNG High-Resolution Spectrograph \\
sdO          &                     O-type Subdwarf \\
s.e.         &                        Standard Error of the Mean \\
SES          &                  STELLA Echelle Spectrograph  \\
StarCAT      &             STIS Stellar Spectral Catalog \\
STELLA       &             Stellar Activity Robotic Observatory, Tenerife \\
\end{tabular}

\begin{tabular}[ht]{ll}
STIS         &                        Space Telescope Imaging Spectrograph ({\em HST}\/ instrument) \\
TAC          &                         Telescope Allocation Committee \\
TNG          &                Telescopio Nazionale Galileo, Roque de los Muchachos Observatory, La Palma \\
T-TAG        &                           Time-Tag (COS data acquisition mode)\\
UV           &                        Ultraviolet  (approx. 1100--3200~\AA) \\
UVES~~~~~~~~~~         &                             Ultraviolet and Visual Echelle Spectrograph (VLT instrument) \\
$V$          &                 Johnson Visual Broad-Band Filter\\
VLT          &                Very Large Telescope, Paranal Observatory \\
x1d          &                 CALSTIS high-level spectral data product \\
x1dsum          &                 CALCOS high-level spectral data product \\
XD           &                             Cross-Dispersion (COS peak-up type) \\
{\em XMM-Newton}   &           X-ray Multi-Mirror Mission (European X-ray Mission) \\
ZDI          &                         Zeeman Doppler Imaging \\
\end{tabular}


\end{document}